\documentclass[12pt]{article}
 \pdfoutput=1
\usepackage{bbm}
\usepackage{color}
\usepackage{authblk}
\usepackage{latexsym}
\usepackage{epsfig,amssymb,euscript, mathrsfs,cite}
\usepackage{amsmath}
\usepackage{mathrsfs} 
\usepackage{enumerate}
\definecolor{MyDarkBlue}{rgb}{0.15,0.15,0.45}
\usepackage[linktocpage=true]{hyperref}
\hypersetup{
colorlinks=true,
citecolor=MyDarkBlue,
linkcolor=MyDarkBlue,
urlcolor=MyDarkBlue,
pdfauthor={},
pdftitle={},
pdfsubject={hep-th}
}
\newsavebox{\ns}
\newsavebox{\dbrane}
\newsavebox{\dbshort}

\def\be{\begin{equation}}
\def\ee{\end{equation}}

\def\gt{\gamma}
\def\v{\vec{v}}
\def\w{\vec{w}}
\def\u{\vec{u}}

\def\b{\vec{b}}

\newcommand{\N}{\nonumber}
\newcommand{\nn}{\nonumber}

\newcommand\R{\mathbb{R}}
\newcommand\Z{\mathbb{Z}}

\newcommand\C{\mathbb{C}}

\newcommand\diff{\mathrm{d}}
\newcommand{\de}{\partial}

\newcommand{\ii}{\mathrm{i}}

\newcommand{\ex}{\mathrm{e}}
\newcommand{\vol}{\mathrm{vol}}
\newcommand{\Vol}{\mathrm{Vol}}

\newcommand{\fm}{\mathfrak{n}}

\newcommand{\cZ}{\mathscr{Z}}
\newcommand{\cS}{\mathscr{S}}
\newcommand{\csugra}{c_{\mathrm{sugra}}}

\newcommand{\BB}{B}

\newcommand{\Ssusy}{S_{\mathrm{SUSY}}}

\newlength{\sswidth}

\newcommand{\pJ}{\mathtt{p}}
\newcommand{\qJ}{\mathtt{q}}

\newcommand{\n}{l}

\numberwithin{equation}{section}       

\begin{document}

\begin{titlepage}

\begin{flushright}
Imperial/TP/2019/JG/01\\
\end{flushright}

\vskip 1.5cm

\begin{center}

{\Large \bf  Toric geometry and the dual of ${\cal I}$-extremization}

\vskip 1cm

{Jerome P. Gauntlett$^{\mathrm{a}}$, 
Dario Martelli$^{\mathrm{b},\dagger}$\renewcommand*{\thefootnote}{\fnsymbol{footnote}}
\footnotetext[2]{On leave at the Galileo Galilei Institute, Largo Enrico Fermi, 2, 50125 Firenze, Italy.}and James Sparks$^{\mathrm{c}}$}

\vskip 0.5cm

${}^{\mathrm{a}}$\textit{Blackett Laboratory, Imperial College, \\
Prince Consort Rd., London, SW7 2AZ, U.K.\\}

\vskip 0.2cm
${}^{\mathrm{b}}$\textit{Department of Mathematics, King's College London, \\
The Strand, London, WC2R 2LS,  U.K.\\}

\vskip 0.2 cm
${}^{\,\mathrm{c}}$\textit{Mathematical Institute, University of Oxford,\\
Andrew Wiles Building, Radcliffe Observatory Quarter,\\
Woodstock Road, Oxford, OX2 6GG, U.K.\\}

\end{center}

\vskip 1.5 cm

\begin{abstract}
\noindent  
We consider $d=3$, $\mathcal{N}=2$  gauge theories arising on membranes sitting at the apex of an arbitrary toric Calabi-Yau 4-fold cone singularity that are then further compactified on a Riemann surface, $\Sigma_g$,
with a topological twist that preserves two supersymmetries.
If the theories flow to a superconformal quantum mechanics
in the infrared, then they have a $D=11$ supergravity dual of the form 
AdS$_2\times Y_9$, with electric four-form flux and 
where $Y_9$ is topologically a fibration of a Sasakian $Y_7$ over $\Sigma_g$. 
These $D=11$ solutions are also expected to arise as the near horizon limit of magnetically charged  black holes in AdS$_4\times Y_7$, with a Sasaki-Einstein metric on $Y_7$. 
We show that an off-shell entropy function for the dual AdS$_2$ solutions may be computed using the toric data and K\"ahler class parameters
of the Calabi-Yau 4-fold,  that are encoded in a \emph{master volume}, as well as 
a set of integers that determine the fibration of $Y_7$ over $\Sigma_g$ and a K\"ahler class parameter for $\Sigma_g$.
We also discuss the class of supersymmetric AdS$_3\times Y_7$ solutions of type IIB supergravity with five-form flux only in the case that $Y_7$ is toric,  
and show how the off-shell central charge of the dual field theory can be obtained from the toric data.
We illustrate with several examples, finding agreement both with explicit supergravity solutions
as well as with some known field theory results concerning ${\cal I}$-extremization.
\end{abstract}

\end{titlepage}

\pagestyle{plain}
\setcounter{page}{1}
\newcounter{bean}
\baselineskip18pt

\newpage


\tableofcontents

\section{Introduction}\label{sec1}

A common feature of supersymmetric conformal field theories (SCFTs) with an abelian R-symmetry
is that the R-symmetry, and hence important physical observables,
can be obtained, in rather general circumstances and in various spacetime dimensions, 
via an extremization principle. In $\mathcal{N}=1$ SCFTs
in $d=4$, for example, the R-symmetry can be obtained via the procedure of $a$-maximization \cite{Intriligator:2003jj},
while for 
$\mathcal{N}=(0,2)$ SCFTs in $d=2$ it can be obtained via $c$-extremization \cite{Benini:2012cz}. In each of these cases
one constructs a trial central charge, determined by the 't Hooft anomalies of the  
theory, which is a function of the possible candidate R-symmetries. After extremizing the trial central charge
one obtains the R-symmetry, and when the trial central charge is evaluated at the extremal point one gets
the exact $a$ central charge and the right moving central charge, $c_R$, for the $d=4$ and $d=2$ SCFTs, respectively.

Next, for $\mathcal{N}=2$ SCFTs in $d=3$, one can use 
$F$-extremization \cite{Jafferis:2010un}. The key quantity now is the free energy of the theory defined on a round three sphere, $F_{S^3}$. After extremizing a 
trial $F_{S^3}$, again calculated as a function of the possible R-symmetries, one finds both the R-symmetry and the free energy at the extremal point. Turning to SCFTs in $d=1$ with two supercharges and an abelian R symmetry, there is not, as far as we know, an analogous
general field theory proposal concerning $F$-extremization, although one has been recently discussed in the context of holography \cite{Couzens:2018wnk},
as we recall below. On the other hand there is a proposed 
``$\mathcal{I}$-extremization'' procedure \cite{Benini:2015eyy}
for the class of such $d=1$ SCFTs that arise after compactifying an $\mathcal{N}=2$ SCFT in $d=3$ on a Riemann
surface, $\Sigma_g$, of genus\footnote{The genus $g=0$ case was discussed in \cite{Benini:2015eyy}; generalizing to $g\ne 0$ was discussed in \cite{Cabo-Bizet:2017jsl} and noted in footnote 5 of \cite{Azzurli:2017kxo}, building on \cite{Hosseini:2016tor,Benini:2016hjo}.} $g$. For this class one considers the topologically twisted index $\mathcal{I}$ for the $d=3$ theory on $\Sigma_g\times S^1$
as a function of the twist parameters and chemical potentials for the flavour symmetries.  After extremization one obtains the index, which
is expected to be the same as the logarithm of the partition function of the $d=1$ SCFT. 
While significant 
evidence for $\mathcal{I}$-extremization has been obtained, it does not yet have the same status
as the $a$-, $c$- and $F$- extremization principles.

For the special subclass of these SCFTs that also have a large $N$ holographic dual, we can
investigate the various extremization principles from a geometric point of view. To do this one first needs to find 
a precise way of taking
the supergravity solutions off-shell in order to set up an appropriate extremization problem. A guiding principle, 
that has been effectively utilised in several different situations, is to identify a suitable class of supersymmetric geometries in which one demands the existence of certain types of Killing spinors, but without imposing the full equations of motion. The best understood examples are those associated with Sasaki-Einstein ($SE$) geometry,
specifically the class of AdS$_5\times SE_5$ solutions of type IIB and the AdS$_4\times SE_7$ solutions of $D=11$ supergravity
that are dual to $\mathcal{N}=1$ SCFTs in $d=4$ and $\mathcal{N}=2$ SCFTs in $d=3$, respectively. Here 
one goes off-shell by relaxing the Einstein condition and considering the space of Sasaki metrics.
It was shown in \cite{Martelli:2005tp,Martelli:2006yb} that the Reeb Killing vector field for the Sasaki-Einstein metric, 
dual to the R-symmetry in the field theory, can be obtained by extremizing
the normalized volume of the Sasaki geometry as a function of the possible Reeb vector fields 
on the Sasaki geometry. Interestingly,
while this geometric extremization problem is essentially the same for $SE_5$ and $SE_7$, and indeed is applicable for arbitrary $SE_{2n+1}$, it is associated with the different physical phenomena of $a$-maximization and $F$-extremization in the $d=4$ and $d=3$ dual field theories, respectively (although see \cite{Giombi:2014xxa}).

In a recent paper \cite{Couzens:2018wnk} an analogous story was presented for the class of AdS$_3\times Y_7$ solutions of type IIB with non-vanishing five-form flux only \cite{Kim:2005ez} and the class of AdS$_2\times Y_9$ solutions of $D=11$ supergravity with purely electric four-form flux \cite{Kim:2006qu}, that are dual to $\mathcal{N}=(0,2)$ SCFTs in $d=2$ and $\mathcal{N}=2$ SCFTs in $d=1$, respectively. The geometry associated with
these solutions was clarified in \cite{Gauntlett:2007ts} where it was also shown that they are examples of an infinite family of ``GK geometries" $Y_{2n+1}$. As explained in \cite{Couzens:2018wnk}, one can take these GK geometries off-shell in such a way to obtain a class of supersymmetric geometries for 
which, importantly, one can still impose appropriate flux quantization conditions. These supersymmetric geometries have an R-symmetry vector which
foliates the geometry with a transverse K\"ahler metric. Furthermore, a {\it supersymmetric action} can be 
constructed which is a function of the R-symmetry vector on $Y_{2n+1}$ as well as the basic cohomology class of the transverse K\"ahler form.
Extremizing this supersymmetric action over the space of possible R-symmetry vectors, for the case of $Y_7$, then gives the R-symmetry vector of the 
dual $(0,2)$ SCFT as well as the central charge, after a suitable normalization. For the case of $Y_9$, it was similarly shown that the on-shell supersymmetric action, again suitably normalized, corresponds to the logarithm of the partition function of the dual $d=1$ SCFT. This is the holographic version of an $F$-extremization principle for such $d=1$ SCFTs that we mentioned above, 
whose field theory formulation remains to be uncovered.

In \cite{Gauntlett:2018dpc} we further developed this formalism for the class of AdS$_3\times Y_7$ solutions in which $Y_7$ arises as a fibration of a toric $Y_5$ over $\Sigma_g$. From a dual point of view such solutions can arise by starting with a quiver gauge theory dual to AdS$_5\times Y_5$, with a Sasaki-Einstein metric on $Y_5$, and then compactifying on $\Sigma_g$ with a topological twist.
Using the toric data of $Y_5$, succinct formulas were presented for how to implement
the geometric version of $c$-extremization for the dual $d=2$ SCFT. A key technical step was to derive a master volume formula for toric $Y_5$ as a function of an R-symmetry vector and an arbitrary transverse K\"ahler class.
Based on various examples, it was conjectured in \cite{Gauntlett:2018dpc} that there is an off-shell agreement between the geometric and field theory versions of $c$-extremization and this was then proven for the case of toric $Y_5$ in \cite{Hosseini:2019use}.

In this paper, we extend the results of \cite{Gauntlett:2018dpc} in two main ways. 
First, we generalise the formalism to the class of AdS$_3\times Y_7$ solutions where $Y_7$ itself is toric. 
This requires generalizing the master volume formula for toric $Y_5$ that was presented in \cite{Gauntlett:2018dpc} to toric $Y_7$. 
These results provide a general framework for implementing the geometric dual of $c$-extremization that applies to $d=2$, $(0,2)$ SCFTs that do not have any obvious connection with a compactification of a $d=4$, 
${\cal N}=1$ SCFT dual to AdS$_5\times SE_5$. In a certain sense these results provide an AdS$_3\times Y_7$ analogue
of the results on AdS$_5\times SE_5$ solutions, with toric $SE_5$ \cite{Martelli:2005tp}. 
As an illustration, we use the formalism to re-derive the central charge of some known explicit AdS$_3\times Y_7$ solutions constructed in \cite{Gauntlett:2006af}, just using the toric data. 

Second, we consider AdS$_2\times Y_9$ solutions where $Y_9$ arises as fibration of a toric $Y_7$ over $\Sigma_g$, 
which allows us to make contact with $\mathcal{I}$-extremization.
These solutions
can be obtained by starting with an ${\cal N}=2$, $d=3$ SCFT dual to an AdS$_4\times Y_7$ solution of $D=11$ supergravity, with a Sasaki-Einstein metric on $Y_7$, and then compactifying on $\Sigma_g$ with 
a topological twist to ensure that two supercharges are preserved. 
Using the master volume formula on $Y_7$ 
we can generalise the results of \cite{Gauntlett:2018dpc} to derive formulae which provide a geometric dual 
of $\mathcal{I}$-extremization.

The principle of $\mathcal{I}$-extremization, introduced in \cite{Benini:2015eyy}, arose from the
programme of trying to reproduce the Bekenstein-Hawking entropy of supersymmetric black holes by carrying out
computations in a dual field theory. Indeed this was achieved for a class of AdS$_4$ black holes with 
AdS$_2\times S^2$ horizons in the context of the ABJM theory in \cite{Benini:2015eyy}, and some interesting extensions
have appeared in \cite{Azzurli:2017kxo,Hosseini:2017fjo,Benini:2017oxt,Bobev:2018uxk,Suh:2018tul,Hosseini:2018usu,Suh:2018szn}, for example.
It is natural to expect that many and perhaps all of the AdS$_2\times Y_9$ solutions that we consider here, with
$Y_9$ a fibration of a toric $Y_7$ over $\Sigma_g$, can arise 
as the near horizon limit of supersymmetric black holes. Such black holes, with $Y_9$ horizon,
would asymptote to AdS$_4\times Y_7$ in the UV, with the conformal boundary having
an $\mathbb{R}\times\Sigma_g$ factor associated with the field theory directions, 
and approach the AdS$_2\times Y_9$ solutions in the IR. 
We will therefore refer to the suitably normalized 
supersymmetric action for this class of $Y_9$ as the {\it entropy function} since, as argued in \cite{Couzens:2018wnk}, it will precisely give the black hole entropy after extremization. 

Now for a general class of $d=3$ quiver gauge theories, using localization techniques it was shown that the large $N$ limit of the topological index can be expressed in terms of a Bethe potential \cite{Hosseini:2016tor}. Furthermore, it was also shown in \cite{Hosseini:2016tor} that the same Bethe potential gives rise to the free energy of the $d=3$ SCFT on the three sphere, $F_{S^3}$. Combining these field theory results with the geometric results of this paper then provides a microscopic derivation of the black hole entropy for each such black hole solution
that actually exists.
This provides a rich framework for extending the foundational example studied in 
\cite{Benini:2015eyy} associated with $Y_7=S^7$ and  the ABJM theory.

An important general point to emphasize is that, as in \cite{Couzens:2018wnk,Gauntlett:2018dpc}, the geometric extremization techniques that we discuss in this paper will give the correct quantities in the dual field theory, provided that the AdS$_3$ and AdS$_2$ and solutions actually
exist. In other words they will give the correct results provided that there are
no obstructions to finding a solution. A related  discussion of obstructions to the existence of Sasaki-Einstein metrics
can be found in \cite{Gauntlett:2006vf} and furthermore, for toric Sasaki-Einstein metrics it is known that, in fact, there are no such obstructions \cite{Futaki:2006cc}. 
No general results are yet available for  AdS$_3\times Y_7$ and  AdS$_2\times Y_9$ solutions, although 
several examples in which the existence of the supergravity solution is obstructed were discussed
in \cite{Couzens:2018wnk,Gauntlett:2018dpc}, showing that this topic is an important one for further study.

The plan of the rest of the paper is as follows. In section \ref{mastvol} we consider toric, complex cone geometries, $C(Y_7)$, in four complex dimensions. In the special case that the metric on the cone is K\"ahler then the 
metric on $Y_7$ is a toric Sasakian metric. Using the toric data we derive a master volume formula for $Y_7$
as a function of an R-symmetry vector and an arbitrary transverse K\"ahler class, generalising a similar analysis for cone geometries in three complex dimensions carried out in section 3 of \cite{Gauntlett:2018dpc}. 
In section \ref{ads3section} we deploy these results to obtain expressions for the geometric dual of $c$-extremization for AdS$_3\times Y_7$ geometries when $Y_7$ is toric and study some examples.

In section \ref{ads2examples} we analyse AdS$_2\times Y_9$ solutions when $Y_9$ is a fibration of a toric $Y_7$ over a Riemann surface $\Sigma_g$, generalising the analysis in section \ref{ads2examples} of \cite{Gauntlett:2018dpc}. We illustrate the formalism for the universal twist solutions of \cite{Gauntlett:2006qw}, in which one fibres a $SE_7$ manifold over
$\Sigma_g$, with $g>1$, in which the fibration is just in the R-symmetry direction of $Y_7$ and in addition the fluxes are all proportional to the R-charges, recovering some results presented in \cite{Azzurli:2017kxo}. 
We also consider some additional generalizations for the special cases when $Y_7=Q^{1,1,1}$ and 
$M^{3,2}$ for which we can compare results obtained using our new formulae with some explicit supergravity solutions first constructed in \cite{Gauntlett:2006ns}. We then consider an example in which 
$C(Y_7)$ is the product of the conifold with the complex plane. Some new features arise for this example, as
the link, $Y_7$, of this cone contains worse-than-orbifold singularities and some care is required in using the master volume formulae. For this example, we are able to make a match between the off-shell entropy function
and the twisted topological index calculated from the field theory side in \cite{Hosseini:2016ume} in the genus zero case. We then revisit the case of $Y_7=Q^{1,1,1}$ and are able to match the off-shell entropy function with the twisted topological index calculated from the field theory side, which was calculated in 
\cite{Hosseini:2016ume} in the genus zero case.
Following this, we consider another example, with similar singularities, associated with $C(Y_7)$ being a certain Calabi-Yau 4-fold singularity, that is closely related to the suspended pinch point 
3-fold singularity. Once again we can match with some field theory results of
\cite{Hosseini:2016tor}. We end section \ref{ads2examples} with some general results connecting our formalism with the index theorem of \cite{Hosseini:2016tor}.
We conclude with some discussion in section \ref{secdisc}.

In appendix \ref{masterdetails}
we have included a few details of how to explicitly calculate
the master volume formula from the toric data in the specific examples discussed in the paper, while 
appendix \ref{apptop} contains a derivation of a homology relation used in the main text. Appendix 
\ref{explicsol} analyses flux quantization 
for the AdS$_2$ solutions of \cite{Gauntlett:2006ns} that we discuss in section \ref{ads2examples}.

\

\noindent {\bf Note added}: as this work was being finalised we became aware that there would be significant overlap with the results of \cite{Hosseini:2019ddy}, which appeared on the arXiv on the same day.

\section{Toric geometry and the master volume formula}
\label{mastvol}

\subsection{General setting}\label{gensetting}
We will be interested in complex cones, $C(Y_7)$, in complex dimension $n=4$ that are Gorenstein, {\it i.e.} they admit a global holomorphic $(4,0)$-form $\Psi$. Furthermore, we demand that there is an Hermitian metric that takes the standard conical form  
\begin{align}\label{6cone}
\diff s^2_{C(Y_7)} = \diff r^2 + r^2\diff s^2_7~,
\end{align}
where the link (or cross-section) of the cone, $Y_7$, is a seven-dimensional manifold. 
The complex structure pairs the radial vector $r\partial_r$ with a canonically defined vector $\xi$. 
Likewise, the complex structure pairs $\diff r/r$ with the dual one-form $\eta$, and $\xi\lrcorner\eta=1$.
The vector $\xi$ has unit norm and defines a foliation $\mathcal{F}_\xi$ of $Y_7$. The 
basic cohomology for this foliation is denoted $H^*_B(\mathcal{F}_\xi)$.

For the class of geometries of interest \cite{Couzens:2018wnk}, we furthermore require the vector $\xi$ to be a Killing vector 
for the metric on $Y_7$, with
\begin{align}\label{splitmetric}
\diff s^2_7 = \eta^2 + \diff s^2_6(\omega)~,
\end{align}
where the metric $\diff s^2_6(\omega)$  transverse to the foliation $\mathcal{F}_\xi$ is \emph{conformally K\"ahler}, with K\"ahler two-form $\omega$. 

Finally, in this paper we will also take the metric to be  invariant under a $U(1)^4$ isometry, with the isometry generated by 
$\xi$ being a subgroup.  Introducing generators $\partial_{\varphi_i}$, $i=1,\dots,4$, 
for each $U(1)$ action, where $\varphi_i$ has period $2\pi$, we may 
then parametrize the vector $\xi$ in terms of $\vec{b}\equiv (b_1,b_2,b_3,b_4)$, with
\begin{align}\label{Reebbasis}
\xi = \sum_{i=1}^4 b_i\partial_{\varphi_i}~.
\end{align} 
For convenience, we choose a basis so that the holomorphic $(4,0)$-form has unit charge under $\partial_{\varphi_1}$ and is uncharged under
$\partial_{\varphi_i}$, $i=2,3,4$. Notice that we then have\footnote{For the case of $SE_7$ geometry we need to take $b_1=4$, as discussed below. For the supersymmetric
AdS$_3$ geometry discussed in section \ref{ads3section} we take $b_1=2$, while for
the supersymmetric AdS$_2$ geometry discussed in section \ref{ads2examples} we need $b_1=1$.} 
\begin{align}\label{liederiv}
\mathcal{L}_\xi\Psi = \ii  b_1 \Psi\,.
\end{align}
This also implies that
\begin{align}
[\diff\eta] = \frac{1}{b_1}[\rho]\in H^2_B(\mathcal{F}_\xi)~,
\end{align}
where $\rho$ denotes the Ricci two-form of the transverse K\"ahler metric, 
and moreover
$[\rho]=2\pi c_1^B$, where $c_1^B$ is the basic first Chern class of the foliation.

\subsection{Toric K\"ahler cones}
We now assume that the cone metric is K\"ahler so that the metric on $Y_7$ is a toric Sasakian metric, as studied in \cite{Martelli:2005tp}. In this case the transverse conformally K\"ahler metric
$\diff s^2_6(\omega)$ in (\ref{splitmetric}) is K\"ahler. 
Denoting the transverse K\"ahler form by $\omega_{\mathrm{Sasakian}}$,
we have
\begin{align}\label{detaSasakian}
\diff\eta = 2\omega_{\mathrm{Sasakian}}~.
\end{align}
Because $\diff\eta$ 
is also a transverse symplectic form in this case, by definition $\eta$ is a \emph{contact} one-form on $Y_7$ and
$\xi$, satisfying $\xi\lrcorner\eta=1$ and $\xi\lrcorner \diff\eta=0$, is then called the \emph{Reeb} vector field. 

Considering now the $U(1)^4$ isometries, we may define the moment map coordinates
\begin{align}\label{yi}
y_i & \equiv  \tfrac{1}{2}r^2\partial_{\varphi_i}\lrcorner \eta~, \qquad i=1,2,3,4~.
\end{align}
These span the so-called moment map polyhedral cone $\mathcal{C}\subset \R^4$, where 
$\vec{y}=(y_1,y_2,y_3,y_4)$ are standard coordinates on $\R^4$. The polyhedral cone $\mathcal{C}$, which is convex, may be written as
\begin{align}
\mathcal{C} = \{\vec{y}\in \R^4\ \mid \ (\vec{y},\v_a)\geq 0~, \quad a=1,\ldots, d\}~,
\end{align}
where $\v_a\in\Z^4$ are the inward pointing primitive normals to the facets, and the index $a=1,\ldots,d\geq 4$ labels the facets. 
Furthermore, $\v_a=(1,\w_a)$, where $\w_a\in\Z^3$, follows from the Gorenstein condition, in the basis for $U(1)^4$ described 
at the end of the previous subsection.
An alternative 
presentation of the polyhedral cone $\mathcal{C}$ is
\begin{align}
\mathcal{C} = \Big\{\sum_\alpha t_\alpha \u_\alpha \ \mid \ t_\alpha\geq 0\Big\}~,
\end{align}
where $\u_\alpha\in\Z^4$ are the outward pointing vectors along each edge of $\mathcal{C}$. 

As shown in \cite{Martelli:2005tp}, for such a K\"ahler cone metric on $C(Y_7)$ the R-symmetry vector 
$\vec{b}=(b_1,b_2,b_3,b_4)$ necessarily lies in the interior of the Reeb cone, $\vec{b}\in \mathcal{C}^*_{\mathrm{int}}$. 
Here the \emph{Reeb cone} $\mathcal{C}^*$ is by definition the dual cone to $\mathcal{C}$. In particular $\vec{b}\in \mathcal{C}^*_{\mathrm{int}}$
is equivalent to 
 $(\vec{b},\u_\alpha)>0$ for all edges $\alpha$. Using $\xi\lrcorner\eta=1$, together with (\ref{Reebbasis}) and (\ref{yi}), the 
image of $Y_7=\{r=1\}$ under the moment map is
hence the compact, convex three-dimensional polytope
\begin{align}\label{Ppoly}
P = P(\vec{b}) \  \equiv \ \mathcal{C}\cap H(\vec{b})~,
\end{align}
where the \emph{Reeb hyperplane} is by definition
\begin{align}\label{ReebH}
H = H(\vec{b}) \ \equiv \ \Big\{\vec{y}\in \R^4\ \mid \ (\vec{y},\vec{b}) = \tfrac{1}{2}\Big\}~.
\end{align}

Later we will frequently refer to the toric diagram (in a minimal presentation) which is obtained by projecting 
onto  $\R^3$ the vertices $\v_a=(1,\w_a)$, with the minimum set of lines drawn between the vertices
to give a convex polytope. When all of the faces of the toric diagram
are triangles the link of the toric K\"ahler cone is either regular or has orbifold singularities.
We will also discuss cases in which some of the faces of the diagram are \emph{not} triangles
and then there are worse-than-orbifold singularities (for some further discussion see \cite{Martelli:2008rt}).

\subsection{Varying the transverse K\"ahler class} 
\label{transversesection}

As in \cite{Gauntlett:2018dpc}, we first fix a choice of toric K\"ahler cone metric on the complex cone $C(Y_7)$. This allows us to introduce 
the moment maps $y_i$ in (\ref{yi}), together with the angular coordinates $\varphi_i$, $i=1,2,3,4$, as coordinates on $C(Y_7)$. 
Geometrically, $C(Y_7)$ then fibres over the polyhedral cone $\mathcal{C}$: over the interior $\mathcal{C}_{\mathrm{int}}$ 
of  $\mathcal{C}$ this is a trivial $U(1)^4$ fibration, with the normal vectors $\v_a\in \Z^4$ to each 
bounding facet in $\partial\mathcal{C}$ specifying which $U(1)\subset U(1)^4$ collapses along that facet.  

For a fixed choice 
of such complex cone, with Reeb vector $\xi$ given by (\ref{Reebbasis}),
we would then like to study a more general class of transversely K\"ahler metrics of the form (\ref{splitmetric}). In particular, we would like to compute the ``master volume'' given by
\begin{align}
\label{defVmaaster}
\mathcal{V} \equiv   \int_{Y_7}\eta\wedge \frac{\omega^3}{3!}~,
\end{align}
as a function both of the vector $\xi$, and transverse K\"ahler class $[\omega]\in H^2_B(\mathcal{F}_\xi)$.
With the topological condition $H^2(Y_7,\mathbb{R})\cong H^2_B(\mathcal{F}_\xi)/[\rho]$, discussed in \cite{Couzens:2018wnk}, which will in fact hold for all the solutions considered in this paper, all closed two-form classes on $Y_7$ can be represented by basic closed two-forms. Following \cite{Gauntlett:2018dpc}, if we take the
$c_a$ to be basic representatives in $H^2_B(\mathcal{F}_\xi)$ that lift to 
 integral classes in $H^2(Y_7,\mathbb{Z})$, which are Poincar\'e dual to 
the restriction of the toric divisors on $C(Y_7)$, then we can write
\begin{align}\label{omegalambda}
[\omega] = -2\pi\sum_{a=1}^d \lambda_a c_a \in  H^2_B(\mathcal{F}_\xi) \, . 
\end{align}
Furthermore, the $c_a$ are not all independent and $[\omega]$ will depend on just $d-3$  of the $d$  parameters $\{ \lambda_a\}$. 
As in \cite{Couzens:2018wnk} it will also be useful to note that the first Chern class of the foliation can be written in terms of the $c_a$ as 
\begin{align}\label{rhoca}
[\rho] &= 2\pi\sum_{a=1}^d c_a  \in H^2_B(\mathcal{F}_\xi)~.
\end{align}
In the special case in which 
\be
\label{lambdaSas}
\lambda_a  =  -\frac{1}{2b_1}\,, \qquad a=1,\dots d\,,
\ee
we recover the Sasakian  K\"ahler class $[\rho]= 2 b_1 [\omega_\mathrm{Sasakian}]$ and the master volume (\ref{defVmaaster})
reduces to the Sasakian volume 
\be
\mathcal{V}  (\b;\{\lambda_a =   -\frac{1}{2b_1}\})\  =  \ \int_{Y_7}\eta\wedge \frac{1}{3!}\omega^3_\mathrm{Sasakian} \ \equiv \ \mathrm{Vol} (Y_7)~.
\ee

Following \cite{Gauntlett:2018dpc}, this volume can be shown to be
\begin{align}\label{VEuc}
\mathcal{V} = \frac{(2\pi)^4}{|\vec{b}|}\vol(\mathcal{P})~.
\end{align}
Here the factor of $(2\pi)^4$ arises by integrating over the torus $U(1)^4$, while $\vol (\mathcal{P})$ is 
the Euclidean volume of the compact, convex three-dimensional polytope
\begin{align}\label{generalP}
\mathcal{P} = \mathcal{P}(\vec{b};\{\lambda_a\}) \ \equiv \ \{\vec{y}\in H(\vec{b}) \ \mid \ (\vec{y}-\vec{y}_0,\v_a) \geq \lambda_a~, \quad a=1,\ldots,d\}~.
\end{align}
Here
\begin{align}\label{originP}
\vec{y}_0 = \left(\frac{1}{2b_1},0,0,0\right) \in H~,
\end{align}
which lies in the interior of $\mathcal{P}$, 
while the $\{\lambda_a\}$ parameters determine the transverse K\"ahler class. 
It will be important to remember that the transverse K\"ahler class $[\omega]\in H^2_B(\mathcal{F}_\xi)$, and hence volume $\mathrm{vol}(\mathcal{P})$, depends 
on only $d-3$ of the $d$ parameters $\{\lambda_a\}$, 
with three linear combinations being redundant.

We may compute the Euclidean volume of $\mathcal{P}$ in (\ref{generalP}) by first finding its vertices $\vec{y}_\alpha$. By construction, these arise as the intersection of 
an edge of $\mathcal{C}$ with the Reeb hyperplane $H(\vec{b})$. Let us fix a specific
two-dimensional facet of $\mathcal{P}$, associated with a specific $(v_a,\lambda_a)$, given by
\begin{align}
\mathcal{P}_a  \equiv  \mathcal{P}\cap \{(\vec{y}-\vec{y}_0,\vec{v}_a) =\lambda_a\}~.
\end{align}
This is a compact, convex two-dimensional polytope, and will have some number $\n_a\geq 3$ of edges/vertices. In turn, each 
edge of $\mathcal{P}_a$ arises as the intersection of $\mathcal{P}_a$ with $\n_a$ other faces 
which we label $\mathcal{P}_{a,k}$, each associated with $(v_{a,k},\lambda_{a,k})$,
with $k=1,\dots,  \n_a$. We choose the ordering of $\mathcal{P}_{a,k}$ cyclically around the $a$th face $\mathcal{P}_a$ and it is then convenient to take the index numbering on $k$ to be understood mod $\n_a$ (hence cyclically). The vertices of $\mathcal{P}_a$ arise 
from the intersection of neighbouring edges in this ordering. We may thus define the vertex $y_{a,k}$ of $\mathcal{P}_a$ 
as the intersection 
\begin{align}
\vec{y}_{a,k} = \mathcal{P}_a\cap \mathcal{P}_{a,k-1}\cap \mathcal{P}_{a,k}~,
\end{align}
where $k=1,\ldots,\n_a$, with the index numbering on $k$ understood mod $\n_a$ (hence cyclically). By definition, $\vec{y}_{a,k}$ then satisfies the four equations
\begin{align}
(\vec{y}_{a,k}-\vec{y}_0,\vec{v}_{a}) &= \lambda_{a}~, \qquad (\vec{y}_{a,k}-\vec{y}_0,\vec{v}_{a,k-1}) =  \lambda_{a,k-1}~,\nn\\
\qquad (\vec{y}_{a,k}-\vec{y}_0,\vec{v}_{a,k}) &=   \lambda_{a,k}~,  \ \, \quad \qquad (\vec{y}_{a,k}-\vec{y}_0,\vec{b}) = 0~,
\end{align}
which we can solve to give
\begin{align}\label{ysol}
\vec{y}_{a,k} - \vec{y}_0 =   \frac{\lambda_a (\vec{E},\v_{a,k-1},\v_{a,k},\vec{b})-
\lambda_{a,k} (\vec{E},\v_{a},\v_{a,k-1},\vec{b})+\lambda_{a,k-1}(\vec{E},\v_{a,k},\v_{a},\vec{b})}{(\v_a,\v_{a,k-1},\v_{a,k},\vec{b})}~.
\end{align}
Here $(\cdot,\cdot,\cdot,\cdot)$ denotes a $4\times 4$ determinant, and we have defined
\begin{align}
\vec{E} \equiv  (\vec{e}_1,\vec{e}_2,\vec{e}_3,\vec{e}_4)^T~.
\end{align}
Here $(\vec{e}_j)^i=\delta_j^i$ and, to be clear, the vector index on the left hand side of \eqref{ysol} corresponds to the vector index on $\vec{E}$ on the right hand side.

We next divide $\mathcal{P}$ up into tetrahedra, as follows. For each face $\mathcal{P}_a$, $a=1,\ldots,d$, we first split the 
face into $\n_a-2$ triangles. Here the triangles have vertices $\{\vec{y}_{a,1},\vec{y}_{a,k} ,  \vec{y}_{a,k+1}\}$, where 
$k=2,\ldots,\n_a-1$. Each of these triangles then forms a tetrahedron by adding the interior 
vertex $\vec{y}_0$. The volume of $\mathcal{P}$ is then simply the sum of the volumes of all of these tetrahedra. 
On the other hand, the volume of the tetrahedron $T_{a,k}$ with vertices $\{\vec{y}_{a,1},\vec{y}_{a,k} ,  \vec{y}_{a,k+1},\vec{y}_0\}$
is given by the elementary formula
\begin{align}
\vol(T_{a,k}) =  \frac{1}{3!|\vec{b}|}(\vec{y}_{a,1}-\vec{y}_0,\vec{y}_{a,k}-\vec{y}_0,\vec{y}_{a,k+1}-\vec{y}_0,\vec{b})~, \qquad 
k=2,\ldots,\n_a-1~.
\end{align}
Thus, the master volume (\ref{VEuc}) can now be written as 
\begin{align}
\mathcal{V}(\vec{b};\{\lambda_a\}) &= \frac{(2\pi)^4}{|\vec{b}|}\sum_{a=1}^d\sum_{k=2}^{\n_a-1}\vol(T_{a,k})\,, \nn\\
&= \frac{(2\pi)^4}{3!(\vec{b},\vec{b})}\sum_{a=1}^d\sum_{k=2}^{\n_a-1}(\vec{y}_{a,1}-\vec{y}_0,\vec{y}_{a,k}-\vec{y}_0,\vec{y}_{a,k+1}-\vec{y}_0,\vec{b})~.
\end{align}
On the other hand, using the explicit formula (\ref{ysol}) for the vertices $\vec{y}_{a,k}$, together with some elementary identities, we find
the master volume formula for $Y_7$ is given by
\begin{align}
\label{nonsimplemaster}
  \boxed{ 
  \mathcal{V} (\b;\{\lambda_a\}) =  -\frac{(2\pi)^4}{3!} \sum_{a=1}^d    \lambda_a  \sum_{k=2}^{\n_a-1}   \frac{X_{a,k}^I X_{a,k}^{II}  }{(\v_a,\v_{a,k-1},\v_{a,k},\b )   (\v_{a,\n_a},\v_a,\v_{a,1},\b )  (\v_{a,k},\v_{a,k+1},\v_a,\b) }
  }
  \nn\\
\end{align}
where we have defined
\begin{align}
X_{a,k}^I  \ \equiv \  &  -\lambda_a (\v_{a,k-1},\v_{a,k},\v_{a,k+1},\b)   +   \lambda_{a,k-1}(\v_{a,k},\v_{a,k+1},\v_a,\b)    \nn\\
& -  \lambda_{a,k} (\v_{a,k+1},\v_a,\v_{a,k-1},\b )   + \lambda_{a,k+1}(\v_a,\v_{a,k-1},\v_{a,k},\b )\,,\nn\\
 X_{a,k}^{II} \  \equiv \ &  -\lambda_a (\v_{a,1},\v_{a,k},\v_{a,\n_a},\b)   +   \lambda_{a,1}(\v_{a,k},\v_{a,\n_a},\v_a,\b)   \nn\\
  &  -  \lambda_{a,k} (\v_{a,\n_a},\v_a,\v_{a,1},\b )   + \lambda_{a,\n_a}(\v_a,\v_{a,1},\v_{a,k},\b )~.
\end{align}
Notice that $\mathcal{V} (\b;\{\lambda_a\})$ is cubic in the $\{\lambda_a\}$, as it should  be. 
When all of the $\lambda_a$ are equal, $\lambda_a = \lambda$, $a=1,\ldots,d$, using a vector product identity these simplify considerably to give
\begin{align}
X_{a,k}^I  =    - \lambda b_1 (\v_{a,k-1},\v_{a,k},\v_{a,k+1},\v_a)\,,\qquad
X_{a,k}^{II} \  =\   - \lambda b_1 (\v_{a,1},\v_{a,k},\v_{a,\n_a},\v_a)\,.
\end{align}
In particular, for the special case of the Sasakian K\"ahler class with $ \lambda_a = -  \frac{1}{2 b_1}$, as in (\ref{lambdaSas}), 
the formula (\ref{nonsimplemaster}) reproduces the known \cite{Hanany:2008fj}  expression  for the volume of toric Sasakian manifolds, namely 
\begin{align}
\mathrm{Vol}(Y_7) \  = \  & \frac{(2\pi)^4}{48 b_1} \sum_{a=1}^d   \sum_{k=2}^{\n_a-1}  \frac{ (\v_{a,k-1},\v_{a,k},\v_{a,k+1},\v_a) (\v_{a,1},\v_{a,k},\v_{a,\n_a},\v_a) }{(\v_a,\v_{a,k-1},\v_{a,k},\b )   (\v_{a,\n_a},\v_a,\v_{a,1},\b )  (\v_{a,k},\v_{a,k+1},\v_a,\b) }~.
\label{sasvolumnosimple}
\end{align}
In \cite{Martelli:2005tp} it was shown that the Reeb vector $\xi\in \mathcal{C}_{\mathrm{int}}^*$ for a Sasaki-Einstein metric on $Y_7$ is the unique minimum of $\mathrm{Vol}(Y_7)$ on $\mathcal{C}_{\mathrm{int}}^*$, considered as a function of $\vec{b}$,
subject to the constraint $b_1=4$.

It will be helpful to present some formulas here that will be useful later.
Using (\ref{omegalambda}) the master volume may be written as
\begin{align}
\label{newmaster}
\mathcal{V} =  - (2\pi)^3 \sum_{a,b,c=1}^d \frac{1}{3!}I_{abc} \lambda_a\lambda_b\lambda_c~,
\end{align}
where the triple intersections $I_{abc}$ are defined as
\begin{align}\label{intersection}
I_{abc} \ \equiv \  \int_{Y_7}\eta\wedge c_a\wedge c_b  \wedge c_c = - \frac{1}{(2\pi)^3}\frac{\partial^3 \mathcal{V}}{\partial\lambda_a \partial\lambda_b \partial \lambda_c}~.
\end{align}
We then have 
\begin{align}
 \int_{Y_7} \eta\wedge \frac{1}{2!}\rho^2 \wedge \omega & =   \frac{1}{2}\sum_{a,b=1}^d \frac{\partial^2 \mathcal{V}}{\partial \lambda_a\partial \lambda_b}  \ =  \ - \frac{(2\pi)^3}{2!} \sum_{a,b,c=1}^d I_{abc} \lambda_c\, ,\nn\\
\int_{Y_7}\eta\wedge \rho\wedge \frac{1}{2!}\omega^2 & =  -\sum_{a=1}^d \frac{\partial \mathcal{V}}{\partial \lambda_a} \  =    \    \frac{(2\pi)^3}{2!}\sum_{a,b,c=1}^d I_{abc} \lambda_b\lambda_c~.
 \label{newformulas}
 \end{align}
Furthermore, the first derivative of the master volume with respect to $\lambda_a$ gives the
 volume of the  $d$ torus-invariant five-manifolds $T_a \subset Y_7$, Poincar\'e dual to the $c_a$,
 via
\be\label{SaVnew}
\int_{T_a} \eta\wedge\frac{1}{2!}\omega^2\  =  \  \frac{(2\pi)^2}{2!}\sum_{b,c=1}^d I_{abc} \lambda_b\lambda_c
= -\frac{1}{2\pi} \frac{\partial \mathcal{V}}{\partial \lambda_a}~.
\ee
Finally, we note that the Sasakian volume $\mathrm{Vol}(Y_7)$ and the Sasakian volume of torus-invariant five-dimensional submanifolds $T_a$,  $\mathrm{Vol}(T_a)$, can be expressed in terms of the $I_{abc}$ as
\be
\mathrm{Vol}(Y_7) =  \frac{\pi^3}{3!b_1^3} \sum_{a,b,c=1}^d I_{abc} \, , \quad \qquad \mathrm{Vol}(T_a) =  \frac{\pi^2}{2b_1^2} \sum_{b,c=1}^d I_{abc} \,  . 
\label{sasakkia}
\ee

For the various examples of $Y_7$ that we consider later which are regular or have orbifold singularities, 
 we have explicitly 
checked that the relation
\begin{align}\label{keyvRrel}
\sum_{a=1}^d\left(\vec{v}_a-\frac{\vec{b}}{b_1}\right) \frac{\partial \mathcal{V}}{\partial \lambda_a}= 0\,,
\end{align}
holds as an identity for all $\vec{b}$ and $\{\lambda_a\}$. We have not yet constructed a proof of this result, but we
conjecture that it will always hold for this class of $Y_7$. When it does hold 
it is simple to see that the master volume formula is invariant under the ``gauge" transformations 
\begin{align}\label{lamgt}
\lambda_a\ \to\ \lambda_a+\sum_{i=1}^4\gt_i(v_a^i b_1-b_i)\,,
\end{align}
for arbitrary constants $\gt_i$, generalising a result of \cite{Hosseini:2019use}. 
Noting that the transformation parametrized by $\gt_1$ is trivial, this explicitly shows that
the master volume only depends on $d-3$ of the parameters $\{\lambda_a\}$, as noted above.

However, we emphasize that \eqref{keyvRrel} does \emph{not} hold for $Y_7$
which have worse-than-orbifold singularities, unless we impose some additional restrictions on the $\{\lambda_a\}$. This is an important point
since many examples whose field theories have been studied in the literature have this property.
We discuss this further for the representative example of the link associated with the product of the complex plane with
 the conifold in appendix~\ref{secctcon}.

To conclude this section we note that the above formulae assume that the polyhedral cone $\mathcal{C}$ is convex, since at the outset we started with a cone that admits a toric K\"ahler cone metric. However, as first noted in \cite{gauntkimwald}, and discussed in \cite{Couzens:2018wnk,Gauntlett:2018dpc}, this convexity condition is, in general, too restrictive for applications
to the classes of AdS$_2$ and AdS$_3$ solutions of interest. Indeed, many such explicit supergravity solutions
are associated with ``non-convex toric cones", as defined in \cite{Couzens:2018wnk}, which in particular have toric data which do not define a convex polyhedral cone. We conjecture that the key formulae in this section are also
applicable to non-convex toric cones and we will assume that this is the case in the sequel. The consistent picture
that emerges, combined with similar results in \cite{Couzens:2018wnk,Gauntlett:2018dpc}, strongly supports the validity of this conjecture.

\section{Supersymmetric AdS$_3\times Y_7$ solutions}
\label{ads3section}

\subsection{General set-up}
In this section 
 the class of supersymmetric AdS$_3\times Y_7$ solutions
of type IIB supergravity that are dual to SCFTs with $(0,2)$ supersymmetry of the form
\begin{align}
\diff s^2_{10} &= L^2 \ex^{-B/2}\left(\diff s^2_{\mathrm{AdS}_3} + \diff s^2_{7}\right)~,\N\\
F_5 &= -L^4\left(\vol_{\mathrm{AdS}_3}\wedge F + *_7 F\right)~.\label{ansatz}
\end{align}
Here $L$ is an overall dimensionful length scale, with $\diff s^2_{\mathrm{AdS}_3}$ being the metric on a 
unit radius AdS$_3$ with corresponding volume form $\vol_{\mathrm{AdS}_3}$. The 
warp factor $\BB$ is a function on the smooth, compact Riemannian internal space $(Y_7, \diff s^2_7)$ and
$F$ is a closed two-form on $Y_7$ with Hodge dual $*_7 F$. 
In order to define a consistent string theory background we must impose the flux quantization condition
\begin{align}
\frac{1}{(2\pi \ell_s)^4 g_s}\int_{\Sigma_A} F_5 = N_A \in \mathbb{Z}~,\label{quantization}
\end{align}
which also fixes $L$. Here $\ell_s$ denotes the string length, $g_s$ is the string coupling constant, and $\Sigma_A\subset Y_7$, with $\{\Sigma_A\}$ forming an integral basis for the free part of $H_5(Y_7,\Z)$.
The geometry of these solutions was first analysed in \cite{Kim:2005ez} and then extended in \cite{Gauntlett:2007ts}.

The geometric dual to $c$-extremization, described in detail in \cite{Couzens:2018wnk}, starts by
considering {\it supersymmetric geometries}. By definition these are
configurations as in \eqref{ansatz} which admit the required Killing spinors. 
These off-shell supersymmetric geometries become supersymmetric solutions when, in addition, we impose the equation of motion for the five-form. Equivalently, we obtain supersymmetric solutions when the equations of motion obtained from extremizing an action, $S$, given explicitly in \cite{Gauntlett:2007ts}
are satisfied.

The supersymmetric geometries have the properties
stated at the beginning of section \ref{gensetting}. In particular, we have
\begin{align}\label{splitmetric2}
\diff s^2_7 = \eta^2 + \ex^B\diff s^2(J)~,
\end{align}
where $\diff s^2(J)$ is a transverse K\"ahler metric with transverse K\"ahler form $J$. This is
exactly as in \eqref{splitmetric} after making the identification $J=\omega$. The transverse K\"ahler metric determines the full supersymmetric geometry, including the fluxes. In particular, the conformal factor is fixed via
$\ex^{B}=R/8$ where $R$ is the Ricci scalar of the transverse K\"ahler metric. We also have
\begin{align}
\diff\eta= \frac{1}{2}\rho\,,
\end{align}
where $\rho$ is the Ricci two-form of the transverse K\"ahler metric, and 
$\mathcal{L}_\xi\Psi= \ii  b_1 \Psi$, with $b_1=2$. The Killing vector $\xi$ is called the R-symmetry vector.

Putting the supersymmetric geometries on-shell implies solving the equations of motion coming from varying a supersymmetric action, $\Ssusy$, which is the action $S$ mentioned above evaluated on a supersymmetric geometry. Explicitly it was shown in \cite{Couzens:2018wnk} that
\begin{align}
\Ssusy=\int_{Y_7}\eta\wedge\rho\wedge\frac{J^{2}}{2!}\,,
\end{align}
which, in fact, just depends on the R-symmetry vector $\xi$ and the transverse K\"ahler class $[J]\in H^2_B(\mathcal{F}_\xi)$ {\it i.e.}
$\Ssusy=\Ssusy(\xi;[J])$. 
Furthermore, in order to impose flux quantization on the five-form the following topological constraint must also be imposed 
\begin{align}
\int_{Y_7}\eta\wedge\rho^2\wedge J =0\,.
\end{align}
Flux quantization is achieved by taking a basis of 5-cycles, $\Sigma_A$, that are tangent to $\xi$
and demanding 
\begin{align}
\int_{\Sigma_A}\eta\wedge\rho\wedge J =\frac{2(2\pi \ell_s)^4g_s}{L^4}N_A\,,
\end{align}
with $N_A\in \mathbb{Z}$.

Assuming now that $Y_7$ is toric, admitting 
a $U(1)^4$ isometry as discussed in section \ref{gensetting}, it is straightforward to generalize section
3 of \cite{Gauntlett:2018dpc} to obtain expressions for $\Ssusy$, the constraint and the flux quantization conditions in terms of the toric data. 
Remarkably, they can all be expressed in terms of the master volume 
$ \mathcal{V}=\mathcal{V} (\b;\{\lambda_a\}) $ given in \eqref{nonsimplemaster}. Specifically,
using the formulas given in section \ref{transversesection},
the off-shell supersymmetric action, the constraint equation and the flux-quantization conditions
are given by 
\begin{align}
\Ssusy  
& =   -\sum_{a=1}^d \frac{\partial \mathcal{V}}{\partial \lambda_a}~,\label{ssusytoricseven}\\
0 & =   \sum_{a,b=1}^d \frac{\partial^2 \mathcal{V}}{\partial \lambda_a\partial \lambda_b}\,, \label{constoricseven}\\
\frac{2(2\pi \ell_s)^4g_s}{L^4}N_a & =  \frac{1}{2\pi}\sum_{b=1}^d \frac{\partial^2 \mathcal{V}}{\partial \lambda_a\partial \lambda_b}~,
\label{Nafluxes}
\end{align}
respectively, where $N_a\in\mathbb{Z}$.
The $N_a$ are not all independent: they are the quantized fluxes through a basis of toric five-cycles $[T_a]\in H_5(Y_7;\mathbb{Z})$. While
the $[T_a]$ generate the free part of $H_5(Y_7;\mathbb{Z})$, they also satisfy 4 linear relations $\sum_{a=1}^d v^i _a[T_a]=0\in H_5(Y_7;\mathbb{Z})$,
and hence we have 
\begin{align}
\sum_{a=1}^d v_a^i N_a = 0 ~,\qquad i=1,2,3,4~.
\end{align}
Notice that the $i=1$ component of this relation is in fact the constraint equation \eqref{constoricseven}.

We also note that when \eqref{keyvRrel} holds, from the invariance of the master volume under the transformations (\ref{lamgt}) it follows that all the derivatives of $\mathcal{V}$ with respect to $\lambda_a$ are also invariant. Therefore, the complete set of 
equations (\ref{ssusytoricseven}), (\ref{constoricseven}), (\ref{Nafluxes}) is invariant under  (\ref{lamgt}) and we could use this  to ``gauge-fix'' three  of the $\lambda_a$ parameters, or alternatively work with gauge invariant combinations. However, in the examples below we will not do this, but instead we will see 
that the results are consistent with the gauge invariance. 
Finally, we also note that we can also write the supersymmetric action in the form
\begin{align}
\Ssusy  
 =   -(2\pi)\frac{(2\pi)^4\ell_s^4g_s}{L^4}\sum_{a=1}^d \lambda_a N_a\,,
\end{align}
where we used the fact the master volume is homogeneous of degree three in the $\lambda_a$ (see \eqref{newmaster}).

We can now state the geometric dual to $c$-extremization of \cite{Couzens:2018wnk}, for toric $Y_7$. We hold $b_1$ fixed to be $b_1=2$,  and then
extremize $\Ssusy $ with respect to $b_2,b_3,b_4$ as well as the $d-3$ independent K\"ahler class parameters determined by $\{\lambda_a\}$, subject
to the constraint \eqref{constoricseven} and flux quantization conditions \eqref{Nafluxes}. Equivalently, we extremize the ``trial central charge", $\cZ$, defined by
\begin{align}\label{cS2Z}
\cZ  \equiv  \frac{3L^8}{(2\pi)^6g_s^2\ell_s^8} \Ssusy~,
\end{align}
which has the property that for an on-shell supersymmetric solution, {\it i.e.} after extremization,
we obtain the central charge of the dual SCFT:
\begin{align}\label{cS23}
\cZ  |_\mathrm{on-shell} =  \csugra~.
\end{align}
In practice, and generically, we have $d-4$ independent flux quantum numbers that we are free to specify. The constraint equation and $d-4$ of the flux quantization conditions \eqref{Nafluxes} can be used to solve for the $d-3$ independent $\{\lambda_a\}$. This 
leaves $\cZ$ as a function of the $d-4$ independent flux numbers as well as $b_2,b_3,b_4$, of which we still need to vary the latter.
We emphasize that \eqref{cS23} will be the central charge of the $(0,2)$ CFT dual to the AdS$_3\times Y_7$ solution,
provided that the latter actually exists ({\it i.e.} when there are no obstructions).

We now illustrate the formalism by considering a class of explicit AdS$_3\times Y_7$ supergravity solutions
presented\footnote{The local solutions were constructed as a special example of a class
of AdS$_3$ solutions of $D=11$ supergravity found in \cite{Gauntlett:2006ns}.} in 
\cite{Gauntlett:2006af}. The construction involves a four-dimensional
K\"ahler-Einstein ($KE$) base manifold with positive curvature, $KE^+_4$. 
Such $KE^+_4$ manifolds are either $\C P^1\times \C P^1$,
$\C P^2$ or a del Pezzo surface $dP_k$ with $k=3,\dots, 8$. 
Of these $\C P^1\times \C P^1$, $\C P^2$ and $dP_3$ are toric. The solutions depend on two integers $p,k$
with $p>0$, $k<0$ and we will label them $\mathscr{Y}^{p,k}(KE^+_4)$. The associated complex cones over 
$\mathscr{Y}^{p,k}(KE^+_4)$ are non-convex toric cones, as defined
in \cite{Couzens:2018wnk,Gauntlett:2018dpc}, 
and the associated compact polytopes are not convex.

The exposition in \cite{Gauntlett:2006qw} illuminated the very close similarity of these $\mathscr{Y}^{p,k}(KE^+_4)$ solutions with a class of seven-dimensional Sasaki-Einstein manifolds, $Y^{p,k}(KE^+_4)$, constructed in \cite{Gauntlett:2004hh}, which utilized exactly the same $KE^+_4$ manifolds. For the latter, using techniques developed in \cite{Martelli:2004wu}, the toric geometry of the associated Calabi-Yau 4-fold singularities for $Y^{p,k} (\C P^2)$ and $Y^{p,k} (\C P^1\times \C P^1)$ 
was discussed in \cite{Martelli:2008rt}. The integers $p,k$ are both positive and satisfy $jp/2<k< jp$, with $j=3$ for 
$Y^{p,k} (\C P^2)$ and $j=2$ for $Y^{p,k} (\C P^1\times \C P^1)$. The associated compact polytopes for these ranges are, of course, convex.
Below we shall analyse these two families in turn.
 Although we will not utilize this below, we note that both of these examples satisfy the relation 
\eqref{keyvRrel} for the master volume.

\subsection{The $\mathscr{Y}^{p,k} (\C P^2)$ and $Y^{p,k} (\C P^2)$ families}

\begin{figure}[h!]
\begin{center}
  \includegraphics[width=6.cm]{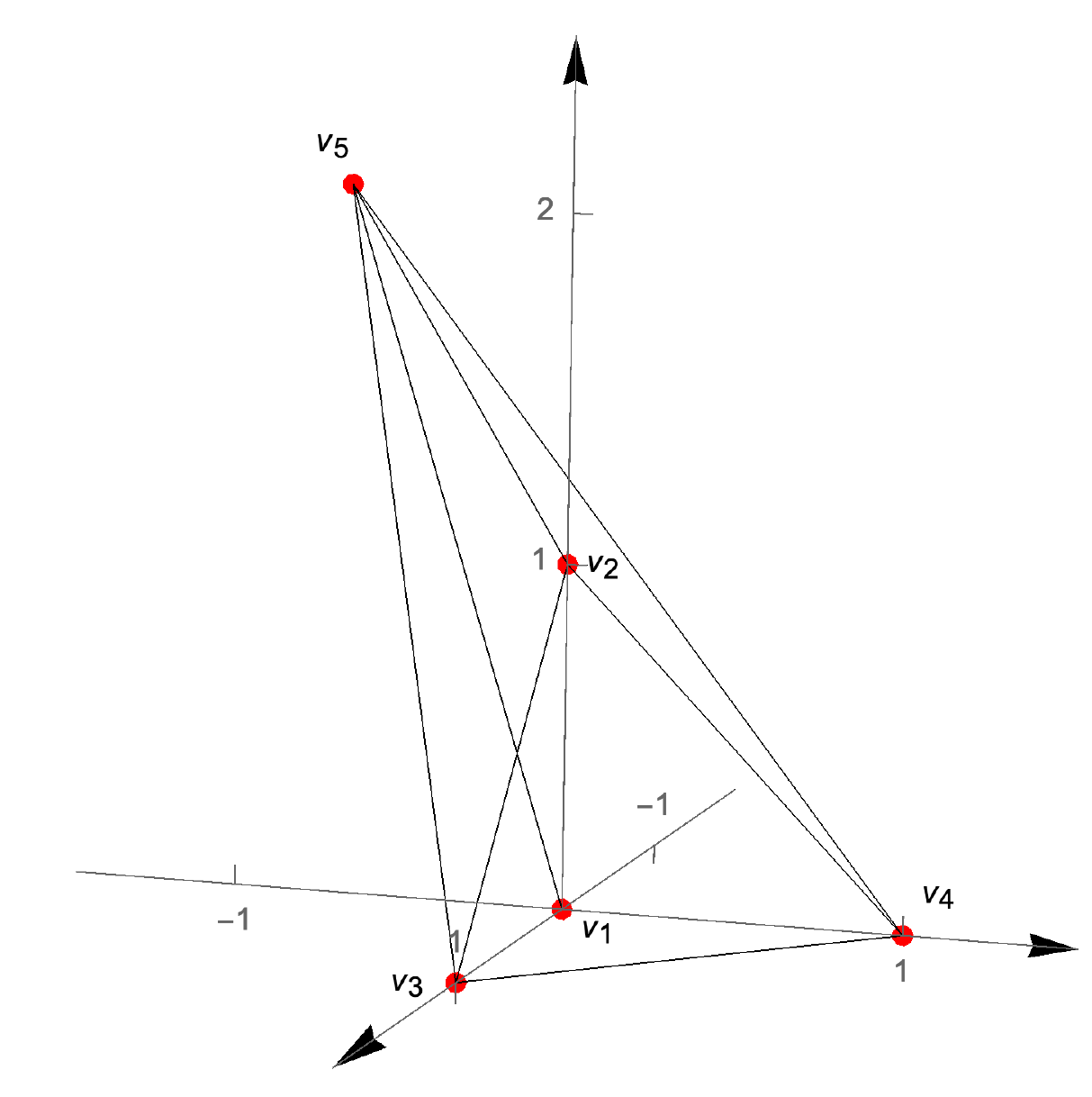}
 \caption{\label{CC2toricdiag} Toric diagram of $Y^{p,k} (\C P^2)$ with $p=1$, $k=2$,  obtained by projecting on $\R^3$ the vertices  in (\ref{Ypkvcp2}).}
\end{center}
\end{figure}

The toric data associated with $Y^{p,k} (\C P^2)$ was given in  \cite{Martelli:2008rt}, in  the context of the discussion of explicit Sasaki-Einstein metrics. We take the $d=5$ inward pointing normal vectors to be given by  
\begin{align}
\label{Ypkvcp2}
\v_1 \,  &=\,  (1,0,0,0)~, \quad \v_2 \, = \, (1,0,0,p)~, \quad \v_3 \, = \, (1,1,0,0)~, \nn\\
\quad \v_4 \, &= \, (1,0,1,0)~,\quad \v_5 \, = \, (1,-1,-1,k) \,.
\end{align}
The associated toric diagram, obtained by projecting on $\R^3$ the vertices in (\ref{Ypkvcp2}), is given in
Figure \ref{CC2toricdiag}.
For $Y^{p,k} (\C P^2)$ we have $0<3p/2<k< 3p$, and we have a convex polytope. However, for
the explicit solutions $\mathscr{Y}^{p,k} (\C P^2)$ we have $k<0$ and $p>0$. We continue with general $p,k$. 

The master volume $ \mathcal{V} (\b;\{\lambda_a\})$ given in (\ref{nonsimplemaster}) can be obtained 
from the toric data \eqref{Ypkvcp2} and some results in appendix \ref{masterdetails}.
In the Sasakian limit, $\{\lambda_a =-\tfrac{1}{2b_1}\}$, setting $b_1=4$ and extremizing  with respect to $\{b_2,b_3,b_4\}$  \cite{Martelli:2005tp} we find that the critical Reeb vector is given by $b_2=b_3=0$, with $b_4$ solving the cubic equation
 \be
 b_4^3 \left(-k^2+3 k p- 3 p^2\right)+ b_4^2 p \left(6 k^2-8 k p+3 p^2\right)+8 b_4 k p^2 (p-2 k)+16 k^2 p^3 =  0\,.
 \label{b4critcp2}
\ee
The fact that $b_2=b_3=0$ is due to the $SU(3)$ symmetry of the $\C P^2$ base space.
 Equivalently, the value of $b_4$ obtained from (\ref{b4critcp2}) 
  can be obtained  from extremizing the Sasakian volume with $b_1=4$ upon setting $b_2=b_3=0$, which reads
  \be
 \mathrm{Vol}(Y_7)  (b_4) =     \frac{\pi ^4 p \left[3 b_4 \left(b_4 \left(k^2-3 k p+3 p^2\right)+4 k p (p-k)\right)+16 k^2 p^2\right]}{3 b_4^3 \left(4 p-b_4\right)^3}~.
  \ee
This expression, with $b_4$ obtained from (\ref{b4critcp2}), can be shown to be precisely equal to the Sasaki-Einstein volume 
\be
{Y}^{p,k} (\C P^2) =  \frac{3 \pi ^4 (x_2-x_1) \left({x_2}^3-{x_1}^3\right)}{256 p (1-x_1) (x_2-1)}~, \qquad \mathrm{with}\quad {x_1}  \ =  \ \frac{{x_2} (k-3 p)}{k-3 p {x_2}}\, , 
\ee
  given in equation (2.13) of \cite{Martelli:2008rt}, where it was computed using the  explicit Sasaki-Einstein metric. The relation between the variables $b_4$ and $x_2$ in the two expressions above 
is simply  $x_2 = \tfrac{4 k} {3b_ 4}$. Note, for example, for the special case $p=2$, $k=3$ we have 
$Y_7=M^{3,2}$ and
  $\mathrm{Vol}(M^{3,2})= 9\pi^4/128$.

We now turn to the AdS$_3\times \mathscr{Y}^{p,k} (\C P^2)$ solutions. We begin by setting 
$b_1=2$ in the formulae \eqref{constoricseven}--\eqref{Nafluxes}.
The transverse K\"ahler class is determined by $d-3=2$ of the parameters $\{\lambda_a\}$.
We use the constraint equation \eqref{constoricseven} and one of the flux equations \eqref{Nafluxes}, which we take to be $N_1$, to solve for two of the $\{\lambda_a\}$ which we take to be
$\lambda_1$ and $\lambda_2$. The remaining fluxes can all be expressed in terms of $N_1$, and 
the flux vector is given by
\begin{align}
\{N_a\}
= \left\{1,-\frac{k }{k-3 p},\frac{ p}{k-3 p},\frac{ p}{k-3 p},\frac{ p}{k-3 p}\right\}N_1\,.
\end{align}

We can then calculate the trial central charge $\cZ$ finding, in particular, that it is independent of 
$\lambda_3$, $\lambda_4$ and $\lambda_5$, 
in agreement with the invariance of the problem under the three independent transformations in (\ref{lamgt}).
Furthermore, $\cZ$ is quadratic in
$b_2,b_3$ and $b_4$, again as expected. It is now straightforward to
extremize $\cZ$ with respect to these remaining variables and we find the unique extremum  has
$\vec{b}=(2,0,0,b_4)$, with
\begin{align}
b_4 = \frac{k p (k-p)}{k^2-3 k p+3 p^2}
\,.
\end{align}
The fact that $b_2=b_3=0$ is again due the $SU(3)$ symmetry of the $\C P^2$ base space. Evaluating $\cZ$ at this extremum we find
the central charge is given by 
\begin{align}
\csugra= \frac{3 k  p^3N_1^2}{(k-3 p) \left(k^2-3 k p+3 p^2\right)}
\,.
\end{align}

This is the central charge for the AdS$_3\times \mathscr{Y}^{p,k} (\C P^2)$ solutions, provided that
they exist. 
We can now compare with the explicit solutions constructed in \cite{Gauntlett:2006af}.
These solutions depended on two relatively prime integers $\pJ,\qJ>0$ 
(which were labelled $p,q$ in \cite{Gauntlett:2006af}).
We first note that in \cite{Gauntlett:2006af} we should set $m=3$, $M=9$, and $h=1$ since we are considering $KE_4^+=\C P^2$.
We then need to make the identifications
\begin{align}
&(k,p)\to (-\pJ,\qJ)\,,\qquad
N_1\to -{ (\pJ+3 \qJ)}{n}\,.
\end{align}
The flux vector is then 
$\{N_a\}
=\{-(\pJ +3 \qJ),\pJ,\qJ,\qJ,\qJ\}{n }$.
In particular, we identify $N_1$, $N_2$ with $N(D_0)$, $N(\tilde D_0)$
 in equation (18) of \cite{Gauntlett:2006af}, respectively, while $(N_3,N_4,N_5)$ are associated with 
 $N(D_a)$.
With these identifications, we precisely recover the result for the central charge given in equation (1) of \cite{Gauntlett:2006af}.
Note that the conditions $\pJ,\qJ>0$, required to have an explicit supergravity solution
\cite{Gauntlett:2006af}, translate into the conditions
\begin{align}
k<0\,\qquad p>0\,,
\end{align}
as mentioned earlier.
In particular the polytope is \emph{not} convex, as observed in \cite{gauntkimwald}.

It is an interesting outstanding problem to identify the $d=2$ $(0,2)$ SCFTs that are dual to these AdS$_3\times \mathscr{Y}^{p,k} (\C P^2)$ solutions.

\subsection{The  $\mathscr{Y}^{p,k} (\C P^1\times \C P^1)$ and $Y^{p,k} (\C P^1\times \C P^1)$ families}\label{s2s2fam}
\begin{figure}[h!]
\begin{center}
  \includegraphics[width=5cm]{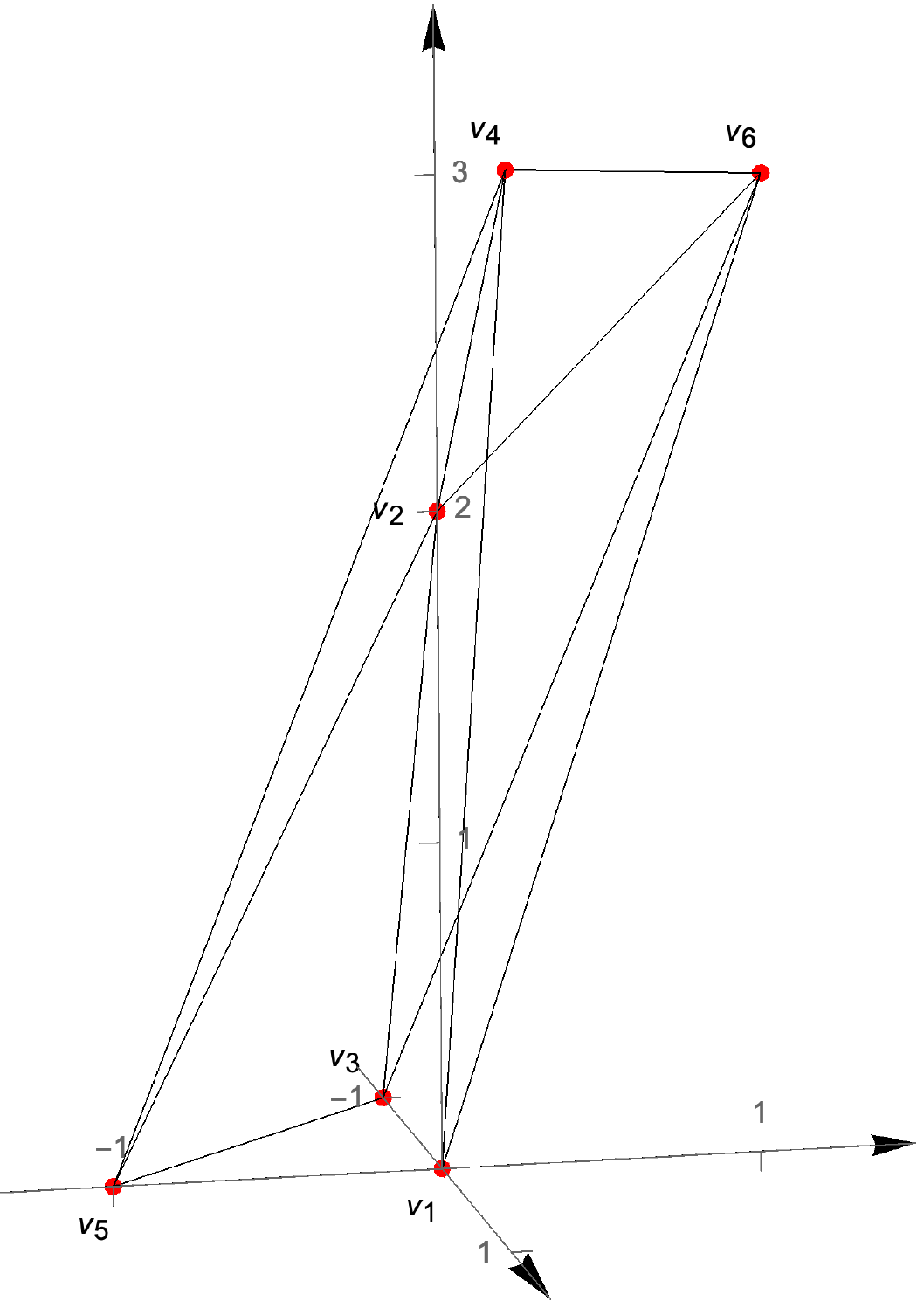}
 \caption{\label{CP1CP1toricdiag} Toric diagram of $Y^{p,k} (\C P^1\times \C P^1)$ with $p=2$, $k=3$,  obtained by projecting on $\R^3$ the vertices  in (\ref{Ypkvcp1cp1}).}
\end{center}
\end{figure}

The toric data associated with $Y^{p,k} (\C P^1\times \C P^1)$ was given in \cite{Martelli:2008rt}.
We take the $d=6$ inward pointing normal vectors to be given by  
\begin{align}
\label{Ypkvcp1cp1}
\v_1 \,  &=\,  (1,0,0,0)~, \quad \v_2 \, = \, (1,0,0,p)~, \quad \v_3 \, = \, (1,-1,0,0)~, \nn\\
\quad \v_4 \, &= \, (1,1,0,k)~,\quad \v_5 \, = \, (1,0,-1,0)\,\quad \v_6 \, = \, (1,0,1,k)\,.
\end{align}
The associated toric diagram, obtained by projecting on $\R^3$ the vertices in (\ref{Ypkvcp1cp1}), is given in
Figure \ref{CP1CP1toricdiag}.
For $Y^{p,k} (\C P^1\times \C P^1)$ we have $0<p<k< 2p$, and there is a convex polytope. For  
the explicit metrics $\mathscr{Y}^{p,k} (\C P^1\times \C P^1)$ we again have $k<0$ and $p>0$.
We continue with general $p,k$. 

The master volume $ \mathcal{V} (\b;\{\lambda_a\})$, given in (\ref{nonsimplemaster}), can be obtained 
from the toric data \eqref{Ypkvcp1cp1} and the results in appendix \ref{masterdetails}.
In the Sasakian limit, $\{\lambda_a =-\tfrac{1}{2b_1}\}$, setting $b_1=4$ and extremizing  with respect to $\{b_2,b_3,b_4\}$  \cite{Martelli:2005tp} we find that the critical Reeb vector is given by $b_2=b_3=0$, with $b_4$ solving the cubic equation
\be
b_4^3 \left(-3 k^2+6 k p-4 p^2\right)+ 2 b_4^2 p \left(9 k^2-8 k p+2 p^2\right)+16 b_4 k p^2 (p-3 k)+48 k^2 p^3 = 0\,.
\label{b4critcp1cp1}
\ee
The fact that $b_2=b_3=0$ is due to the $SU(2)\times SU(2)$ symmetry of the $\C P^1\times \C P^1$ base space.
 Equivalently, the value of $b_4$ obtained from (\ref{b4critcp1cp1}) 
  can be obtained  from extremizing the Sasakian volume with $b_1=4$ upon setting $b_2=b_3=0$, which reads
  \be
 \mathrm{Vol}(Y_7)  (b_4) =    \frac{2 \pi ^4 p \left(b_4^2 \left(3 k^2-6 k p+4 p^2\right)+4 b_4 k p (2 p-3 k)+16 k^2 p^2\right)}{3 b_4^3 \left(4 p-b_4\right){}^3}~.
 \ee
Again, this expression, with $b_4$ obtained from (\ref{b4critcp1cp1}), can be shown to be precisely equal to the Sasaki-Einstein volume 
\be
{Y}^{p,k} (\C P^1\times \C P^1)=  \frac{ \pi ^4 (x_2-x_1) \left({x_2}^3-{x_1}^3\right)}{96 p (1-x_1) (x_2-1)}~, \qquad \mathrm{with}\quad {x_1} = \frac{{x_2} (k-2 p)}{k-2 p {x_2}}\, , 
\ee
  given in equation (2.13) of \cite{Martelli:2008rt}, where it was computed using the  explicit Sasaki-Einstein metric. The relation between the variables $b_4$ and $x_2$ in the two expressions above 
  is  $x_2 = \tfrac{2k} {b_ 4}$. Note, for example, for the special case $p=k=1$ we have $Y_7=Q^{1,1,1}$ and
  $\mathrm{Vol}(Q^{1,1,1})= \pi^4/8$.

We now turn to the AdS$_3\times \mathscr{Y}^{p,k} (\C P^1\times \C P^1)$ solutions. We begin by setting 
$b_1=2$ in the formulae \eqref{constoricseven}--\eqref{Nafluxes}.
The transverse K\"ahler class is determined by $d-3=3$ of the parameters $\{\lambda_a\}$.
We use the constraint equation \eqref{constoricseven} and two of the flux equations \eqref{Nafluxes}, which we take to be $N_1$ and $N_3$, to solve for three of the $\{\lambda_a\}$ which we take to be
$\lambda_1$, $\lambda_2$ and $\lambda_3$. The remaining fluxes are then expressed in terms of $N_1$, $N_3$, and we have
\begin{align}
\{N_a\}
= \left\{N_1,-\frac{kN_1}{k-2p},N_3,N_3,\frac{pN_1}{k-2 p}-N_3,\frac{pN_1}{k-2 p}-N_3\right\}\,.
\end{align}
It will be useful in a moment to notice that if we restrict the fluxes by imposing
$N_3=\frac{pN_1}{2 (k-2 p)}$, then 
$\{N_a\}
=\{1,-\frac{k}{k-2p},\frac{p}{2 (k-2 p)},\frac{p}{2 (k-2 p)},\frac{p}{2 (k-2 p)},\frac{p}{2 (k-2 p)}\}N_1$.

We next calculate the trial central charge $\cZ$ and find, in particular, that it is independent of 
$\lambda_4$, $\lambda_5$ and $\lambda_6$, 
in agreement with the invariance of the problem under the three independent transformations in (\ref{lamgt}).
Furthermore, $\cZ$ is quadratic in
$b_2,b_3$ and $b_4$, again as expected. It is now straightforward to
extremize $\cZ$ with respect to these remaining variables and we find the unique extremum  has
$\vec{b}=(2,0,0,b_4)$, with
\begin{align}
b_4 = 
\frac{k p \left[N_1^2 p^2 (k-p)+N_3^2 (k-2 p)^3-N_1 N_3 p (k-2 p)^2\right]}{N_1^2 p^2 (k-p)^2+k N_3^2 (k-2 p)^3-k N_1 N_3 p (k-2 p)^2}\,.
\end{align}
The fact that $b_2=b_3=0$ is due to the $SU(2)\times SU(2)$ symmetry of the $\C P^1\times \C P^1$ base space. Evaluating $\cZ$ at this extremum we find
the central charge is given by
\begin{align}\label{cccp1cp1}
\csugra = 
\frac{6 k N_3 p \left[N_1^3 p^3-2 N_3 N_1^2 p^2 (k-2 p)+2 N_3^2 N_1 p (k-2 p)^2-N_3^3 (k-2 p)^3\right]}{N_1^2 p^2 (k-p)^2+k N_3^2 (k-2 p)^3-k N_1 N_3 p (k-2 p)^2}\,.
\end{align}

This is the central charge for the AdS$_3\times \mathscr{Y}^{p,k} (\C P^1\times \C P^1)$ solutions, provided that they exist. We can now compare the above results, for the special case that the fluxes are restricted via $N_3=\frac{pN_1}{2 (k-2 p)}$ as mentioned above, with the explicit solutions constructed in \cite{Gauntlett:2006af}.
These solutions depended on two relatively prime integers $\pJ,\qJ>0$ (which were labelled $p,q$ in \cite{Gauntlett:2006af}).
Since we are considering $KE_4^+=\C P^1\times \C P^1$, we need to set $m=2$ and $M=8$
in the formulae in \cite{Gauntlett:2006af}.
We also need to make the identifications
\begin{align}
&(k,p)\to (-\pJ,\qJ)\,,\qquad N_1\to -2(\pJ + 2 \qJ)\frac{n }{h}\,.
\end{align}
The flux vector is then 
$\{N_a\}
=\{-2(\pJ + 2 \qJ),2\pJ,\qJ,\qJ,\qJ,\qJ\}\frac{n }{h}$.
In particular, we identify $N_1$, $N_2$ with $N(D_0)$, $N(\tilde D_0)$
in equation (18) of \cite{Gauntlett:2006af}, respectively,
while $(N_3,N_4,N_5,N_6)$ are associated with 
 $N(D_a)$.
With these identifications, we precisely recover the result for the central charge given in equation (1) of \cite{Gauntlett:2006af}.
The analysis of \cite{Gauntlett:2006af} shows that the supergravity solutions exist for $\pJ,\qJ>0$, which translates into the conditions
\begin{align}
k<0\, ,\qquad p>0\,.
\end{align}
In particular the polytope is not convex, as observed in \cite{gauntkimwald}.

It is interesting that the central charge for these AdS$_3\times \mathscr{Y}^{p,k} (\C P^1\times \C P^1)$
can also be obtained in another way. Indeed, by selecting one of the
$\C P^1$ factors, we can view $\mathscr{Y}^{p,k} (\C P^1\times \C P^1)$ as a fibration of
$Y^{\bar p,\bar q}$ over the other $\C P^1$ factor, as discussed in section~6.1 and 7.2 of 
\cite{Gauntlett:2018dpc} (and we note that $(\bar p,\bar q)$ were denoted $(p,q)$ in \cite{Gauntlett:2018dpc}). The fibration in \cite{Gauntlett:2018dpc} was specified by three
integers $n_1,n_2,n_3$, with $n_1=2$, as demanded by supersymmetry, and for simplicity $n_2$ and $n_3$ were taken to be equal with $n_2=n_3\equiv -s$. In addition, the solutions were specified
by an additional two integers, $m,N$, which determined the fluxes. To compare to the solutions discussed here we should first restrict the solutions so that the fibration has $s=\bar q-\bar p$. We then need to make the identifications $(p,k)=(\bar p,\bar p -\bar q)$ as well as $(m,N)=(-N_1/N_3,N_3)$. Having done this, one finds that the central charge in \eqref{cccp1cp1} agrees exactly
with equation (6.7) of \cite{Gauntlett:2018dpc}. 

It is an interesting outstanding problem to identify the $d=2$ $(0,2)$ SCFTs that are dual to these explicit AdS$_3\times \mathscr{Y}^{p,k} (\C P^1\times \C P^1)$ solutions. In particular, as discussed in
\cite{Gauntlett:2018dpc}, viewing them as a fibration of $Y^{\bar p,\bar q}$ over $\C P^1$ we have $\bar p <\bar q$ and hence  
they are not associated, at least in any simple way, with compactifying the $d=4$ quiver gauge theories dual to 
AdS$_5\times Y^{\bar p,\bar q}$, with Sasaki-Einstein metric on $Y^{\bar p,\bar q}$, since the latter have 
$\bar p >\bar q$.

Finally, we note that the explicit supergravity solutions in \cite{Gauntlett:2006af} with $KE_4^+=\C P^1\times \C P^1$
can be generalised, allowing the relative sizes of the two $\C P^1$ to be different. Indeed such local solutions can be obtained by T-dualising the solutions in section 5 of \cite{Gauntlett:2006qw}. It is natural to conjecture that
regular solutions with properly quantized flux can be obtained with independent $N_1$, $N_3$, 
and central charge as in \eqref{cccp1cp1}.
  
\section{Supersymmetric AdS$_2\times Y_9$ solutions}\label{ads2examples}
\subsection{General set-up}\label{gensetup}
We now consider supersymmetric AdS$_2\times Y_9$ solutions of $D=11$ supergravity that are dual to superconformal quantum mechanics with two supercharges of the form
\begin{align}
\diff s^2_{11} &=  L^2 \ex^{-2B/3}\left(\diff s^2_{\mathrm{AdS}_2} + \diff s^2_{9}\right)~,\nn\\
G &= L^3\vol_{\mathrm{AdS}_2}\wedge F ~.\label{ansatzd11}
\end{align}
Here $L$ is an overall length scale and $\diff s^2_{\mathrm{AdS}_2}$ is the metric on a 
unit radius AdS$_2$ with volume form $\vol_{\mathrm{AdS}_2}$. The 
warp factor $\BB$ is a function on the compact Riemannian internal space $(Y_9, \diff s^2_9)$ and
$F$ is a closed two-form on $Y_9$. We also need to impose flux quantization. Since $G\wedge G=0$ for the above ansatz, we need to impose
\begin{align}
\frac{1}{(2\pi \ell_p)^6}\int_{\Sigma_A} *_{11}G &= N_A \in \mathbb{Z}~,\label{quantization11}
\end{align}
where $\ell_p$ is the Planck length and $\Sigma_A\subset Y_9$, with $\{\Sigma_A\}$ forming an integral basis for the free part of $H_7(Y_9,\Z)$. The geometry of these solutions was first analysed in \cite{Kim:2006qu} and then extended in \cite{Gauntlett:2007ts}.

We again consider off-shell supersymmetric geometries, as described in detail in \cite{Couzens:2018wnk}.
These are configurations of the form \eqref{ansatzd11} which admit the required Killing spinors and become supersymmetric solutions 
when we further impose the equation of motion for the four-form. The complex cone $C(Y_9)$, in complex dimension $n=5$ and
with Hermitian metric $\diff s^2_{C(Y_9)} = \diff r^2 + r^2\diff s^2_9$,
admits a global holomorphic $(5,0)$-form $\Psi$. The 
complex structure pairs the radial vector $r\partial_r$ with the R-symmetry vector field $\xi$. 
Likewise, the complex structure pairs $\diff r/r$ with the dual one-form $\eta$, and $\xi\lrcorner\eta=1$.
The vector $\xi$ has unit norm and defines a foliation $\mathcal{F}_\xi$ of $Y_9$. The 
basic cohomology for this foliation is denoted $H^*_B(\mathcal{F}_\xi)$.

The supersymmetric geometries have a metric of the form
\begin{align}\label{splitmetric11}
\diff s^2_9 = \eta^2 + \ex^B\diff s^2(J)~,
\end{align}
where $\diff s^2(J)$ is a transverse K\"ahler metric with transverse K\"ahler form $J$. 
The transverse K\"ahler metric determines the full supersymmetric geometry including the fluxes. In particular, the conformal factor is fixed via
$\ex^{B}=R/2$, where $R$ is the Ricci scalar of the transverse K\"ahler metric. We also have
\begin{align}
\diff \eta= \rho\,,
\end{align}
where $\rho$ is the Ricci two-form of the transverse K\"ahler metric, and 
$\mathcal{L}_\xi\Psi= \ii  b_1 \Psi$, with $b_1=1$. 
It was shown in \cite{Couzens:2018wnk} that there is a supersymmetric action $\Ssusy=\Ssusy[\xi;[J]]$, whose extremum allows
one to determine the effective two-dimensional Newton's constant, $G_2$, with $1/(4G_2)$ giving the logarithm
of the partition function of the dual superconformal quantum mechanics.

In this paper we are interested in the specific class of $Y_9$ which are fibred over a Riemann surface $\Sigma_g$:
\begin{align}\label{fibred}
Y_7 \ \hookrightarrow \ Y_9 \ \rightarrow  \Sigma_g\,.
\end{align}
The R-symmetry vector $\xi$ is assumed to be tangent to $Y_7$. While the general class of supersymmetric
AdS$_2\times Y_9$ solutions might arise as the near horizon limits of black hole solutions of $D=11$ supergravity,
this seems particularly likely in the case that $Y_9$ is of the fibred form \eqref{fibred}. Indeed we expect that such solutions
can arise as the near horizon limit of black holes, with horizon topology $Y_9$, in an asymptotically AdS$_4\times Y_7$ background
with a Sasaki-Einstein metric on $Y_7$. In fact this is known to be the case for the so-called universal twist fibration with genus $g>1$ \cite{Romans:1991nq,Caldarelli:1998hg,Gauntlett:2001qs,Gauntlett:2006qw,Gauntlett:2007ma}.
As shown in \cite{Couzens:2018wnk} the entropy of the black holes, $S_{BH}$, should be related to the effective 
two-dimensional Newton's constant, $G_2$, via $S_{BH}=1/(4G_2)$. In the following we will refer to the
supersymmetric action $\Ssusy$, with a suitable normalization given below, as the {\it entropy function}.

We now further consider $Y_7$ to be toric with an isometric $U(1)^4$ action, as described in section \ref{mastvol}.
In order to obtain $\Ssusy$, we can generalise the analysis of section 4 of \cite{Gauntlett:2018dpc}. 
The fibration structure is specified by four integers $(n_1,n_2,n_3,n_4)$ and we have
\begin{align}
n_1= 2(1-g)\,,
\end{align}
since we have chosen a basis for the $U(1)^4$ vectors satisfying \eqref{liederiv} with $b_1=1$. Furthermore,
up to an irrelevant exact basic two-form, the transverse K\"ahler form on $Y_9$ may be taken to be 
\begin{align}\label{Jtwisted}
J = \omega_{\mathrm{twisted}} + A\, \vol_{\Sigma_g} + \mbox{basic exact}~.
\end{align}
Here $\omega_{\mathrm{twisted}}$ is a K\"ahler form on the complex cone over $Y_7$ that is suitably twisted over
$\Sigma_g$. We have normalized $\int_{\Sigma_g}\vol_{\Sigma_g}=1$, and $A$ is effectively a K\"ahler class parameter for the Riemann surface. 

By directly generalizing the arguments in section 4.2 of \cite{Gauntlett:2018dpc}, we find that the key quantities can all be expressed in terms of $n_i$ and $A$ as well as the master volume 
$\mathcal{V}(\vec{b};\lambda_a)$. The supersymmetric action is given by
\begin{align}\label{rhoomegaintseven}
\Ssusy  \ &
= -A \sum_{a=1}^d \frac{\partial\mathcal{V}}{\partial \lambda_a} - 2\pi b_1 \sum_{i=1}^4 n_i \frac{\partial\mathcal{V}}{\partial b_i}~.
 \end{align}
The constraint equation that must be imposed, in order that flux quantization is well-defined, is given by
\begin{align}\label{constraintagain}
0 \ 
 &= \ A \sum_{a,b=1}^d \frac{\partial^2\mathcal{V}}{\partial\lambda_a\partial\lambda_b} - 2\pi n_1 \sum_{a=1}^d\frac{\partial \mathcal{V}}{\partial\lambda_a} + 2\pi b_1 \sum_{a=1}^d \sum_{i=1}^4 n_i\frac{\partial^2 \mathcal{V}}{\partial \lambda_a\partial b_i}~.
\end{align}
Finally, we consider flux quantization, and there are two types of seven-cycle to consider. First, there is a distinguished seven-cycle, 
$\Sigma$, which is a copy of $Y_7$ obtained by picking a point on $\Sigma_g$, and we have
\begin{align}\label{Nnice}
\frac{(2\pi \ell_p)^6}{L^6}N \ 
 & = \ -\sum_{a=1}^d\frac{\partial\mathcal{V}}{\partial\lambda_a}~.
\end{align}
We can also consider the seven-cycles $\Sigma_a$, $a=1,\dots, d$, obtained by fibreing 
a toric five-cycle $T_a$ on $Y_7$, over $\Sigma_g$, and we have
\begin{align}\label{Maint}
\frac{(2\pi \ell_p)^6}{L^6}M_a \ 
&=\ \frac{A}{2\pi} \sum_{b=1}^d \frac{\partial^2\mathcal{V}}{\partial\lambda_a\partial\lambda_b} + b_1 \sum_{i=1}^4 n_i \frac{\partial^2 \mathcal{V}}{\partial\lambda_a\partial b_i}~.
\end{align}
We find it convenient to also introduce the equivalent notation for the fluxes $M_a$:
\begin{align}\label{fluxredefgoth}
\fm_a\ \equiv \ -\frac{M_a}{N}\,.
\end{align}
The toric five-cycles $[T_a]\in H_5(Y_7,\mathbb{Z})$ are not all independent. The $[T_a]$ generate the free part of $H_5(Y_7;\mathbb{Z})$, 
but they also satisfy 4 linear relations $\sum_{a=1}^d v^i _a[T_a]=0\in H_5(Y_7;\mathbb{Z})$. This gives rise to the corresponding homology relation 
in $Y_9$, $\sum_{a=1}^d v^i _a[\Sigma_a]=-n_i[Y_7]\in H_7(Y_9;\mathbb{Z})$,
which implies the useful relation\footnote{A topological proof of (\ref{vMrelation}) may be found in 
appendix \ref{apptop}. 
It would be nice to prove this relation more 
 directly,  using a similar method to that given in (4.37)--(4.39) of \cite{Gauntlett:2018dpc}.}
\begin{align}
\sum_{a=1}^dv^i_aM_a= -n_i N\quad \Leftrightarrow\quad 
\sum_{a=1}^dv^i_a\fm_a= n_i\,,\qquad 
 i=1,2,3,4\,.
\label{vMrelation}
\end{align} 
We thus have a total of $d-3$ independent flux numbers $N$ and $\{M_a\}$.
In all of the above formulae we should set 
\begin{align}
b_1= 1\,,
\end{align}
after taking any derivatives with respect to the $b_i$.
Finally, we note that we can also express the supersymmetric action in the following compact
form
\begin{align}\label{susyactneat}
\Ssusy  = \frac{(2\pi \ell_p)^6}{L^6}  \frac{2\pi N}{3}\left(\frac{A}{2\pi}+\sum_{a=1}^d\lambda_a\fm_a\right)\,.
 \end{align}
To prove this we first multiply \eqref{Maint} by $\lambda_a$ and then sum over $a$. Recalling that
the master volume is homogeneous of degree three in the $\lambda_a$ and using Euler's theorem we deduce
that 
\begin{align}
\frac{(2\pi \ell_p)^6}{L^6} \sum_{a=1}^d \lambda_aM_a  =  \frac{A}{2\pi} 2 \sum_{b=1}^d  \frac{\partial \mathcal{V}}{\partial\lambda_b} + 3 b_1 \sum_{i=1}^4 n_i \frac{\partial \mathcal{V}}{\partial b_i}\, . 
\end{align}
Using this and
\eqref{Nnice} we then obtain \eqref{susyactneat}.

For a given fibration, specified by $(n_1,n_2,n_3,n_4)$ with $n_1=2(1-g)$,
the on-shell action is obtained by extremizing $\Ssusy$. {\it A priori} with $b_1=1$, there are $d+1$ parameters comprising 
$(b_2,b_3,b_4)$, along with the $(d-3)+1$ independent K\"ahler class parameters $\{\lambda_a\}$ and $A$. The procedure is to impose \eqref{constraintagain}, \eqref{Nnice} and \eqref{Maint}, which, as we noted, is generically $d-2$ independent
conditions, and hence $\Ssusy$ will generically be a function of three remaining variables. 
We then extremize the action with respect to these variables,
or equivalently extremize the ``trial entropy function", $\cS$, defined by
\begin{align}\label{cS2}
\cS \equiv   \frac{8\pi^2 L^9}{(2\pi\ell_p)^9}\, \Ssusy~,
\end{align}
which has the property that for an on-shell supersymmetric solution, {\it i.e.} after extremization,
we obtain the two-dimensional Newton's constant
\begin{align}\label{cS23next}
\cS  |_\mathrm{on-shell} =  \frac{1}{4G_2}~.
\end{align}
As explained in \cite{Couzens:2018wnk} this should determine the logarithm of the partition function 
of the dual supersymmetric quantum mechanics. Moreover, when the AdS$_2\times Y_9$ solution
arises as the near horizon limit of a black hole solution, it gives the entropy of the black hole,
$S_{BH}=\cS  |_\mathrm{on-shell}$.
The entropy of such black holes should be accounted for by the  
microstates of the dual $d=3$, ${\cal N}=2$ field theories when placed on $S^1\times \Sigma_g$; the number of these microstates is expected to be captured by the corresponding supersymmetric topological twisted index.  

We may also compute the geometric R-charges $R_a=R[T_a]$ associated with the operators dual to M5-branes wrapping
the toric divisors $T_a\subset Y_7$ at a fixed point on the base $\Sigma_g$.
The natural expression\footnote{We have not verified this formula by explicitly checking the $\kappa$-symmetry of an M5-brane wrapped on the toric divisors $T_a$.
It is analogous to the corresponding expression for AdS$_3$ solutions, where it was also motivated by computing the dimension of baryonic operators dual to D3-branes wrapping supersymmetric cycles in $Y_5$ \cite{Couzens:2017nnr}. We will indirectly verify this normalization below.} is given by 
\begin{align}
R_a= R[T_a]= \frac{4\pi L^6}{(2\pi \ell_p)^6}\int_{T_a}\eta\wedge \tfrac{1}{2!}\omega^2\,.
\label{defwrapped}
\end{align}
Following similar arguments to those of section 4 in \cite{Gauntlett:2018dpc} we then deduce that
\begin{align}
R_a=  - \frac{2L^6}{(2\pi \ell_p)^6}\frac{\de \mathcal{V}}{\de \lambda_a}\, .
\label{defgeomRcharges}
\end{align}
As for the fluxes in \eqref{fluxredefgoth}, we find it convenient to strip out a factor of $N$ and define
\begin{align}\label{defcapDel}
\Delta_a\equiv \frac{R_a}{N}\,.
\end{align}
In particular, using \eqref{Nnice}, notice that we have 
\begin{align}
\label{sumrchargesistwo}
\sum_{a=1}^d R_a  =   2 N
\quad \Leftrightarrow\quad 
\sum_{a=1}^d \Delta_a  =    2\,.
\end{align}
We also note that for the generic examples,  with toric data satisfying \eqref{keyvRrel} we have, equivalently,
\begin{align}\label{keyvRrel2}
\sum_{a=1}^d{v}_a^iR_a= \frac{2b^i}{b_1}N
\quad \Leftrightarrow\quad 
\sum_{a=1}^d{v}_a^i\Delta_a= \frac{2b^i}{b_1}\,,
\qquad 
 i=1,2,3,4\,,
\end{align}
from which the relation \eqref{sumrchargesistwo} is the $i=1$ component. 
Recall that this relation implies that the master volume $\mathcal{V}$ is invariant
under the ``gauge transformation'' \eqref{lamgt} acting on the $\lambda_a$. As we noted in the previous section, this implies that all of the derivatives of 
$\mathcal{V}$ with respect to $\lambda_a$
are also invariant under this gauge transformation. 
However, this is not the case after taking derivatives with respect to $b_i$ (since the gauge transformation involves the vector $b_i$) 
and so we now 
discuss the effect of \eqref{lamgt} on the extremal problem in the case of fibered geometries\footnote{This analysis applies also to the $Y_5  \hookrightarrow Y_7  \rightarrow  \Sigma_g $ geometries discussed in \cite{Gauntlett:2018dpc}.}.

The variation of $\frac{\de \mathcal V}{\de b_j}$ under  \eqref{lamgt} is given by 
 \begin{align}
 \delta \frac{\de \mathcal V}{\de b_j} & \ =  \  \sum_{a=1}^d   \frac{\de^2 \mathcal V}{\de \lambda_a \de b_j} \delta \lambda_a =   \sum_{i=1}^4\gt_i  \sum_{a=1}^d  (v_a^i b_1-b_i)  \frac{\de^2 \mathcal V}{\de \lambda_a \de b_j} \, .
  \end{align}
On the other hand, assuming that \eqref{keyvRrel} holds and taking a derivative of this with respect to $b_j$, a short computation leads to the identity  \cite{Gauntlett:2018dpc}
 \begin{align}\label{litiden}
 \sum_{a=1}^d \left( b_1 v_a^i    - b_i  \right)   \frac{\partial^2  \mathcal{V}}{\partial b_j \partial \lambda_a }   & \ =  \  \left( \delta_{ij}    - \frac{b_i \delta_{1j}}{b_1} \right)\sum_{a=1}^d  \frac{\partial  \mathcal{V}}{\partial \lambda_a }  ~,
\end{align}
and hence we have 
 \begin{align}
 \label{thischanges}
 \delta \frac{\de \mathcal V}{\de b_j} &    
 =   -\left(\gt_j-\frac{1}{b_1}\delta_{1j}\sum_{i=1}^4 \gt_i b_i  \right)\frac{(2\pi \ell_p)^6}{L^6} N \, , 
  \end{align} 
where we used \eqref{Nnice}. 
A similar computation for the variation $\frac{\de^2 \mathcal V}{\de \lambda_a\de b_j}$, and using the expression obtained
by differentiating \eqref{litiden} with respect to $\lambda_b$, we deduce that
 \begin{align}
 \label{thischanges2}
 \delta \frac{\de^2 \mathcal V}{\de b_j\de\lambda_a} & 
 =   \left(\gt_j-\frac{1}{b_1}\delta_{1j}\sum_{i=1}^4 \gt_i b_i  \right) \frac{\de^2 \mathcal V}{\de b_j\de\lambda_a}  \, .
  \end{align} 
Using these results we find that if we extend the gauge transformation to also allow for a variation of the
K\"ahler class parameter $A$ via
\begin{align}\label{lamgt2}
\delta \lambda_a&\ =  \ \sum_{i=1}^4\gt_i(v_a^i b_1-b_i)\,,\nn\\
\delta A & \  \equiv \  - 2\pi    \sum_{a=1}^d \delta \lambda_a \fm_a  \  =  \   - 2\pi      \sum_{i=1}^4\gt_i ( n_i  b_1- b_in_1 )\,,  
\end{align}
where the second expression in the second line follows from \eqref{vMrelation}, then in addition to
$N$ being invariant then so are the fluxes $M_a$ as well as the supersymmetric action $\Ssusy$, as one
can easily see from the expression \eqref{susyactneat}. 

While these gauge transformations are certainly interesting and useful, they are constrained. This follows from the
fact that since $\{\lambda_a\}$ and $A$ parametrize K\"ahler classes they must satisfy some positivity constraints.
For example, the transformations \eqref{lamgt2} naively suggest that we might choose a gauge with $A=0$, but this should not be possible. In fact in some of the examples we study, one finds $b_i  \ =   \  \frac{b_1}{n_1} n_i $, on-shell, which also indicates
the problem with such a putative gauge choice. It would certainly be interesting to determine
the positivity constraints on the K\"ahler class parameters
and hence the restrictions on the gauge transformations.

\subsection{Entropy function in terms of $\Delta_a$ variables}\label{chgevars}

Before discussing some explicit examples of AdS$_2\times Y_9$ solutions with $Y_9$ obtained as a fibration of toric $Y_7$ over $\Sigma_g$, we first show that the above variational problem incorporates some 
general features concerning $\mathcal{I}$-extremization discussed  in \cite{Hosseini:2016tor}. 
We will further develop the connection of our formalism to $\mathcal{I}$-extremization, 
in the subsequent subsections, especially
section \ref{SISI}.

The master volume $\mathcal{V}$ is defined to be a function of $(d-3)+3=d$ independent variables $(\lambda_a, b_2,b_3,b_4)$. We want to consider a change of variables in which $\mathcal{V}$ is, instead, a function of 
the $d$ variables $\Delta_a$ (see \eqref{defcapDel}) given by
\begin{align}
\Delta_a=  - \frac{2L^6}{N(2\pi \ell_p)^6}\frac{\de \mathcal{V}}{\de \lambda_a}\,\,,
\label{defgeomRcharges2}
\end{align}
where at this stage $N$ is a free parameter ({\it i.e.} not yet given by \eqref{Nnice} so we don't yet impose $\sum_a\Delta_a=2$.)
Assuming that this is an invertible change of variables,
using the chain rule, we then have
\begin{align}\label{chainrule}
\frac{\de \mathcal{V}}{\de b_i } &= \sum_a \frac{\de \mathcal{V}}{\de \Delta_a}    \frac{\de \Delta_a}{\de b_i}
= - \frac{2L^6}{N(2\pi \ell_p)^6}\sum_a \frac{\de \mathcal{V}}{\de \Delta_a}  \frac{\de^2 \mathcal{V}}{\de b_i\de \lambda_a}\,,\nn\\
\frac{\de \mathcal{V}}{\de \lambda_a } &= \sum_b \frac{\de \mathcal{V}}{\de \Delta_b}    \frac{\de \Delta_b}{\de \lambda_a }=  - \frac{2L^6}{N(2\pi \ell_p)^6}\ \sum_b \frac{\de \mathcal{V}}{\de \Delta_b}    \frac{\de^2 \mathcal{V}}{\de \lambda_a \de \lambda_b}\,. \end{align}

Using this, and also \eqref{Nnice}, we can then write the supersymmetric action \eqref{rhoomegaintseven} as
\begin{align}
\Ssusy  \ &
= \frac{(2\pi \ell_p)^6}{L^6}A    N \ 
   + 
     \frac{4\pi b_1}{N} \frac{L^6}{(2\pi \ell_p)^6}\sum_{i,a} n_i \frac{\de \mathcal{V}}{\de \Delta_a}     \frac{\de^2 \mathcal{V}}{\de b_i\de \lambda_a}\,.
 \end{align}
We next multiply the expression for the fluxes $M_a$, given in (\ref{Maint}), by $\frac{\de \mathcal{V}}{\de \Delta_a}$ and then sum over $a$ to get
\begin{align}\label{mock}
\frac{(2\pi \ell_p)^6}{L^6} \sum_a M_a \frac{\de \mathcal{V}}{\de \Delta_a}
&=\ \frac{A}{2\pi} \sum_{a,b} \frac{\partial^2\mathcal{V}}{\partial\lambda_a\partial\lambda_b} \frac{\de \mathcal{V}}{\de \Delta_a}+ b_1 \sum_{i,a} n_i \frac{\partial^2 \mathcal{V}}{\partial\lambda_a\partial b_i}\frac{\de \mathcal{V}}{\de \Delta_a}~.
\end{align}
Using the second line of \eqref{chainrule} as well as \eqref{Nnice}, we can recast this as
\begin{align}
\frac{4\pi}{N}\sum_a M_a \frac{\de \mathcal{V}}{\de \Delta_a} 
= \frac{(2\pi \ell_p)^6}{L^6}A    N \ 
   + 
   \frac{4\pi b_1}{N} \frac{L^6}{(2\pi \ell_p)^6}\sum_{i,a} n_i \frac{\de \mathcal{V}}{\de \Delta_a}     \frac{\de^2 \mathcal{V}}{\de b_i\de \lambda_a}\,.
 \end{align}
Hence the {\it off-shell} supersymmetric action can be written in the remarkably simple form
\begin{align}\label{ssss}
\Ssusy  (b_i,\fm_a)= -  4\pi \sum_{b=1}^d \fm_b \left. \frac{\de \mathcal{V}}{\de \Delta_b} \right |_{\Delta_b(b_i,\fm_a)} \,,
\end{align}
where $\fm_a$ are the normalized fluxes $\fm_a\equiv -M_a/N$ that were introduced 
in \eqref{fluxredefgoth}. 
Here on the right hand side recall that originally the master volume
$\mathcal{V}$ is a function of $(\lambda_a,b_2,b_3,b_4)$, which we then express as a function 
of $\Delta_b=\Delta_b(\lambda_a,b_2,b_3,b_4)$, assuming this is invertible. However, one can then eliminate the 
K\"ahler parameters $\{\lambda_a\}$ in terms of the flux quantum numbers  $\fm_a\equiv -M_a/N$ by imposing (\ref{Maint}) as a 
final step, so that $\Delta_b=\Delta_b(b_i,\fm_a)$.

\subsection{The universal twist revisited}
\label{unitwist}

As our first example, we apply our general formalism of section \ref{gensetup}
to the case called the universal twist. 
Specifically, we consider a nine-dimensional manifold $Y_9$ that is a fibration of a toric $Y_7$ 
over a Riemann surface $\Sigma_g$, with genus $g>1$, where the twisting is only along the $U(1)_R$ R-symmetry. 
The corresponding supergravity solutions exist for any $Y_7=SE_7$ that is a quasi-regular Sasaki-Einstein manifold;
these solutions, generalising \cite{Gauntlett:2001qs}, were mentioned in footnote 5 of \cite{Gauntlett:2006qw} and in section 6.3 of \cite{Gauntlett:2006ns}. Furthermore, the magnetically charged black hole solutions of 
\cite{Romans:1991nq,Caldarelli:1998hg} can be uplifted on an arbitrary $SE_7$ using the results of \cite{Gauntlett:2007ma} to obtain
solutions which interpolate between AdS$_4\times SE_7$ in the UV and the AdS$_2\times Y_9$ solutions in the IR.
These solutions and the associated field theories were recently discussed in \cite{Azzurli:2017kxo}. 
We will use the formalism of section \ref{gensetup} to recover some of the results of \cite{Azzurli:2017kxo} as well
as extend them by discussing the geometric R-charges associated with wrapped M5-branes.

We closely follow the analysis in section 5 of \cite{Gauntlett:2018dpc} which considered the analogous universal twist in the context of AdS$_3$ solutions. From a geometric point of view the universal twist corresponds to choosing the fluxes $n_i$ to be
aligned with the R-symmetry vector, and so we impose 
\begin{align}\label{twistb}
n_i  \ =   \  \frac{n_1}{b_1} b_i~,
\end{align}
with $n_1=2(1-g)$. We also need to impose that the R-charges are proportional to the fluxes as is clear from the construction
of the supergravity solutions. Note that we will need to check, {\it a posteriori}, that after carrying out extremization
the on-shell value of $\vec{b}$ is consistent with the left hand side of (\ref{twistb})  being integers. 
Inserting this into the formulas for the action (\ref{rhoomegaintseven}),  the constraint (\ref{constraintagain})    and the flux quantization conditions (\ref{Nnice}), \eqref{Maint}, and using the fact that the master volume $\mathcal{V}$ is homogeneous of degree minus one in $\vec{b}$,
these reduce respectively to 
\begin{align}
\label{susyactuni}
\Ssusy & =  A  \frac{(2\pi \ell_p)^6}{L^6}  N   + 2\pi n_1   \mathcal{V} ~,\\
\label{constraintformulauniverse}
0 &=  A\sum_{a,b=1}^d \frac{\partial^2\mathcal{V}}{\partial{\lambda_a}\partial{\lambda_b}}+ 4\pi n_1
   \frac{(2\pi \ell_p)^6}{L^6}  N~,\\
\label{formuletta}
\frac{(2\pi \ell_p)^6}{L^6}  M_a & =    \frac{A}{2\pi}\sum_{b=1}^d\frac{\partial^2\mathcal{V}}{\partial\lambda_a\partial\lambda_b}-n_1 \frac{\partial \mathcal{V}}{\partial\lambda_a}~,\\
\label{enntwo}
\frac{(2\pi \ell_p)^6}{L^6} N &=  -\sum_{a=1}^d \frac{\partial \mathcal{V}}{\partial\lambda_a}~.
\end{align}

In contrast to \cite{Gauntlett:2018dpc}, the above equations are now quadratic in $\lambda_a$ instead of linear. In general we may also freely specify 
the flux quantum numbers $M_a$, subject to the constraint (\ref{vMrelation}) that follows because the seven-cycles $\Sigma_a$ 
are not all independent in homology on $Y_9$. However, by definition the universal twist has a specific choice of the fluxes $M_a$, 
proportional to the R-charges $R_a$ (see equation (\ref{itslate}) below). In order to solve (\ref{susyactuni})--(\ref{enntwo}), we will instead 
make the ansatz that the $\lambda_a$ parameters are all equal, and then \emph{a posteriori} check that this correctly
reproduces the universal twist solutions. 
Thus setting $\lambda_a=\lambda$ for $a=1,\dots,d$, from 
(\ref{newmaster})
and (\ref{sasakkia})
we have 
\begin{align}
\mathcal{V}   =  & -8 b_1^3 \lambda^3  \mathrm{Vol} (Y_7)\,,
 \label{putin}
\end{align}
and from 
(\ref{newformulas}) 
we also have
\be
\sum_{a=1}^d\frac{\partial \mathcal{V}}{\partial\lambda_a}= -24 b_1^3 \lambda^2  \mathrm{Vol} (Y_7)\quad  \mathrm{and} \quad  
\sum_{a,b=1}^d\frac{\partial^2 \mathcal{V}}{\partial\lambda_a\lambda_b} = -48 b_1^3 \lambda  \mathrm{Vol} (Y_7)\, . 
\ee
We can next use the constraint equation \eqref{constraintformulauniverse} to solve for $A$ to obtain
\begin{align}
A= \frac{(2\pi \ell_p)^6}{L^6} \frac{4\pi n_1N}{48\lambda b_1^{3}\mathrm{Vol} (Y_7)}\,.
\end{align}
Since $A>0$ is the volume of the Riemann surface $\Sigma_g$ (see (\ref{Jtwisted})), we deduce that $n_1N$ has the same sign as $\lambda$. Without loss of generality 
we continue with $N>0$, and since we are assuming $g>1$ we must have $\lambda<0$.
From \eqref{enntwo} we next solve for $\lambda$ to get
\begin{align}
\lambda  =  -\frac{(2\pi \ell_p)^3}{L^3} \frac{N^{1/2}}{2\sqrt{6}b_1^{3/2}\mathrm{Vol} (Y_7)^{1/2}}~.
\label{trump}
\end{align}
Inserting these results into 
the supersymmetric action \eqref{susyactuni} we find that we can write the off-shell entropy function \eqref{cS2} as
\begin{align}
\cS  & \  = \ \frac{32\pi^3 (g-1)N^{3/2}}{3\sqrt{6}b_1^{3/2}\mathrm{Vol} (Y_7)^{1/2}}\,.
\end{align}

This action has to be extremized with respect to $b_2, b_3,b_4$, holding $b_1$ fixed to be 1. 
On the other hand, the Sasaki-Einstein volume can be obtained by varying over 
$b_2, b_3,b_4$ while holding $b_1$ fixed to be 4. To proceed we 
define $\vec{b} = \frac{1}{4}\vec{r}$ and use the fact that $\Vol (Y_7)$ is homogeneous of degree minus 
four in $\vec{b}$, to rewrite the action as
\begin{align}
\cS  (\vec{r}) \   =  \  \frac{2\pi^3 (g-1)N^{3/2}}{3\sqrt{6}b_1^{3/2}\mathrm{Vol} (Y_7)^{1/2}}~.
\end{align}
Since $\mathrm{Vol} (Y_7) (\vec{r})$ with $r_1=4$ is extremized by the critical Reeb vector $\vec{r}=\vec{r}_*$, with $\mathrm{Vol} (Y_7) (\vec{r}_*)$ being the Sasaki-Einstein volume, we conclude that $\Ssusy  (\vec{r})$ is extremized 
for the critical R-symmetry vector given by
\begin{align}\label{critbv}
\vec{b}_* = \frac{1}{4}\vec{r}_*\,.
\end{align}
The value of the  entropy function  at the critical point is then
\begin{align}
\cS  |_\mathrm{on-shell}
\   = \    \frac{(g-1)N^{3/2}\pi^3 \sqrt{2}}{\sqrt{27\mathrm{Vol} (Y_7)}}~.
\label{entropytovolumeuniversal}
\end{align}
Recalling that the holographic free energy on $S^3$ associated with the AdS$_4\times SE_7$ solutions is given by
\cite{Herzog:2010hf,Martelli:2011qj,Cheon:2011vi,Jafferis:2011zi}
\begin{align}
F_{S^3} = N^{3/2} \sqrt{\frac{2\pi^6}{27\mathrm{Vol}(Y_7)}}~,
\label{FS3tovol}
\end{align}
we finally obtain 
\begin{align}
\cS  |_\mathrm{on-shell}
 \  = \  (g-1)F_{S^3}~,
  \label{entropytoF}
\end{align}
in  agreement with the general field theory result  derived  in Section 2 of \cite{Azzurli:2017kxo}. In particular, the latter result follows from restricting 
the topological twist performed in the index computation  \cite{Benini:2015noa,Benini:2016hjo} to coincide with the twist along the exact superconformal R-symmetry of the three-dimensional theory.
In the field theory, implementing the universal twist amounts to identifying the R-charges of the fields $\Delta_I$ with their topological fluxes $\mathfrak{n}_I$, where $I$ labels the fields in the field theory, as
\be
\Delta_I \ =  \ \frac{\mathfrak{n}_I} { 1-g} \, ,
\label{unitwistrel}
\ee
which we can indeed reproduce in our set up, as we discuss further below. We also note 
that using (\ref{putin}), (\ref{trump}), as well as the above rescaling argument,
the off-shell master volume is also related 
simply to the off-shell geometric free energy in this case, as
\begin{align}\label{commasvex1}
\mathcal{V} =  \frac{(2\pi \ell_p)^9}{L^9} \frac{F_{S^3}}{64\pi^3}\, . 
\end{align}

Next it is straightforward to compute the geometric R-charges defined in (\ref{defwrapped}).  In particular, we have
\be
\frac{\de \mathcal{V} }{\de \lambda_a} =  -\lambda^2  8\pi b_1^2 \mathrm{Vol}(T_a)\, ,
\ee
and using the rescaling argument above, we obtain
\be
R_a =  \frac{\pi N \mathrm{Vol}(T_a)}{6\mathrm{Vol}(Y_7)} \ \equiv \  N \Delta^{3d}_a\, , 
\label{R1dtoR3d}
\ee
where $\Delta^{3d}_a$ denote the geometric R-charges of the three-dimensional theories \cite{Berenstein:2002ke}.
The equations \eqref{formuletta}, (\ref{enntwo}) then imply
that the fluxes $M_a$ are related to the geometric R-charges via
\begin{align}
M_a= (g-1)R_a\,.
\label{itslate}
\end{align}
Using \eqref{R1dtoR3d} we deduce that the R-charges of the parent three-dimensional field theory, $\Delta^{3d}_a$, 
are rational numbers, as expected from the fact that the Sasaki-Einstein seven-manifolds in the dual supergravity solutions 
must be quasi-regular. This is analogous to what was found in \cite{Gauntlett:2018dpc}. The relation 
(\ref{itslate}) between fluxes $M_a$ and R-charges $R_a$ is part of the definition of the universal twist solution, 
and thus this equation also confirms, \emph{a posteriori}, that our ansatz earlier for the $n_i$ and 
$\lambda_a$ correctly reproduces the universal twist.

To make further contact with the field theory discussion of \cite{Azzurli:2017kxo}, 
it is convenient to use the geometric R-charges and fluxes stripped of the overall factor of $N$, as
in \eqref{defcapDel} and \eqref{fluxredefgoth},
namely 
\be\label{ftdeldef}
R_a\equiv N\Delta_a     \,  ,  \qquad M_a \   \equiv \   - N \fm_a  \, , 
\ee
which are related in the present context via
\be
\Delta_a\  = \  \frac{  \fm_a }{1-g}\, . 
\label{Deltatonuniversal}
\ee
From \eqref{sumrchargesistwo} and the $i=1$ component of (\ref{vMrelation}) we have
\be
\sum_{a=1}^d \Delta_a =  2 \, , \qquad \quad \sum_{a=1}^d \fm_a  \ =  \ 2 (1-g)~.
\label{univtwistconst}
\ee
More generally, using \eqref{twistb}, from (\ref{vMrelation}) we deduce
\be
(1-g) \sum_{a=1}^d \v_a \Delta_a  =   \sum_{a=1}^d \v_a \fm_a \ =  \  2(1-g) \b \, . 
\label{ifoundareeb}
\ee
Note that the relation (\ref{Deltatonuniversal}) has exactly the same form as the field theory result (\ref{unitwistrel}).
However, the index $a$ in (\ref{Deltatonuniversal}) runs over all $d$ toric divisors, while the index $I$ in (\ref{unitwistrel}) labels the chiral fields of the field theory. For the special case of ABJM theory, with $d=4$, these two indices can be identified, and in this case  the relations in  (\ref{univtwistconst}) can be 
directly interpreted as the conditions that the superpotential of the quiver gauge theory
has R-charge 2 and flux $2(1-g)$ \cite{Azzurli:2017kxo}, respectively. More generally, the  
fields\footnote{In the class of ${\cal N}=2$ superconformal quiver theories of interest, 
the $\Phi_I$ are the chiral fields transforming in the adjoint and bi-fundamental representations of the gauge groups  as well as certain chiral monopole operators that arise in the description of the quantum corrected vacuum moduli space \cite{Benini:2009qs}. Note that the index label $I$ does \emph{not} include chiral ``flavour" fields transforming in the (anti-)fundamental representations. We also note that since the fields $\Phi_I$
have definite charges under the flavour group, and in particular under the abelian subgroup, setting
a field to zero in the abelian quiver gauge theory picks out a particular toric divisor as in \eqref{fieldtodivisors}.} $\Phi_I$  are 
associated
to  linear combinations of the toric divisors $T_a$, through a ``field-divisors'' map 
\begin{align}
\Phi_I \ \longleftrightarrow \ \sum_{a}^d c^a_I T_a~,
\label{fieldtodivisors}
\end{align}
which induces the relations $\Delta_I   =\sum_{a}^d c^a_I \Delta_a$, and $ \mathfrak{n}_I  =\sum_{a}^d c^a_I  \fm_a  $. 
Since these are linear relations, from (\ref{Deltatonuniversal}) we can deduce that
for every field in the quiver we must have $\mathfrak{n}_I = (1-g)\Delta_I$, as in \cite{Azzurli:2017kxo}.

\subsection{Comparing with some explicit supergravity solutions}\label{exsgsols}

In this section we will make some additional checks of our new formulae by comparing with
some other explicit AdS$_2\times Y_9$ supergravity solutions, with $Y_9$ a toric 
$Y_7$ fibred over $\Sigma_g$, first constructed in \cite{Gauntlett:2006ns}. The construction of
interest here utilises an eight-dimensional transverse K\"ahler manifold which is a product of a four-dimensional K\"ahler-Einstein space, $KE_4^+$, with the product of two two-dimensional K\"ahler-Einstein spaces,
taken to be $\C P^1\times \Sigma_g$, with $g>1$. Focusing on toric $Y_7$, the $KE_4^+$ is either
$\C P^1\times \C P^1$, $\C P^2$ or the third del Pezzo surface. For simplicity, we just discuss the first two cases.
When $KE_4^+=\C P^1\times \C P^1$ we have $Y_7=Q^{1,1,1}$ and when $KE_4^+=\C P^2$  
we have $Y_7=M^{3,2}$ (although not, in general, with their Sasaki-Einstein metrics). 
The solutions are specified by a positive number, $x$, and in the case $x=1$
we have special instances of the universal twist solutions considered in the last subsection.

In appendix \ref{explicsol} we have extended the results of 
\cite{Gauntlett:2006ns} by carrying out the analysis of flux quantization
for the AdS$_2\times Y_9$ solutions. Combined with some results of this paper we can then
extract the four integers $\vec{n}$,  determining the fibration of $Y_7$ over $\Sigma_g$, as well
as the R-symmetry vector $\vec{b}$, the R-charges, $R_a$, the fluxes $M_a$ and the entropy function $\cS$. 
Ideally we would like to
recover all of these results by carrying out the extremization procedure described in section \ref{gensetup}.
However, it turns out that this is algebraically somewhat involved and so instead we show that if we assume the
R-symmetry vector $\vec{b}$ of the explicit solutions is indeed the critical, on-shell vector of the extremal problem, then
we precisely recover the remaining results of appendix \ref{explicsol}. 

We first consider the case when $KE_4^+=\C P^1\times \C P^1$. We take the twisting parameters to be
given by $\vec{n}=n_1(1,0,0,1/2)$, with $n_1=2(1-g)$ as in the explicit solutions. We also take
the R-symmetry vector to be $\vec{b}=(1,0,0,1/2)$, which we notice is proportional to $\vec{n}$,
and assume that it is the critical vector, as just mentioned.
The toric data can be  obtained from that of $Y^{p,k} (\C P^1\times \C P^1)$ in (\ref{Ypkvcp1cp1}) with $k=p=1$ (for $k=p\neq 1$ one has $Q^{1,1,1}/\Z_p$) and is given by the following six inward pointing normal vectors
\begin{align}
\label{Q111toricdatafirst}
\v_1 \,  &=\,  (1,0,0,0)~, \quad \v_2 \, = \, (1,0,0,1)~, \quad \v_3 \, = \, (1,-1,0,0)~, \nn\\
\quad \v_4 \, &= \, (1,1,0,1)~,\quad \v_5 \, = \, (1,0,-1,0)\, , \quad \v_6 \, = \, (1,0,1,1)\,.
\end{align}
The toric diagram is shown in Figure \ref{Q111toricdiag} in section \ref{gotit}.
Of the six K\"ahler class parameters, $\lambda_a$, only three are independent and, after 
some analysis, one can show that these can be taken to be $\lambda_1+\lambda_2$,
$\lambda_3+\lambda_4$ and $\lambda_5+\lambda_6$. 
With the given R-symmetry vector, we find that the 
constraint equation \eqref{constraintagain} and the flux quantization conditions in \eqref{Nnice}, \eqref{Maint} 
are all
satisfied providing that
\begin{align}\label{lamsxq111}
\lambda_1+\lambda_2\, = \, \lambda_3+\lambda_4&\, = \, -\frac{1}{8\pi^2}\frac{x^{1/2}}{(2+x)^{1/2}}\left(\frac{2\pi \ell_p}{L}\right)^3N^{1/2}\,,\nn\\
\lambda_5+\lambda_6& \, = \, -\frac{1}{4\pi^2}\frac{1}{x^{1/2}(2+x)^{1/2}}\left(\frac{2\pi \ell_p}{L}\right)^3N^{1/2}\,,
\end{align}
where $x>0$ and $A=\frac{(2\pi \ell_p)^3}{L^3}\frac{x^{1/2}(2+x)^{1/2}}{4\pi(1+2x)}(g-1)N^{1/2}$.
Indeed we find that the fluxes $M_a$ are given by 
\begin{align}\label{m32msbulk}
M_1 \, = \, M_2 \, = \, M_3& \, = \, M_4 \, = \ \frac{1+x+x^2}{(2+x)(1+2x)}(g-1)N\,,\nn\\
M_5& \, = \, M_6 \, = \, \frac{3x}{(1+2 x)(2+x)}(g-1)N\,.
\end{align}
To ensure that these are integers we demand that $x+1/x\in\mathbb{Q}$.
Furthermore, the R-charges are given by
\begin{align}
R_1 \, = \, R_2 \, = \, R_3 \, = \,  R_4 \, = \, \frac{1}{2+x}N,\qquad R_5 \, = \,  R_6 \,  = \,  \frac{x}{2+x}N\,,
\end{align}
with $\sum_a R_a=2N$. It is interesting to point out that while the geometry is quasi-regular for all values of $x$ (since $\vec{b}=(1,0,0,1/2)$)
the R-charges can be irrational.
Notice also that when $x=1$ the R-charges are proportional to the fluxes,  
as in the universal twist solutions in section \ref{unitwist}.
Finally, after calculating the on-shell supersymmetric action \eqref{rhoomegaintseven}, \eqref{cS2}
we obtain
\begin{align}
\cS  |_\mathrm{on-shell} =  2\pi (g-1)\frac{3+2x+x^2}{(1+2x)}\frac{x^{1/2}}{(2+x)^{3/2}}N^{3/2}\,.
\end{align}
These expressions precisely agree with their counterparts in appendix \ref{explicsol}
obtained by analysing the explicit supergravity solutions.

As an aside we note that given the K\"ahler class parameters in \eqref{lamsxq111}
and our choice of $\vec{b}$, the master volume as a function of $x$ takes the  simple form 
\begin{align}\label{veexexpqone}
\mathcal{V}=\frac{(2\pi \ell_p)^9}{L^9}\frac{x^{1/2}}{16\pi^2(2+x)^{3/2}}N^{3/2}\,.
\end{align}
As we recalled in section \ref{unitwist}, a dual quiver gauge theory for $Y_7=Q^{1,1,1}$
was proposed in \cite{Benini:2009qs} and a calculation of 
the large $N$ topologically twisted index on $S^1\times S^2$ was presented in \cite{Hosseini:2016ume}. Indeed for $x=1$ (which corresponds to the universal twist) we have already noted that the geometric results are in agreement with 
the field theory results\footnote{Which requires setting $\Delta_m=0$ in \cite{Hosseini:2016ume}.}.  
It would be interesting to find a dual field theory interpretation of the $x$-deformed geometry 
that we discussed above.

We now consider the case when $KE_4^+=\C P^2$, which is very similar. 
We take the twisting parameters to be
given by $\vec{n}=n_1(1,0,0,1)$, with $n_1=2(1-g)$ as in the explicit solutions. We also take
the R-symmetry vector to be $\vec{b}=(1,0,0,1)$, which is again proportional to $\vec{n}$,   
and we again assume that it is the critical vector.
The toric data for $M^{3,2}$ can be obtained from $Y^{p,k} (\C P^2)$ in \eqref{Ypkvcp2}
with $p=2$ and $k=3$:
\begin{align}
\label{M32toricdata23}
\v_1 \,  &=\,  (1,0,0,0)~, \quad \v_2 \, = \, (1,0,0,2)~, \quad \v_3 \, = \, (1,1,0,0)~, \nn\\
\quad \v_4 \, &= \, (1,0,1,0)~,\quad \v_5 \, = \, (1,-1,-1,3)\, \,.
\end{align} 
Of the five K\"ahler class parameters, $\lambda_a$, only two are independent and, after 
some analysis, one can show that these can be taken to be $\lambda_1+\lambda_2$ and
$\lambda_3+\lambda_4+\lambda_5$. With the given R-symmetry vector, we find that the 
constraint equation \eqref{constraintagain} and the flux quantization conditions \eqref{Nnice}, \eqref{Maint} 
are all
satisfied providing that
\begin{align}
\lambda_1+\lambda_2& \, = \, -\frac{1}{6\pi^2}\frac{1}{x^{1/2}(2+x)^{1/2}}\left(\frac{2\pi \ell_p}{L}\right)^3N^{1/2}\,,\nn\\
\lambda_3+\lambda_4+\lambda_5&\, = \, -\frac{1}{4\pi^2}\frac{x^{1/2}}{(2+x)^{1/2}}\left(\frac{2\pi \ell_p}{L}\right)^3N^{1/2}\,,
\end{align}
where $x>0$ and 
$A=\frac{(2\pi \ell_p)^3}{L^3}\frac{x^{1/2}(2+x)^{1/2}}{3\pi(1+2x)}(g-1)N^{1/2}$.
The fluxes $M_a$ are given by 
\begin{align}\label{m32msbulk2}
M_1& \, =  \, M_2 \, =  \, \frac{3x}{(1+2 x)(2+x)}(g-1)N\,,\nn\\
M_3 \, = \, M_4& \, = \, M_5 \, =  \, \frac{4}{3}\frac{1+x+x^2}{(2+x)(1+2x)}(g-1)N\,.
\end{align}
We again demand $x+1/x\in \mathbb{Q}$ in order that these are all integers.
The R-charges are
\begin{align}R_1 \, = \, R_2 \, = \, \frac{x}{2+x}N\,,\qquad
R_3 \, = \, R_4 \, =  \, R_5 \, = \, \frac{4}{3}\frac{1}{2+x}N\,,
\end{align}
with $\sum_a R_a=2N$, and these can be irrational. One can again check that the R-charges are proportional to the fluxes
when $x=1$, which is the case of the universal twist solution.
Finally, for the on-shell supersymmetric action \eqref{rhoomegaintseven}, \eqref{cS2} 
we obtain
\begin{align}
\cS  |_\mathrm{on-shell} =  \frac{8\pi}{3} (g-1)\frac{3+2x+x^2}{(1+2x)}\frac{x^{1/2}}{(2+x)^{3/2}}N^{3/2}\,.
\end{align}
These expressions precisely agree with their counterparts in appendix \ref{explicsol}
obtained by analysing the explicit supergravity solutions. 

\subsection{$\mathbb{C}\times$Conifold example}\label{cconsection}

\begin{figure}[h!]
\begin{center}
  \includegraphics[height=4cm]{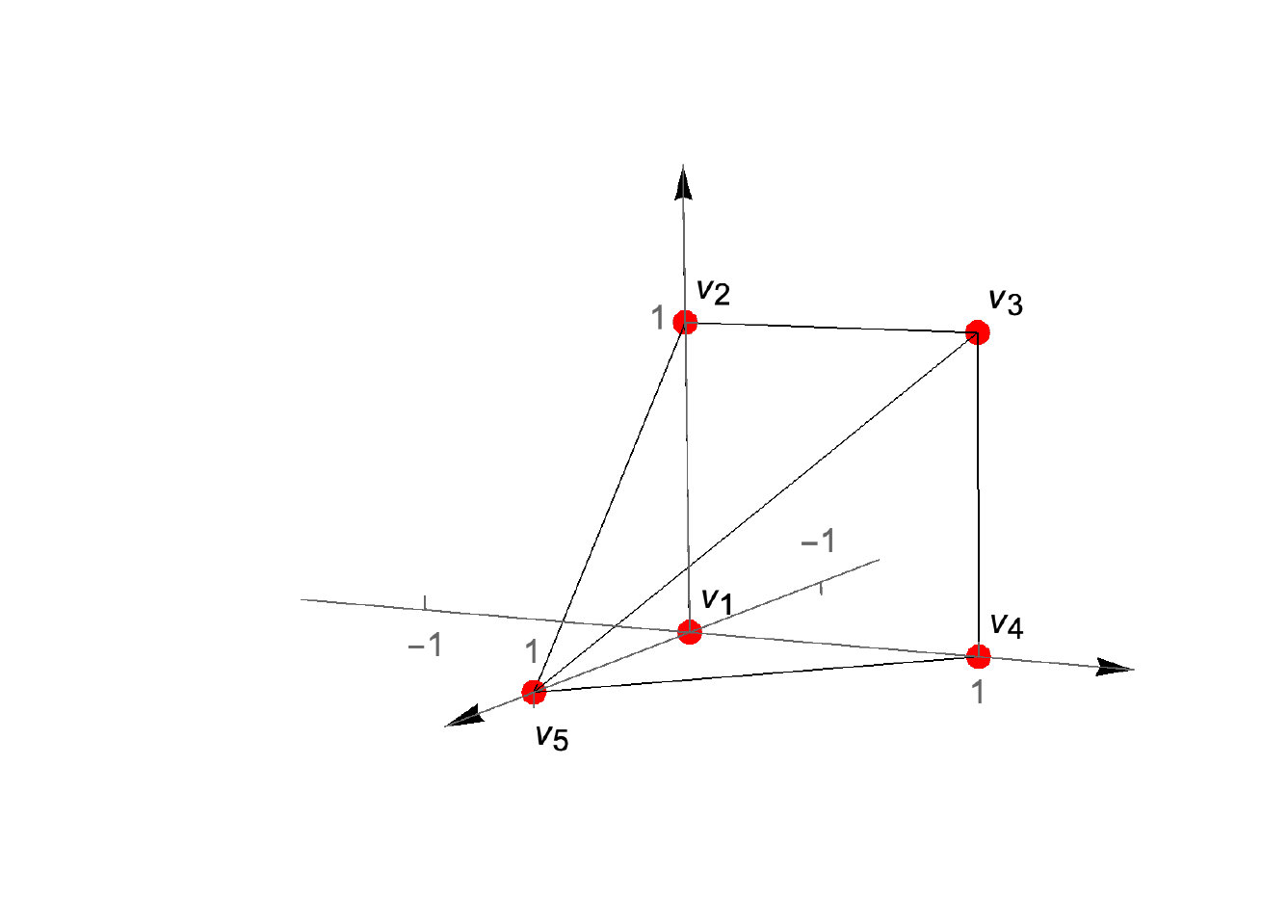}
 \caption{\label{Cconfigure} Toric diagram for the link of the $\mathbb{C}\times$Conifold singularity.}
\end{center}
\end{figure}

In the reminder of this section we will study 
examples of the form $Y_7 \ \hookrightarrow \ Y_9 \ \rightarrow  \Sigma_g $, with toric $Y_7$, with known dual ${\cal N}=2$ three-dimensional field theories. 
Specifically, we start here considering  as $Y_7$  the link of the complex cone obtained by taking the product of the complex plane with the conifold singularity.
This complex cone is specified by five inward pointing normal vectors given by
\begin{align}
\label{Cconapptext}
\v_1 \,  &=\,  (1,0,0,0)~, \quad \v_2 \, = \, (1,0,0,1)~, \quad \v_3 \, = \, (1,0,1,1)~, \nn\\
\quad \v_4 \, &= \, (1,0,1,0)~,\quad \v_5 \, = \, (1,1,0,0) \,.
\end{align}
The toric diagram is obtained by projecting on $\R^3$ the vertices  in (\ref{Cconapptext}) and 
is shown in Figure \ref{Cconfigure}. 

The presence of the square face in the toric diagram (as opposed to a triangle), indicates that the link
$Y_7$ of $\mathbb{C}\times$Conifold has worse-than-orbifold singularities. Specifically, the divisor associated
with $\vec{v}_5$ is a copy of the conifold, sitting at the origin of the complex plane $\C$, and this gives rise to an associated singularity on $Y_7$. 
As we explain in more detail in appendix \ref{secctcon} some care is required in using the master volume formula. The diagnostic that the master volume formula is not, in general, calculating a volume is
that the relation \eqref{keyvRrel} is not satisfied unless we impose that the K\"ahler class parameters satisfy
$X\equiv \lambda_1-\lambda_2+\lambda_{3}-\lambda_{4}=0$.

A procedure one can follow is to resolve the singularity by adding an extra line either from $\v_2$ to $\v_4$ or 
from $\v_1$ to $\v_3$ as illustrated in Figure \ref{Cconfigurediagres} in appendix \ref{secctcon}. 
In both of these resolutions \eqref{keyvRrel} is satisfied and from the $\{\lambda_a\}$
one can construct two gauge invariant variables given by
\begin{align}
X&=\lambda_{1}-\lambda_{2}+\lambda_{3}-\lambda_{4}\,,\nn\\
Y&=(b_1-b_2-b_4) \lambda_{1}+(b_4-b_3) \lambda_{2}+b_3 \lambda_{3}+b_2 \lambda_{5}\,.
\end{align}
Furthermore, when one sets $X=0$ in the associated master volume formulae one finds that the two expressions are
equal and moreover they are equal to the master volume for the toric diagram in Figure \ref{Cconfigure}, associated with the singular $Y_7$, after setting $X=0$.
Thus, we conclude that one can use the master volume formulae for $Y_7$ associated with Figure
\ref{Cconfigure} provided that one sets
$X=0$, and then checks \emph{a posteriori} that one has a set-up consistent with flux quantization. An additional
subtlety is that for the singular geometry $Y_7$ we should not impose that all fluxes $M_a$ are integer, but instead only certain linear combinations, associated with the fact that it is these linear combinations that correspond to \emph{bona fide} cycles of $Y_7$. We expect that this procedure should yield the same results
as starting with the non-singular resolved geometries, associated with Figure \ref{Cconfigurediagres}, and then imposing an additional condition on the quantised fluxes, but we have not checked this in detail\footnote{It is
difficult to explicitly carry out the extremization procedure at the algebraic level.}.

Proceeding with $X=0$ and with the master volume for Figure \ref{Cconfigure}, we first solve the constraint equation \eqref{constraintagain} 
for $A$, finding a long expression that we don't record\footnote{One finds that after substituting for
$Y$, $A$ still has some dependence on $\lambda_1$, $\lambda_2$ and $\lambda_3$. This is expected, because $A$ is not invariant under gauge transformations but it transforms as in (\ref{lamgt2}).}.
We then solve the flux quantization condition \eqref{Nnice} for $Y$ finding
\begin{align}\label{ysolcon}
Y^2=
\frac{(2\pi \ell_p)^6}{L^6}\frac{b_2 b_3b_4 (b_2+b_3-1)  (b_2+b_4-1)}{8 \pi ^4 (1-b_2) }N\,,
\end{align}
where we have now set $b_1=1$. The sign ambiguity in solving for $Y$ will get resolved after extremization
and demanding that the entropy is positive. This issue arises in generic examples and we will not explicitly keep track of it.
One can then use \eqref{ysolcon} to obtain expressions for the fluxes
$M_a$ from \eqref{Maint} obtaining 
\begin{align}\label{exemmscconc}
M_1&=\frac{(b_1^2+b_2^2-2b_1b_2) (-n_1+n_2+n_3+n_4)+(b_2-b_1) (b_4 n_3+b_3 n_4)+b_3 b_4 (n_1-n_2)}{(b_1-b_2)^2}N\nn\\
M_2&= \frac{b_3b_4  (n_2-n_1)+b_4(b_1-b_2) n_3+(b_1-b_2) (-b_1+b_2+b_3) n_4 }{(b_1-b_2)^2}N  \,,\nn\\
M_3&= \frac{ b_3 b_4 (n_1-n_2)+b_4(b_2-b_1)  n_3+b_3(b_2-b_1) n_4}{(b_1-b_2)^2} N \,,\nn\\
M_4&= -\frac{ (b_1-b_2) (b_1-b_2-b_4) n_3+b_3 b_4 (n_1-n_2)+b_3(b_2-b_1) n_4}{(b_1-b_2)^2}N \,,\nn\\
M_5&=  -n_2N \,,
\end{align}
and one can check that \eqref{vMrelation} is satisfied.
Apart from $M_5$ these are not, in general integers. However, various linear combinations are, for example:
\begin{align}\label{emmmscon}
M_1+M_2&=(-n_1+n_2+n_3)N\,,\qquad
M_3+M_4=-n_3N\,,\nn\\
M_2+M_3&=-n_4N\,,\qquad
M_1+M_4-M_2-M_3=(-n_1+n_2+2n_4)N\, .
\end{align}
We can also work out the R-charges from \eqref{defgeomRcharges}
and we find  
\begin{align}\label{rchceccon}
R_1&=-\frac{2 (b_2+b_3-1) (b_2+b_4-1)}{b_2-1}N,\qquad
R_2=\frac{2 (b_2+b_3-1) b_4 }{b_2-1}N\,,\nn\\
R_3&=-\frac{2 b_3 b_4 }{b_2-1}N,\qquad
R_4=\frac{2 b_3 (b_2+b_4-1) }{b_2-1}N,\qquad
R_5=2 b_2 N\,, 
\end{align}
which satisfy (\ref{keyvRrel2}). 
Various linear combinations of these expressions simplify, echoing the expressions in \eqref{emmmscon}.
Finally, we can then obtain an explicit form for the off-shell entropy function $\cS$, using
\eqref{rhoomegaintseven} \eqref{cS2} (or equivalently (\ref{susyactneat}))
which is expressed in terms of $b_2,b_3,b_4$, $n_1=2(1-g),n_2,n_3,n_4$ and $N$. 
Up to an overall sign ambiguity (arising from \eqref{ysolcon}) we obtain
\begin{align}\label{osentccon}
\cS=&\frac{2 \sqrt{2} \pi N^{3/2}  }{3 (1-b_2)^{3/2}[b_2 b_3 b_4(b_2+b_3-1) (b_2+b_4-1)]^{1/2}}\times\nn\\
 &\Big[b_2 b_3 b_4 \left(-b_2^2+2 b_2+b_3 b_4-1\right) n_1\nn\\
 &+b_3 b_4 \left(2 b_2^3+(b_3+b_4-5) b_2^2-2 (b_3+b_4-2) b_2-(b_3-1) (b_4-1)\right) n_2\nn\\
 &+(b_2-1) b_2 (b_2+2 b_3-1) b_4 (b_2+b_4-1) n_3\nn\\&+(b_2-1) b_2 b_3 (b_2+b_3-1) (b_2+2 b_4-1) n_4\Big]\, . 
\end{align}

We can now compare these results with the field theory analysis, for genus $g=0$, carried out in
\cite{Hosseini:2016ume}. We first recall various aspects of the three-dimensional quiver gauge theory
discussed in section 6.1 of \cite{Benini:2009qs}.
This is an instance of a general family of ``flavoured'' quiver gauge theories with gauge group $SU(N)$ and three adjoint chiral fields $\phi_1,\phi_2,\phi_3$.   
There are also three sets of fields $q^{(f)},\tilde q^{(f)}$, $f=1,2,3$  transforming in the fundamental and anti-fundamental representation of $SU(N)$ 
and associated with $U(k_f)$ global symmetries. The superpotential reads 
 \begin{align}
 W \ =  \ \mathrm{Tr} \left[\phi_1[\phi_2,\phi_3]+\sum_{i=1}^{k_1}q_i^{(1)}\phi_1\tilde q^{(1)}_i+     \sum_{i=1}^{k_2}q_i^{(2)}\phi_2\tilde q^{(2)}_i+     \sum_{i=1}^{k_3}q_i^{(3)}\phi_3\tilde q^{(3)}_i \right]\, ,
 \label{superpot}
 \end{align}
and the quiver diagram can be found in (5.48) of \cite{Hosseini:2016ume}, whose notation we will follow below.  As discussed in  \cite{Benini:2009qs}
the $\C \times$Conifold geometry corresponds to the theory with $k_1=k_2=1$ and $k_3=0$ (see Figure 3(b) of \cite{Benini:2009qs}). An important aspect of these models is that there is a quantum correction to the moduli space of vacua, due to the presence of monopole operators $T$ and $\tilde T$, which satisfy the relation 
\begin{align}
T\tilde T = \phi_1^{k_1}\phi_2^{k_2}\phi_3^{k_3}~.
\end{align}
When $k_1=k_2=1$, $k_3=0$ this gives the  $\C \times$Conifold geometry. 

For generic values of $k_1,k_2,k_3$ these three-dimensional theories flow to a SCFT in the IR, with gravity dual 
AdS$_4\times Y_7$, where $Y_7$ is the Sasaki-Einstein base of the  Calabi-Yau cone singularity. 
In  \cite{Jafferis:2011zi} it was shown that the large $N$ limit of the free energy, $F_{S^3}$,
obtained from the exact localized partition function on $S^3$, takes the form
(\ref{FS3tovol}), where   $\mathrm{Vol}(Y_7)$ is the Sasakian volume.

To compare the field theory with the geometry, we need to relate the fields of the quiver with the toric data of the singularity. 
In particular, the fields $\phi_1,\phi_2,\phi_3, T,\tilde T$ 
correspond to linear combinations of the toric divisors and the field-divisors map 
(\ref{fieldtodivisors}) may be obtained by employing  the perfect matching variables \cite{Benini:2009qs}. 
This map was explicitly given in  \cite{Jafferis:2011zi} for the above class of theories and for the case of
the $\C\times$Conifold model reads, in the notation of \cite{Jafferis:2011zi}, 
\begin{align}\label{cconmap}
\phi_1 =  a_0 a_1 \, , \qquad \phi_2 =  b_0b_1 \, , \qquad \phi_3 = c_0 \, , \qquad
T  = a_0 b_0 \, , \qquad \tilde T   =  a_1b_1 \, ,
\end{align}
where the perfect matching variables $(a_0,a_1,b_0,b_1,c_0)$ are associated to the toric 
data (\ref{Cconapptext})  as in  Table \ref{tablevpfcon} below.
\begin{table}[h!]
\begin{center}
\begin{tabular}{|c|c|c|c|c|c|c|}
\hline
 $\v_1$ & $\v_{2}$ &$\v_{3}$& $\v_{4}$ &$\v_{5}$   \\ 
\hline
$a_0$ & $a_1$ & $b_1$ & $b_0$ &$c_0$    \\
\hline
\end{tabular}
\caption{Relation between toric data  (\ref{SPPtoricdata}) and perfect matchings  for the
$\C\times$Conifold  singularity  \cite{Jafferis:2011zi}.} 
\label{Cxcpftable}
\label{tablevpfcon}\end{center}
\end{table}
With this map, we can parametrize the R-charges of the fields in the quiver in terms of the geometric R-charges $\Delta^{3d}_a$, 
defined using the  volumes of supersymmetric five-dimensional toric submanifolds $T_a$, through the  relation 
 (\ref{R1dtoR3d}).  

We now consider compactifying this $d=3$ quiver gauge theory on a Riemann surface $\Sigma_g$, with a twist that is parametrized by integer valued flavour magnetic fluxes for the fields $\{\mathfrak{n}_{\phi_1}, \mathfrak{n}_{\phi_2},\mathfrak{n}_{\phi_3}\}$ with  
$\mathfrak{n}_{\phi_1}+ \mathfrak{n}_{\phi_2}+\mathfrak{n}_{\phi_3}=2(1-g)$ units of flux, as required for supersymmetry.   
Assuming that the theory flows to a SCQM in the IR, we expect that the dual supergravity solution will be an AdS$_2\times Y_9$ solution of $D=11$ supergravity with $Y_9$ a fibration of a toric  $Y_7$ over $\Sigma_g$ and we can compare with our geometric results above.
To proceed, we can use the map \eqref{cconmap} to relate the R-charges of the fields, $\Delta_I$, with 
the geometric R-charges, $\Delta_a\equiv R_a/N$ (see \eqref{defcapDel}),
via
\begin{align}\label{firstdeldicccontrue}
\Delta_{\phi_1}&  = \Delta_1+\Delta_2=2(1-b_2-b_3)\,,\nn\\
\Delta_{\phi_2}& = \Delta_3+\Delta_4=2b_3\,,\nn\\
\Delta_{\phi_3}& = \Delta_5=2b_2\,,
\end{align}
and 
\begin{align}
\Delta_{T}& = \Delta_1+\Delta_4 =  2 (1- b_2 - b_4)   , \nn\\
\Delta_{\tilde T} & = \Delta_2+\Delta_3= 2b_4 \, , 
\end{align}
where in the last equalities we used the parametrization (\ref{rchceccon}) coming from the geometry.
We also define (see (2.3) of \cite{Jafferis:2011zi}) 
\begin{align}
\Delta_m&\equiv \frac{1}{2} (\Delta_{T} - \Delta_{\tilde T} ) = 1-b_2-2b_4 \, .
\end{align}
Notice that the R-charges of the adjoint fields satisfy $\Delta_{\phi_1}+\Delta_{\phi_2}+\Delta_{\phi_3}=2$, as implied by supersymmetry. 
 
Similarly, the  fluxes of the fields can be identified with a set of geometric flux parameters $ \fm_a\equiv -M_a/N$ (see \eqref{fluxredefgoth})
in an entirely analogous manner, namely 
\begin{align}
\label{nicerisntit}
\fm_{\phi_1}&= \fm_1+\fm_2=n_1-n_2-n_3\,,\nn\\
\fm_{\phi_2}&= \fm_3+\fm_4=n_3\,,\nn\\
\fm_{\phi_3}&= \fm_5=n_2\,,
\end{align}
and 
\begin{align}\label{lastdeldicccon}
\fm_{T}& = \fm_1+\fm_4 =  \, n_1 - n_2 -n_4 , \nn\\
\fm_{\tilde T} & = \fm_2+\fm_3= n_4  \, , 
\end{align}
where in the last equalities we used the parametrization (\ref{emmmscon}) coming from the geometry. 
Notice that the fluxes of the adjoint fields satisfy $\fm_{\phi_1}+\fm_{\phi_2}+\fm_{\phi_3}=n_1$, as implied by supersymmetry. 

Finally, for the case of $g=0$, we can compare with the 
large $N$ limit for the off-shell index on $S^1\times S^2$, $\mathcal{I} (\Delta,\mathfrak{n})$,
that was computed in   \cite{Hosseini:2016ume}. Specifically, equation (5.56) of this reference\footnote{To compare with the expression in
 \cite{Hosseini:2016ume} one should relate the variables used here to that used in     
 \cite{Hosseini:2016ume}  as  $\Delta_{HM} = \pi \Delta $.} gives 
\begin{align}
\mathcal{I} (\Delta_{\phi_I},\mathfrak{n}_{\phi_I})=-\frac{\pi}{3}\sqrt{  \frac{\hat \Delta}{2 \bar \Delta} (\bar\Delta^2 - 4\Delta_m^2) }\left[\hat\fm +\frac{\bar \fm(\bar\Delta^2+4\Delta_m^2)}{\bar\Delta(\bar\Delta^2-4\Delta_m^2)}  -\frac{8\Delta_m}{\bar\Delta^2-4\Delta_m^2} \right] \, ,
\label{finalentropyccon2}
\end{align}
with
\begin{align}
\hat\fm &\equiv \frac{\fm_{\phi_1}}{\Delta_{\phi_1}}+\frac{\fm_{\phi_2}}{\Delta_{\phi_2}}+\frac{\fm_{\phi_3}}{\Delta_{\phi_3}}\,,\qquad
\bar\fm \equiv \fm_{\phi_1}+\fm_{\phi_2}\,,\nn\\
\hat\Delta&\equiv \Delta_{\phi_1}\Delta_{\phi_2}\Delta_{\phi_3}\,,\qquad
\bar\Delta\equiv \Delta_{\phi_1}+\Delta_{\phi_2}\,.
\end{align}
Using the dictionary given in \eqref{firstdeldicccontrue}--\eqref{lastdeldicccon}, we see that
the off-shell entropy function (\ref{osentccon}) calculated from the geometry side
 \emph{cannot} agree with the expression given in (\ref{finalentropyccon2}), since the former depends on $n_4$ whereas the latter does not (only the monopole fluxes $\fm_T, \fm_{\tilde T}$ depend on $n_4$). 
However, remarkably, if we impose\footnote{The relation (\ref{extrarel}) is equivalent to the particular relation among monopole charges $\fm_{T} - \fm_{\tilde T} = 2$ in the field theory variables.
Interestingly, we also find agreement of (\ref{osentccon}) and (\ref{finalentropyccon2})
 if we restrict to the subspace of  $\Delta_m =1-b_2-2b_4=0$, without imposing any relation among the fluxes $n_i$.} 
the additional constraint on the geometric fluxes that
\begin{align}
n_2   =n_1  - 2n_4 -2 \, ,
\label{extrarel}
\end{align}
then we find our off-shell entropy geometric result $\cS  (\b,n_1,n_3,n_4)$,
obtained from \eqref{osentccon}, agrees with the expression
$\mathcal{I} (\Delta_{\phi_a},\mathfrak{n}_{\phi_a})$ in (\ref{finalentropyccon2}). 
The result reported in  \cite{Hosseini:2016ume}  corresponds to setting $g=0$.

We can make a further connection between  geometry and  field theory by relating
our master volume $\mathcal {V}$ with the  function $\mu$  that is
 proportional to the large $N$ limit of the matrix model Bethe potential and
 determines the $S^1\times \Sigma_g$ index.
 This was shown   \cite{Hosseini:2016ume} to coincide with the large $N$ limit of the  free energy on $S^3$ of the $d=3$ field theory, namely 
\begin{align}\label{Ftomu}
\mu =  \frac{3\pi}{4N^{3/2}}F_{S^3} . 
\end{align}
Recall from \eqref{commasvex1} that in the universal twist case we found that the off-shell master volume is related to the large $N$ free energy as $\mathcal{V} =  \frac{(2\pi \ell_p)^9}{L^9} \frac{F_{S^3}}{64\pi^3}$. We 
 can show that this relation also holds in the $\C\times$Conifold setting. 
To see this, from \cite{Hosseini:2016ume} we have that 
\begin{align}
\label{jedi}
\mu =\pi^2\sqrt{  \frac{\hat\Delta}{2\bar\Delta}(\bar\Delta^2-4\Delta_m^2)}\,,
\end{align}
and using the dictionary above we find
\begin{align}
\label{jedi2}
\mu (b_i) =\pi^2\sqrt{\frac{32  b_2 b_3 b_4 (b_2+b_3-1)  (b_2+b_4-1)}{1-b_2}}\, .
\end{align}
On the other hand, evaluating the master volume with $X=0$ and
$Y$ obtained from \eqref{ysolcon}, we find 
\begin{align}
\frac{L^9}{(2\pi \ell_p)^9}\frac{48\pi^4}{N^{3/2}}\mathcal{V} (b_i) =  \mu (b_i) \, ,
 \end{align}
where both sides are regarded as functions of $(b_2,b_3,b_4$). 

We conclude this subsection by considering the expression (\ref{ssss})
 for the supersymmetric action
in the context of the present example. 
Recall that when the change of variables (\ref{defgeomRcharges2})
between the $\{\Delta_a\}$ and the $\{\lambda_1,\dots,\lambda_{d-3},b_2,b_3,b_4\}$ 
is invertible,  we can write the master volume as a function of the  $\{\Delta_a\}$ and the off-shell 
supersymmetric action takes  the form   (\ref{ssss}).
For the $\C\times$Conifold example, we imposed $X=0$ on the K\"ahler classes, leaving 
us with four variables $Y,b_2,b_3,b_4$ (before imposing the constraint or flux quantisation conditions), 
implying that we cannot carry out such an invertible change of variables. 
Nevertheless, we can re-write 
the off-shell master volume in terms of the $\Delta_a$ variables, where an ambiguity is fixed by requiring that this is a homogeneous function of degree two. Namely we have
\begin{align}\label{expcconlast}
\mathcal{V} (\Delta_a)
&=\frac{(2\pi \ell_p)^9}{L^9}\frac{N^{3/2}}{24\sqrt{2}\pi^2}\frac{ [ (\Delta_1+\Delta_2) (\Delta_2+\Delta_3) (\Delta_1+\Delta_4) (\Delta_3+\Delta_4) \Delta_5          ]^{1/2}}{[  \Delta_1+\Delta_2+\Delta_3+\Delta_4  ]^{1/2}}\, .
\end{align}
We can now take derivatives of $\mathcal{V}$ in (\ref{expcconlast}) with respect to $\Delta_a$ 
and after substituting for $\Delta_a=R_a/N$ and 
$\fm_a=-M_a/N$ from \eqref{rchceccon} and \eqref{exemmscconc}, we find that (\ref{ssss}) gives
$\cS \equiv   \frac{8\pi^2 L^9}{(2\pi\ell_p)^9}\, \Ssusy (b_i,\fm_a)$. 

\subsection{$Q^{1,1,1}$ example}
\label{gotit}

\begin{figure}[h!]
\begin{center}
  \includegraphics[width=5.5cm]{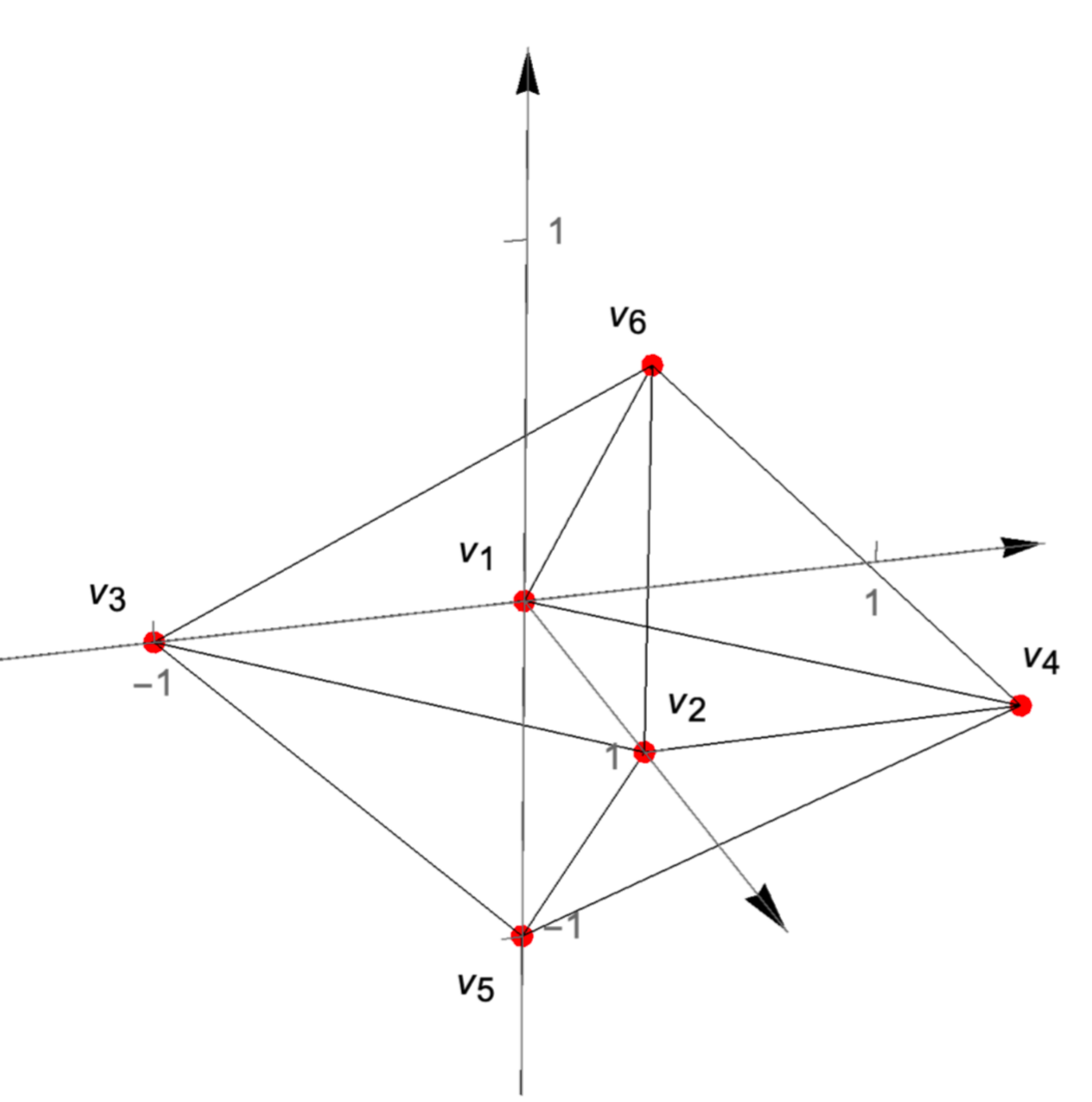}
 \caption{\label{Q111toricdiag} Toric diagram of $Q^{1,1,1}$.}
\end{center}
\end{figure}

In this section, we will revisit the case of $Y_7=Q^{1,1,1}$, that we already studied in section \ref{exsgsols} 
in the context of explicit supergravity solutions. In particular here we will be able to make a connection with a field theory
result for the twisted topological index that was given in \cite{Hosseini:2016ume}. Recall that the toric data is specified by the vectors
\begin{align}
\label{Q111toricdata}
\v_1 \,  &=\,  (1,0,0,0)~, \quad \v_2 \, = \, (1,0,0,1)~, \quad \v_3 \, = \, (1,-1,0,0)~, \nn\\
\quad \v_4 \, &= \, (1,1,0,1)~,\quad \v_5 \, = \, (1,0,-1,0)\, , \quad \v_6 \, = \, (1,0,1,1)\,.
\end{align}
The corresponding toric diagram, obtained by projecting the vertices in \eqref{Q111toricdata} onto $\mathbb{R}^3$,  
is given in
Figure \ref{Q111toricdiag}.

To connect with the field theory analysis of \cite{Hosseini:2016ume} we will consider the fibration to be of the form
\begin{align}\label{ennansq1}
\vec{n}=(n_1,0,n_3, \frac{1}{2} \left(n_1 +2 n_3 \right))\,,
\end{align}
where $n_3$ is an arbitrary integer and $n_1=2(1-g)$.
Due to the algebraic complexity of the extremal problem, to proceed we will make a simplifying assumption on the Reeb vector, 
consistent with the symmetries associated with \eqref{ennansq1}, 
which then needs to be justified {\it a posteriori}. Specifically, 
we assume
\begin{align}
\vec{b} = (1,0,b_3, \frac{1}{2}+b_3)\,.
\end{align}
This implies, via \eqref{keyvRrel2}, that we are assuming that the R-charges satisfy 
$R_3=R_4$, $R_6-R_5=2b_3N$ and $R_2+R_4+R_6=(1+2b_3)N$ in addition to $\sum_aR_a=2N$. 
There are similar conditions for the fluxes $M_a$ due to \eqref{vMrelation} and \eqref{ennansq1}.

Within this ansatz we can construct three linear combinations of the K\"ahler parameters, invariant under
the gauge transformations \eqref{lamgt}, given by
\begin{align}
 X= b_3 ( \lambda_2-  \lambda_5 )+  \frac{1}{2}(\lambda_1+\lambda_2 )\,,  \nn\\
 Y = b_3 (\lambda_2-\lambda_5)+ \frac{1}{2}(\lambda_3+\lambda_4)\,, \nn\\
 Z= b_3 (\lambda_2-\lambda_5)+ \frac{1}{2}(\lambda_5+\lambda_6)\,. 
\end{align}
In terms of these variables, the master volume reads
\begin{align}
\mathcal{V} = \frac{256 \pi ^4}{3} \left[ - 3 XYZ+ \frac{b_3X^3}{\left(2 b_3+1\right){}^2}+\frac{3 b_3X^2 Y}{2 b_3+1}-\frac{Z^2b_3 \left(-6 b_3 Y+3 Y+Z\right)}{\left(1-2 b_3\right){}^2}-b_3(X-Z)^3\right]\,.
\label{masterXYZ}
\end{align}
Note if we set $b_3=0$, $X=Y$ and also\footnote{Including the 
parameter $n_3$ in the analysis, should be associated with the more general explicit supergravity solutions
discussed in section 6.3 of \cite{Gauntlett:2006ns}.}
$n_3=0$ then we are within the framework of the explicit supergravity solutions that we discussed in section
\ref{exsgsols}. We continue with $b_3\ne0$.

Next we can solve the constraint equation \eqref{constraintagain} for $A$. We also find that
the expression for $N$ in \eqref{Nnice} is linear in $Y$ and hence can be simply solved
for $Y$. At this point we would next like to solve two of the equations for the fluxes $M_a$ given in
\eqref{Maint} for $X$ and $Z$. However, it is difficult to solve the simultaneous polynomial equations 
in closed form.
However, we can get results matching with the field theory results using some inspired guesswork.
Specifically, we make the further assumption that $Y=X=Z$.  With the given solution for $Y$ we then have
\begin{align}\label{mvolq1}
X=Y=Z=\frac{(2\pi\ell_p)^3}{L^3}\frac{ 4 b_3^2-1}{16\pi^2 \sqrt{3-4 b_3^2} }N^{1/2}\,.
\end{align}
Substituting this into the master volume we find
\begin{align}\label{mvolq1fin}
\mathcal{V} = \frac{(2\pi\ell_p)^9}{L^9}\frac{1-4 b_3^2}{48\pi^2 \sqrt{3-4 b_3^2}}N^{3/2}\,.
\end{align}
The R-charges take the form
\begin{align}
R_1= \left(\frac{1}{3-4 b_3^2}-b_3\right) N, \quad R_2 = \left(\frac{1}{3-4 b_3^2}+b_3\right) N , \quad R_3 = \left( \frac{2 }{4 b_3^2-3}+1 \right) N\, ,\nn\\
R_4 = \left(\frac{2 }{4 b_3^2-3}+1\right) N, \quad R_5 = \left(\frac{1}{3-4 b_3^2}-b_3\right) N ,\quad R_6 =\left(\frac{1}{3-4 b_3^2}+b_3\right) N\,,
\end{align}
while the fluxes are given by
\begin{align}\label{emmsfinq1}
M_1 &= M_5 = \frac{\left(2 b_3-1\right)  \left[\left(6 b_3+3\right) n_1+\left(8 b_3^3+4 b_3^2-10 b_3-9\right) n_3\right]}{2\left(3-4 b_3^2\right){}^2} N\,,\nn\\
M_3 &= M_4 = \frac{ \left(16 b_3 n_3-\left(16 b_3^4+3\right) n_1\right)}{2\left(3-4 b_3^2\right){}^2} N\,,\nn\\
M_2 &= M_6 = \frac{\left(2 b_3+1\right)  \left[\left(6 b_3-3\right) n_1+\left(-8 b_3^3+4 b_3^2+10 b_3-9\right) n_3\right]}{2\left(3-4 b_3^2\right){}^2} N\,.
\end{align}
Finally, 
we find the following off-shell expression for the entropy 
\begin{align}\label{brxt}
\cS =  4\pi(g-1)\frac{\left(8b_3^4-6b_3^2+3\right)-n_3 b_3 \left(4b_3^2-5\right)}{3 \left(3-4b_3^2\right){}^{3/2}}N^{3/2}~.
\end{align}
Remarkably, for $g=0$ this  agrees 
 precisely computation of the large $N$ limit of the $S^1\times S^2$ index presented in  (5.47) of  \cite{Hosseini:2016ume}, after identifying $n_3=\mathfrak{t} + \mathfrak{\tilde t}$ and $b_3=\Delta_m/2 $, 
 as we discuss further below. Furthermore, there
is also agreement between the large $N$ free energy and the expression for the master volume given in \eqref{mvolq1}.

An important point is that we have a consistent framework provided that the $M_a$ are all integer. This is possible 
provided that the extremal point of the entropy function is such that the expression for the $M_a$ in 
\eqref{emmsfinq1} are all
rational multiples of $N$. 
We leave further investigation of this point for the future.
It is worth noting, though, that if we set 
$b_3=0$ then this condition is satisfied.
In addition, when $b_3=0$ both the master volume and the entropy do not depend on $n_3$ and the expressions agree with
the corresponding expressions for the universal twist. However, noting that the R-charges are not proportional
to the fluxes, we see that these solutions are \emph{not} associated with the universal twist, but instead can be interpreted 
as a marginal deformation, parametrised by $n_3$.

As in the previous subsection,  we can  compare these geometric results with the field theory analysis that was presented in
\cite{Hosseini:2016ume},  for genus $g=0$. The relevant  three-dimensional quiver gauge theory was 
discussed in section 6.2 of \cite{Benini:2009qs}; it is an instance of a family of ``flavoured ABJM'' theories, with gauge group $SU(N)\times SU(N)$ and four 
bi-fundamental  chiral fields $A_1,A_2,B_1,B_2$.  The flavour fields consist of
four sets of fields $q^{(f)},\tilde q^{(f)}$, $f=1,2,3,4$  transforming in the fundamental or  anti-fundamental representations of one of the two  $SU(N)$ nodes. 
The associated quiver diagrams are drawn in Figure 6(a) of \cite{Benini:2009qs}, and the superpotential is given by
 \begin{align}
 W \ =  \  \mathrm{Tr} \left[A_1B_1A_2 B_2   - A_1 B_2 A_2 B_1  + \sum_{i=1}^{k_1}       q^{(1)}_i A_1  \tilde q^{(1)}_i   + \sum_{i=1}^{k_2}       q^{(2)}_i A_2  \tilde q^{(2)}_i    \right. \nn\\
 \left.  + \sum_{i=1}^{k_3}       q^{(3)}_i B_1  \tilde q^{(3)}_i  + \sum_{i=1}^{k_4}       q^{(4)}_i B_2  \tilde q^{(4)}_i \right]\, . 
 \label{superpotq1}
 \end{align}
In particular, the theory\footnote{Interestingly, the case $k_1=1$ and $k_2=k_3=k_4=0$ corresponds to the $\C\times$conifold geometry that we discussed in the previous section.} 
with $k_1=k_2=1$ and $k_3=k_4=0$ (see Figure 9 of \cite{Benini:2009qs}) 
corresponds to the $C(Q^{1,1,1})$  geometry of relevance here. 
In this family of theories the monopole operators $T$ and $\tilde T$ satisfy the quantum relation 
\begin{align}
T\tilde T = A_1 A_2\,.
\end{align}
The large $N$ free energy on $S^3$ for the $Q^{1,1,1}$ case  was first  computed in \cite{Cheon:2011vi} and later extended to the full class of theories with arbitrary number of flavours in  \cite{Jafferis:2011zi}. In this reference it was
also shown that the free energy agrees with the expression \eqref{FS3tovol} in terms 
 of the Sasakian volume in the dual AdS$_4\times Y_7$ supegravity solution.  The large $N$ topologically twisted index on $S^1\times S^2$ of these theories was calculated in \cite{Hosseini:2016ume}. 

Let us now focus on the $Q^{1,1,1}$ model.
The  field-divisors map (\ref{fieldtodivisors}) that is needed to read off the charges of fields in the quiver is obtained using the perfect matching variables which were given in
 \cite{Benini:2009qs}. In the notation of that reference we have
\begin{align}
A_1 =  a_{-1} a_0 \, , \quad A_2  = c_0 c_1 \, , \quad B_1 \ =  \ b_0 \, , \quad B_2 \ =  \ d_0 \,,    \quad 
T = a_{-1}c_0  \, , \quad   \tilde T = a_0 c_1 \, , 
\label{Q111map}
\end{align}
where the perfect matching variables $(a_0,b_0,c_0,d_0,a_{-1},c_1)$ are associated to the toric 
data\footnote{The toric data given in \eqref{Q111toricdata} and
that associated with Figure 9 of \cite{Benini:2009qs}  are related 
via an $SL(3;\Z)$ transformation  given by 
\begin{align}
\left(\begin{tabular}{ccc}
$-1$ & $-1$ & 1 \\
$0$ & $-1$ & $1$ \\
$0$ & $1$ & $0$ \\
\end{tabular}
\right)
\end{align}
acting on the $\v_a$  in (\ref{Q111toricdata}), followed by a reflection of the third coordinate $z\to -z$.}
 (\ref{Q111toricdata}) as in Table \ref{tableQ111vpf} below. 
\begin{table}[h!]
\begin{center}
\begin{tabular}{|c|c|c|c|c|c|c|}
\hline
 $\v_1$ & $\v_{2}$ &$\v_{3}$& $\v_{4}$ &$\v_{5}$  &$\v_{6}$    \\ 
\hline
$a_0$ & $c_0$ & $d_0$ & $b_0$ &$c_1$ & $a_{-1}$   \\
\hline
\end{tabular}
\caption{Relation between toric data  (\ref{Q111toricdata}) and perfect matchings  for the $Q^{1,1,1}$ singularity \cite{Benini:2009qs}.} 
\label{tableQ111vpf}\end{center}
\end{table}

We now consider compactifying this $d=3$ quiver gauge theory on a Riemann surface $\Sigma_g$, with a twist that is parametrized by integer valued flavour magnetic fluxes for the fields
 $\{\mathfrak{n}_{A_1}, \mathfrak{n}_{A_2},\mathfrak{n}_{B_1},\mathfrak{n}_{B_1},\fm_T,\fm_{\tilde T}\}$ with  
 $\mathfrak{n}_{A_1}+ \mathfrak{n}_{A_2}+\mathfrak{n}_{B_1} +\mathfrak{n}_{B_2} = 2(1-g)$, as required for supersymmetry.   

Assuming that the theory flows to a SCQM in the IR, we expect that the dual supergravity solution will be an AdS$_2\times Y_9$ solution of $D=11$ supergravity with $Y_9$ a fibration of a toric  $Y_7$ over $\Sigma_g$ and we can compare with our geometric results above.
Using the relations  \eqref{Q111map}  we can express the R-charges of the fields, $\Delta_I$, in terms of 
the geometric R-charges, $\Delta_a$,
via
\begin{align}\label{firstdeldicccon}
\Delta_{A_1} &  =  \  \Delta_1 + \Delta_6  \   =      \frac{2 }{3-4 b_3^2}\, ,    \qquad  \Delta_{A_2} =     \Delta_2 + \Delta_5   =   \frac{2 }{3-4 b_3^2}\, ,\nn\\
\Delta_{B_1}& = \Delta_4 =   \frac{1-4 b_3^2 }{3 -4 b_3^2}\, ,  \qquad\qquad\qquad \Delta_{B_2} = \Delta_3 =  \frac{1-4 b_3^2 }{3 -4 b_3^2}\, ,  
\end{align}
and 
\begin{align}
\Delta_{T}  =  \Delta_2 + \Delta_6 =   2 \left(b_3+\frac{1}{3-4 b_3^2}\right)\,  , \qquad \Delta_{\tilde T}  =  \Delta_1 + \Delta_5 =   2 \left(-b_3+\frac{1}{3-4 b_3^2}\right)\,  .
\end{align} 
Here the equalities $\Delta_{A_1}   =    \Delta_{A_2}$, $\Delta_{B_1}   =    \Delta_{B_2}$   follow from our initial restriction of $\vec{b}= (1,0,b_3,\frac{1}{2}+b_3)$.    
Notice that the $\Delta_m$ monopole charge  is simply given by
\begin{align}
\label{monoppol}
\Delta_m \equiv \frac{1}{2} (\Delta_{T} - \Delta_{\tilde T} ) =  2 b_3  \, ,
\end{align}
and of course the R-charges satisfy $\Delta_{A_1}+ \Delta_{A_2}+\Delta_{B_1}+ \Delta_{B_2}=  2$. 
Analogously, the  fluxes associated to the fields can be identified with a set of geometric flux parameters $ \fm_a\equiv -M_a/N$ (see \eqref{fluxredefgoth}) via
\begin{align}
\fm_{A_1}= \fm_1+\fm_6 &=    \frac{  -3 \left(4 b_3^2-1\right) n_1+8 b_3 n_3}{\left(3-4 b_3^2\right)^2} \, \,,\nn\\
\fm_{A_2}= \fm_2+\fm_5&=    \frac{  -3 \left(4 b_3^2-1\right) n_1+8 b_3 n_3}{\left(3-4 b_3^2\right)^2} \, \,,\nn\\
\fm_{B_1}= \fm_4&=  \frac{ -16 b_3 n_3+\left(16 b_3^4+3\right) n_1}{2\left(3-4 b_3^2\right)^2} \   , \nn\\
\fm_{B_2}= \fm_3&=-\frac{ -16 b_3 n_3+\left(16 b_3^4+3\right) n_1}{2\left(3-4 b_3^2\right)^2} \,,
\label{urgy}
\end{align}
and 
\begin{align}\label{lastiko}
\fm_{T}& = \fm_2+\fm_6 =\frac{3\left(1-4 b_3^2\right)n_1+(9+8b_3-24b_3^2-16b_3^4)n_3}{\left(3-4 b_3^2\right)^2} \nn\\
\fm_{\tilde T} & = \fm_1+\fm_5= \frac{3\left(1-4 b_3^2\right)n_1+(-9+8b_3+24b_3^2-16b_3^4)n_3}{\left(3-4 b_3^2\right)^2} \, , 
\end{align}
Notice that  the fluxes of the adjoint fields satisfy $\fm_{A_1}+\fm_{A_2}+\fm_{B_1}+\fm_{B_2}=n_1$, as implied by supersymmetry, while $\fm_T-\fm_{\tilde T} =2n_3$, mirroring \eqref{monoppol}
and $\fm_{T}+\fm_{\tilde T}=\frac{6\left(1-4 b_3^2\right)n_1+16b_3n_3}{\left(3-4 b_3^2\right)^2}$. Again, 
the equalities $\fm_{A_1}   =    \fm_{A_2}$, $\fm_{B_1}   =    \fm_{B_2}$  follow from our initial restriction of $\vec{n}= (n_1,0,n_3,\frac{1}{2}+n_3)$.
As already noted above the fluxes given in (\ref{urgy}) are \emph{not} rational \emph{a priori} and their values 
depend on the $b_3$, which a dynamical variable.  These should be held fixed while extremizing the index, given below, as a function of $b_3$. This is to be contrasted with the example discussed in the previous subsection, where the fluxes \eqref{nicerisntit} were manifestly integer and independent of the $b_i$. It would be
interesting to determine the precise conditions when a corresponding supergravity solution exists.

For the case of $g=0$, we can compare our results with the 
large $N$ limit for the off-shell index on $S^1\times S^2$, $\mathcal{I} (\Delta,\mathfrak{n})$,
that was computed in   \cite{Hosseini:2016ume}. Specifically, equation (5.46) of this reference\footnote{To compare with the expression in
 \cite{Hosseini:2016ume} one should relate the variables used here to that used in     
 \cite{Hosseini:2016ume}  as  $\Delta_{HM} = \pi \Delta $.} gives 
 \begin{align}
\mathcal{I} =   -  \frac{2\pi N^{3/2}}{3}\frac{ \left(\Delta _m^4-3 \Delta _m^2+6\right) +  (\mathfrak{t} + \mathfrak{\tilde t}) \Delta _m \left(\Delta _m^2-5\right) }{ \left(3-\Delta _m^2\right)^{3/2}}~,
\end{align}
which remarkably agrees with (\ref{brxt}), after identifying   $n_3 = \mathfrak{t} + \mathfrak{\tilde t}$ and using 
\eqref{monoppol}.

As for the $\C\times$Conifold example, we find that the master volume in \eqref{mvolq1fin} 
is related to the large $N$ free energy and Bethe potential $\mu$ as 
\begin{align}
\mathcal{V} (b_3)= \frac{(2\pi \ell_p)^9}{L^9} \frac{F_{S^3} (b_3)}{64\pi^3} = \frac{(2\pi \ell_p)^9}{L^9} \frac{N^{3/2}}{48\pi^4} \mu (b_3)\,,
\end{align}
which we can also write as a homogeneous function of the geometric R charges, namely  
\begin{align}
\mathcal{V} (\Delta_a) = \frac{32 \pi ^7 l_p^9 N^{3/2}\left( ( \Delta_{1}+ \Delta_{2}+\Delta_{3}+ \Delta_{4} +\Delta_{5}+ \Delta_{6} )^2 -4\Delta_m^2\right)}{6 L^9 \sqrt{3( \Delta_{1}+ \Delta_{2}+\Delta_{3}+ \Delta_{4} +\Delta_{5}+ \Delta_{6} )^2-4\Delta_m^2}}\, , 
\end{align}
with $2\Delta_m \equiv\Delta_2 +\Delta_6  - \Delta_1 - \Delta_5$. Using this, we find that indeed 
\begin{align}
\Ssusy  (b_i,\fm_a)= -  4\pi \sum_{a=1}^d \fm_a  \frac{\de \mathcal{V}}{\de \Delta_a}   \,,
\end{align}
holds true.

We conclude by making contact with the results for the universal twist discussed in
section  \ref{unitwist}. Setting $b_3=0$ we find the following 
\begin{align}
& \Delta_{A_1}   =    \Delta_1 + \Delta_6  \   = \  \frac{2}{3}\,  ,   & \Delta_{A_2} \  = \   \Delta_2 + \Delta_5  \  = \ \frac{2}{3}\,  , \nn\\
&\Delta_{B_1} =   \Delta_4  \  =  \ \frac{1}{3}\,  , \,\, \qquad  & \Delta_{B_2} \ =  \  \Delta_3  \  = \  \frac{1}{3}\,  ,  
\end{align}
with $\Delta_m  \equiv \frac{1}{2} (\Delta_{T} - \Delta_{\tilde T} ) =  \frac{1}{2}(\Delta_2 +  \Delta_6 - \Delta_1-   \Delta_5)=0$, and 
the values of the entropy function and master volume reduce to 
\be
\cS  |_\mathrm{on-shell} =  (g-1) \frac{4\pi}{3\sqrt{3}}N^{3/2} \ =  \ (g-1)F_{S^3}~,
\ee
which are the values obtained in section \ref{unitwist}.
Although the parameter $n_3$ does not enter these expressions, nor the following
field theory fluxes, 
\begin{align}
\mathfrak{n}_{A_1} = \mathfrak{n}_{A_2}=  \frac{2}{3}(1-g)\, ,
\qquad 
\mathfrak{n}_{B_1} = \mathfrak{n}_{B_2} \  = \  \frac{1}{3}(1-g) \, ,
\end{align}
we have $ \fm_{T}=2/3(1-g)+n_3$ and $\fm_{\tilde T} =2/3(1-g)-n_3$. When $n_3=0$ we precisely recover 
the universal twist. However, when $n_3\ne 0$, as noted earlier we don't
have $\Delta_a = \frac{  \fm_a }{1-g}$ and, $n_3$  corresponds 
to a marginal deformation of the universal twist.
 
\subsection{SPP example}\label{sppex}

In this section we consider another example of the form $Y_7 \ \hookrightarrow \ Y_9 \ \rightarrow  \Sigma_g $, with toric $Y_7$.
Specifically, $Y_7$ is the link of a toric Calabi-Yau 4-fold singularity that is closely related to the 
3-fold singularity known as the suspended pinch point (SPP). In a slight abuse of terminology, we will
refer to it as the SPP 4-fold singularity. The dual $d=3$ field theory is a known quiver 
Chern-Simons theory which we will recall momentarily, and its Abelian ({\it i.e.} rank $N=1$) mesonic vacuum moduli space is
precisely the SPP 4-fold singularity. In the genus zero case, $g=0$, the large $N$ limit of the twisted topological index on
$S^1\times S^2$ index was 
computed in \cite{Hosseini:2016tor}. In this subsection we will use our new formalism to recover
the results of \cite{Hosseini:2016tor} from the gravitational point of view. 

\begin{figure}[h!]
\begin{center}
  \includegraphics[height=4.5cm]{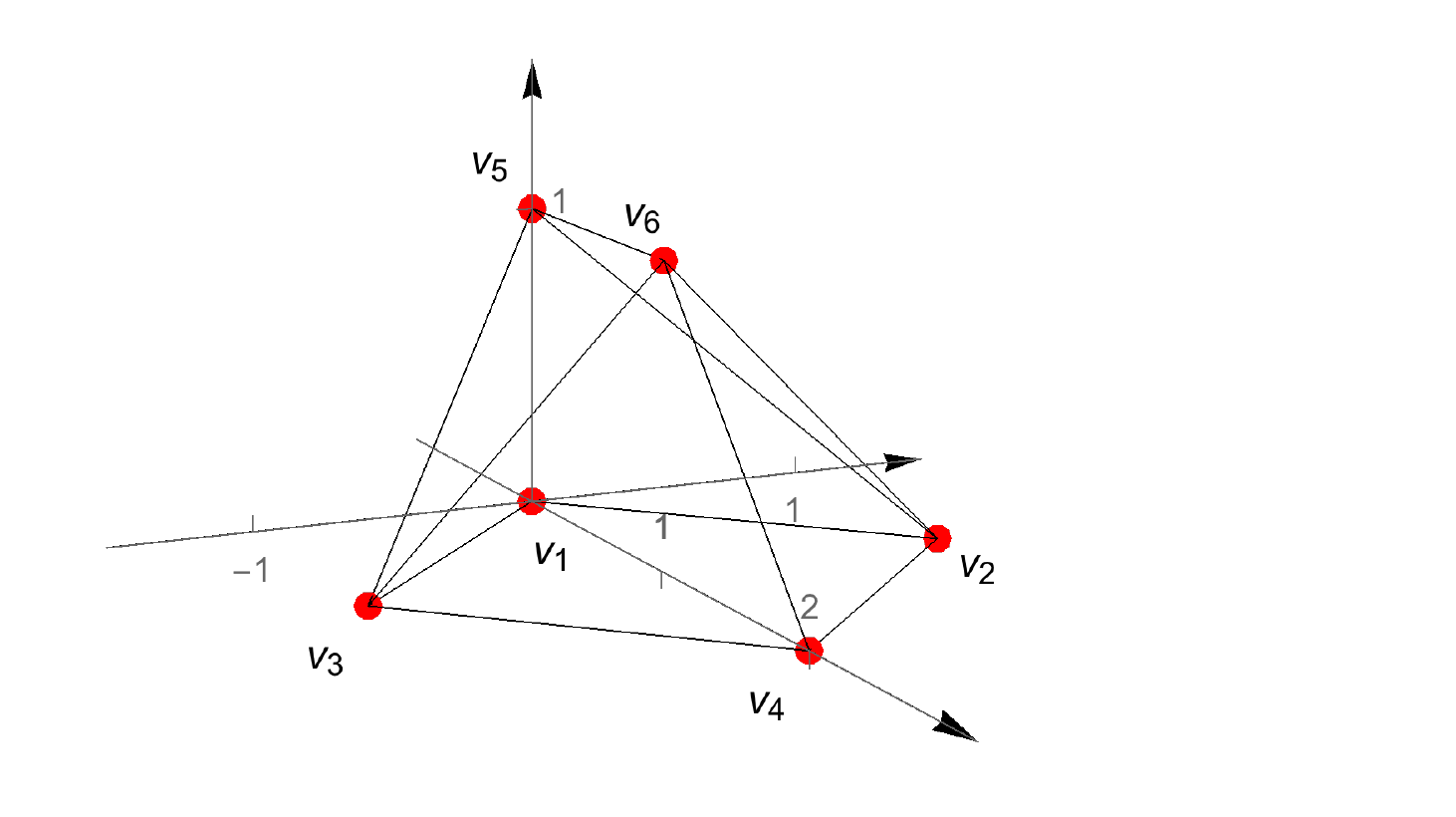}
 \caption{\label{spptoricdiag} Toric diagram for the link of the SPP 4-fold singularity.}
\end{center}
\end{figure}

The SPP 4-fold singularity is labeled by an integer $k$, which parametrizes the Chern-Simons levels in the dual gauge theory, and for simplicity we will set $k=1$ in the following.  
The toric diagram\footnote{The model is a special case of a family of quiver theories labeled by two integers $a,b$, known as $L^{a,b,a}$, for which the corresponding toric diagram has eight vertices, given in (\ref{Labaapp}). The case of  SPP is $L^{1,2,1}$ for which the eight vertices degenerate
to six; see for example \cite{Farquet:2013cwa}.} 
has six vertices associated with six
inward pointing normal vectors which we write as 
\begin{align}
\label{SPPtoricdata}
\v_1 \,  &=\,  (1,0,0,0)~, \,\,\quad \qquad \v_2 \, = \, (1,1,1,0)~, \quad\quad \v_3 \, = \, (1,1,-1,0)~, \nn\\
\quad \v_4 \, &= \, (1,2,0,0)~,\,\quad \qquad \v_5 \, = \, (1,0,0,1)\, , \,\,\quad\quad \v_6 \, = \, (1,1,0,1)\,.
\end{align}
The toric diagram  is obtained by projecting on $\R^3$ the vertices  in (\ref{SPPtoricdata}) and 
is shown in Figure \ref{spptoricdiag}. Notice that this has a manifest $\Z_2$ symmetry along the $y$ axis of the Figure 
(the third entry in \eqref{SPPtoricdata}). The presence of the square face in the diagram reveals that, like 
in the preceding subsection, this is a case with worse-than-orbifold singularities.

As in the section \ref{cconsection} we can use our master volume formulae for the toric diagram
in Figure \ref{spptoricdiag}, but we must suitably restrict the K\"ahler class parameters.
By resolving the toric diagram by adding in an extra line either from $\v_1$ to $\v_4$ or from
$\v_3$ to $\v_2$ we can obtain master volume formulae that satisfy
\eqref{keyvRrel} and hence we can introduce gauge invariant variables $X,Y$ and $Z$ that are explicitly given in \eqref{xyzspp}.
We can proceed with the toric diagram in Figure \ref{spptoricdiag} provided that we set $X=0$.
In fact the algebraic expressions are still rather unwieldy so we will make some additional
assumptions in order to connect with the results of \cite{Hanany:2008fj}. We will assume that
the genus $g\ne 1$ and consider a one-parameter family of fibrations, parametrised 
by $\mathfrak{n}\in\mathbb{Z}$, given by
\begin{align}
\vec{n}=(1,1  -  \tfrac{1}{2}  \mathfrak{n},0, \mathfrak{n})n_1\,,
\label{begiven}
\end{align}
with $n_1=2(1-g)$, as usual.
Associated with these fibrations, which preserves certain symmetries, 
we take the trial R-symmetry vector to be 
\begin{align}\label{beespp}
\b  =   (1,1-\tfrac{1}{2}b_4,0,b_4)~.
\end{align}
It is worth noting that we are not in the universal twist 
class\footnote{To see this, note that a necessary requirement for the universal twist is that $b_1n_i=n_1 b_i$, as
in (\ref{twistb}), but this is satisfied if and only if $b_4=\mathfrak{n}$.
However, $b_4=\mathfrak{n}\in\mathbb{Z}$ is not a critical point of the Sasaki volume 
(\ref{sasvolumeapp}), whose extremum is found at an irrational value of 
$b_4$ \cite{Martelli:2011qj}, in agreement with the fact that the SPP is not associated with
a quasi-regular Sasaki-metric.}.
With this choice of fibration parameters, we find that is consistent, \emph{a posteriori}, 
to not only set $X=0$, which we must do to use the master volume formula, but also to impose  
$\lambda_1=\lambda_2=\lambda_3=\lambda_4$ and $\lambda_5=\lambda_6$ which then implies that 
$X=0$, $Y=-2Z$ and $Z=(1-b_4)\lambda_1+b_4 \lambda_5$.

We can now implement our general procedure to compute the off-shell entropy function $\cS$. 
We first solve the constraint equation 
\eqref{constraintformulauniverse} for $A$. We next solve \eqref{enntwo} for $Z$ finding
\begin{align}\label{zsolspp}
Z^2=
\frac{(2\pi \ell_p)^6}{L^6}\frac{ (b_4 -2)^2 (b_4 -1)^2 b_4 }{64 \pi ^4 (4-3 b_4) }N\,.
\end{align}
The fluxes $M_a$ in (\ref{formuletta}) are then given by
\begin{align}\label{flxspp}
   M_1 &=  M_4 = -\frac{2(g-1)(b_4-1)(4-2b_4+\mathfrak{n}(3b_4-5))}{(3b_4-4)^2}N\,, \nn\\
      M_2 &=  M_3 =-  \frac{(g-1)(-8+12b_4 -5b_4^2+ \mathfrak{n}(6-8b_4+3b_4^2))}{(3b_4-4)^2}N\,,  \nn\\ 
 M_5 &=  M_6   =   (g-1)\fm  N~.
\end{align}
While $M_5$ and $M_6$ are integers, $M_1,\dots,M_4$ are not, in general. However, we do have
\begin{align}
M_1+M_2=M_3+M_4=(g-1)(1-\fm)N\,.
\end{align}
We will see that these fluxes precisely agree with those in the field theory; we expect that from
a geometric point of view these can also be directly justified by determining the \emph{bona fide} cycles in
$Y_7$ and then demanding that the associated fluxes are all integers.
Furthermore, the geometric R-charges can be computed using formula
(\ref{defgeomRcharges}) to get
\begin{align}\label{sppdelts2}
R_1&=  R_4=\frac{2(1-b_4)^2}{4-3b_4}N\,,\nn\\
R_2&=  R_3=\frac{(2-b_4)(1-b_4)}{4-3b_4}N\,,
\nn\\  
R_5 &=  R_6=b_4 N\,,
\end{align}
and one can check that $R_1+ R_2=(1- b_4)N$.
Finally, the off-shell entropy function is computed to be
\begin{align}
\cS= \frac{4\pi }{3} (g-1)  N^{3/2}  \frac{  b_4  \left(7 b_4^2-18   b_4 +12 \right)  + \mathfrak{n}  \left(-6 b_4^3+19   b_4^2-18  b_4 +4 \right)   }{ (4  -3 b_4 )^{3/2}b_4^{1/2}}  \, .
\label{finalentropyspp}
\end{align}
As in previous examples there is an overall sign ambiguity, not explicitly displayed, associated with solving for $Z$ in \eqref{zsolspp}, and can be fixed by demanding that the on-shell value, obtained by extremizing with respect to $b_4$,  is positive. 

We can now compare these results with the field theory analysis, for $g=0$, carried out in
\cite{Hosseini:2016tor}. We first recall various aspects of the three-dimensional quiver gauge theory.
The gauge group is $SU(N)^3$ and there are three doublets of bi-fundamental chiral fields $A_i,B_i,C_i$.   The fields 
 $A_i$ and $C_i$
 transform  in the fundamental and anti-fundamental of a residual global $SU(2)$ symmetry, while the bi-fundamental $B_i$ and the adjoint field 
 $\phi$ 
 are singlets under this $SU(2)$ symmetry. The quiver diagram can be found in Figure 1 of \cite{Hosseini:2016tor}
 and the superpotential reads 
 \be
 W \ =  \ \mathrm{Tr} \left[\phi (A_1A_2 -C_1 C_2) - A_2 A_1 B_1 B_2  + C_2 C_1 B_2 B_1\right]\, .
 \label{superpotspp}
 \ee
 The R-charges of the fields, $\Delta_I$, all depend on one parameter, which we denote by $\Delta$ (and will shortly be identified as $\Delta=b_4$ in the geometry), via
\begin{align}
\Delta_{A_i}=  \Delta_{C_i}= 1 - \Delta \, , \qquad  \Delta_{B_i}= \Delta\, , \qquad \Delta_{\phi}= 2\Delta\, .
\label{fieldscharges}
\end{align}
 To see this we use the fact that each monomial in $W$ must have R-charge equal to 2, and also that 
 the $SU(2)$ symmetry implies that fields in a doublet have equal R-charges. This implies the conditions on 
 $\Delta_{A_i}$, $\Delta_{C_i}$ given in 
 \eqref{fieldscharges} as well as $\Delta_{B_1}+\Delta_{B_2}=2\Delta$. To deduce that $\Delta_{B_1}= \Delta_{B_2}$ one can  
 invoke a $\Z_2$ symmetry of the quiver and superpotential that acts on the fields as $A_i\leftrightarrow C_i$, $B_1\leftrightarrow - B_2$, $\phi\to -\phi$.

This three-dimensional theory flows to a SCFT in the IR, with gravity dual 
AdS$_4\times Y_7$, where $Y_7$ is the Sasaki-Einstein base of the SPP Calabi-Yau cone singularity.  
In  \cite{Martelli:2011qj} it was shown that the large $N$ limit of the free energy, $F_{S^3}$,
obtained from the exact localized partition function on $S^3$, takes the form
(\ref{FS3tovol}), with the Sasakian volume $\mathrm{Vol}(Y_7)$   given in (\ref{sasvolumeapp}).
Moreover the R-charges of the fields in the quiver may be expressed in terms of  the geometric R-charges $\Delta^{3d}_a$, 
 defined using the  volumes of supersymmetric five-dimensional toric submanifolds $T_a$, through the  relation 
 (\ref{R1dtoR3d}).
 
The fields in the $d=3$ quiver field theory
correspond to linear combinations of these toric divisors and, furthermore, the field-divisors map 
(\ref{fieldtodivisors})  may be obtained by employing  the perfect matching variables. For the SPP singularity this map
was given in \cite{Hanany:2008fj}, and in the notation of that reference reads
\begin{align}\label{sppmap}
A_1 &=  p_1 q_1 \, , \qquad A_2 =  p_2 q_2 \, , \qquad B_1 = p_3 \, , \qquad B_2 =  p_4 \, , \nn\\
C_1 & = p_1 q_2 \, , \qquad C_2\  =\  p_2 q_1 \, , \qquad \phi \  = \ p_3 p_4 \, ,
\end{align}
where the perfect matching variables $(p_1,p_2,p_3,p_4,q_1,q_2)$ are associated to the toric 
data\footnote{The toric diagram here and that used in  \cite{Hanany:2008fj}  are related via an $SL(3;\Z)$ transformation  given by 
 \begin{align}
\left(\begin{tabular}{ccc}
0 & 0 & $-1$ \\
1 & 0 & 0 \\
0 & $-1$ & 0 \\
\end{tabular}
\right)
\end{align}
acting on the $\v_a$    in (\ref{SPPtoricdata}).}
(\ref{SPPtoricdata})  as in  Table \ref{tablevpf} below.
\begin{table}[h!]
\begin{center}
\begin{tabular}{|c|c|c|c|c|c|c|}
\hline
 $\v_1$ & $\v_{2}$ &$\v_{3}$& $\v_{4}$ &$\v_{5}$  &$\v_{6}$    \\ 
\hline
$p_1$ & $q_2$ & $q_1$ & $p_2$ &$p_3$ & $p_4$   \\
\hline
\end{tabular}
\caption{Relation between toric data  (\ref{SPPtoricdata}) and perfect matchings  for the SPP singularity   \cite{Hanany:2008fj}.} 
\label{tablevpf}\end{center}
\end{table}

We now consider compactifying this $d=3$ quiver gauge theory on a Riemann surface $\Sigma_g$, with a twist that is parametrized by integer valued flavour magnetic fluxes for the fields $\{\mathfrak{n}_{A_i}, \mathfrak{n}_{B_i}, \mathfrak{n}_{C_i},\mathfrak{n}_\phi\}$, respecting the global 
 symmetries of the theory.  
 Assuming that the theory flows to a SCQM in the IR, we expect that the dual supergravity solution will be an AdS$_2\times Y_9$ solution of $D=11$ supergravity with $Y_9$ a fibration of a toric  $Y_7$ over $\Sigma_g$ and we can compare with our geometric results above.
 To compare with the known field theory results of  \cite{Hosseini:2016tor}
we restrict our considerations to the following fluxes 
\be
\mathfrak{n}_{A_i}   =   \mathfrak{n}_{C_i}=   (1-g)(1- \mathfrak{n} )\, ,  \qquad  \mathfrak{n}_{B_i}   \ =   (1-g) \mathfrak{n}\, ,  \qquad  \mathfrak{n}_\phi    =   (1-g)2 \mathfrak{n} \, ,
\label{sppfieldthfluxes}
\ee
where $\mathfrak{n}\in\mathbb{Z}$.
In particular, one can check that every term in the superpotential  (\ref{superpot}) has  $2(1-g)$ units of flux, as required for supersymmetry.   
Next, we can use the map \eqref{sppmap} to relate the R-charges of the fields, $\Delta_I$, with 
the geometric R-charges, $\Delta_a\equiv R_a/N$ (see \eqref{defcapDel}),
via
 \begin{align}
\label{fieldstogeometric}
\Delta_{A_1} &  = \Delta_1 + \Delta_3\, , \quad   \Delta_{A_2} =   \Delta_2 + \Delta_4  \, , \quad  \Delta_{B_1} = \Delta_5  \, , \quad  \Delta_{B_1} =    \Delta_6  \, , \nn\\
\Delta_{C_1} & =  \Delta_1 + \Delta_2\, , \quad   \Delta_{C_2}=  \Delta_3 + \Delta_4  \, , \quad \Delta_{\phi}  =  \Delta_5 + \Delta_6\, .
\end{align}
Then from \eqref{fieldscharges} we have
\begin{align}\label{sppdelts}
&\Delta_4=  \Delta_1\,,\quad  \Delta_3=  \Delta_2\,,\quad  \Delta_6 =  \Delta_5\,,\nn\\
&\Delta_1+  \Delta_2=1- \Delta\,,\quad \Delta_5=\Delta\,,
\end{align}
exactly as in the geometric expressions \eqref{sppdelts2}. 
Indeed, comparing to the latter allows us to identify
\begin{align}
\Delta=b_4\, .
\end{align}
Similarly, the  fluxes of the fields can be identified with a set of geometric flux parameters $ \fm_a\equiv -M_a/N$ (see \eqref{fluxredefgoth})
in an entirely analogous manner
{\it e.g.} $\mathfrak{n}_{A_1}   =  \fm_1 + \fm_3$ etc.
For the specific fluxes that we want to consider, given in \eqref{sppfieldthfluxes}, we conclude that
the geometric flux parameters are
\begin{align}\label{conspppfracs}
&\fm_4=  \fm_1\,,\quad  \fm_3=  \fm_2\,,\quad  \fm_6 =  \fm_5\,,\nn\\
&\fm_1+  \fm_2=(1-g)(1- \mathfrak{n})\,,\quad \fm_5=(1-g)\mathfrak{n}\,.
\end{align}
Notice in particular, that there is not a quantization condition on $\fm_1$ and $\fm_2$ individually, but just on their sum. This is again in exact agreement with the geometric expressions \eqref{flxspp}.

To further compare with the geometric results we need
to identify the twist parameters $n_i$, $i=1,\dots,4$, 
that define the geometric fibration of $Y_7$ over $\Sigma_g$, and
which appear in (\ref{rhoomegaintseven})-\eqref{Maint}. To do this, we can use the relation 
(\ref{vMrelation}):
\be\label{frakmarel}
  \sum_a^d v_a^i \fm_a =  n_i\, .
\ee
Using the toric data \eqref{SPPtoricdata}, along with the relations \eqref{conspppfracs} obtained from field theory
we deduce that the relationship between the geometric twist $n_i$ and the field theory twist $\mathfrak{n}$ is 
given by $\vec{n}=(1,1  -  \frac{1}{2}  \mathfrak{n},0, \mathfrak{n})n_1$, which is precisely
what we assumed on the geometry side in \eqref{begiven}.

Finally, for the case of $g=0$, with the field theory data for the R-charges $\Delta_I$, determined by
$\Delta$ in \eqref{fieldscharges}, and the
magnetic fluxes $\fm_I$, determined by $\fm$ in \eqref{sppfieldthfluxes}, the 
large $N$ limit for the off-shell index on $S^1\times S^2$, $\mathcal{I} (\Delta,\mathfrak{n})$,
that was computed in Appendix B of  \cite{Hosseini:2016tor}; see equation (B.19) of this reference\footnote{To match the expression in (B.19) of   \cite{Hosseini:2016tor} one should relate the variable $\Delta$ used here to that used in  \cite{Hosseini:2016tor}, $\Delta_{HZ}$, via
$\Delta_{HZ} = \pi \Delta $.},  namely
\begin{align}
\mathcal{I} (\Delta,\mathfrak{n})=-\frac{4\pi }{3}  N^{3/2}  \frac{  \Delta  \left(7 \Delta ^2-18   \Delta +12 \right)  + \mathfrak{n}  \left(-6 \Delta ^3+19   \Delta ^2-18  \Delta +4 \right)   }{ (4  -3 \Delta )^{3/2}\Delta^{1/2}}  \, .
\label{finalentropyspp2}
\end{align}
Remarkably, this exactly agrees with the geometric result $\cS  (\b,n_i)$ in \eqref{finalentropyspp}, for the restrictions
on $(\b,n_i)$ given in \eqref{begiven}, \eqref{beespp}
 after setting
$g=0$, identifying $\Delta=b_4$, and taking the upper sign.

From our analysis of the geometry we can obtain the following off-shell expression for the master volume
\begin{align}
\mathcal{V}(\Delta)   = \frac{(2\pi\ell_p)^9}{L^9} \frac{N^{3/2}}{24 \pi ^2 }\frac{(1-\Delta) (2-\Delta ) \sqrt{\Delta }}{\sqrt{4-3 \Delta }}\, .
\end{align}
To obtain this we used \eqref{beespp}, the conditions on the K\"ahler class parameters,
$X=0$, $Y=-2Z$ and $Z=(1-\Delta)\lambda_1+\Delta \lambda_5$, 
and we have fixed the sign ambiguity arising from solving \eqref{zsolspp}. Using \eqref{sasvolumeapp}, \eqref{FS3tovol} we can again
relate this to the off-shell free energy of the dual $d=3$ SCFT on $S^3$ via
\begin{align}
\mathcal{V} (\Delta)  = \frac{(2\pi\ell_p)^9}{L^9} \frac{1}{64\pi^3}F_{S^3} (\Delta)\, .
\label{mastertoF}
\end{align}
Furthermore, the following relation between the master volume and the trial entropy function holds
\be
\cS  \ = \  - \frac{(2\pi\ell_p)^9}{L^9}  32\pi^3(1-g) \left( 2 \mathcal{V}  + (\mathfrak{n} - \Delta) \frac{\de\mathcal{V}}{\de \Delta}\right)  \, , 
\label{sppentropyfunction}
\ee
consistent with the field theory results in appendix B of \cite{Hosseini:2016tor}. 

\subsection{Connection to the index theorem of  \cite{Hosseini:2016tor}}
\label{SISI}

In the above examples we have seen that our entropy function $\cS$ 
coincides, off-shell, with the large $N$ limit of the topologically twisted index, ${\mathcal I}$, after 
using the dictionary between the geometric and field theory quantities.
In this section, we will place these results in a more general context,
making a closer comparison with the index theorem presented in  \cite{Hosseini:2016tor}. 
A key ingredient is the result for the entropy function given in \eqref{ssss} in terms of the variables $\Delta_a$, that we discussed in section \ref{chgevars}.

The main results of \cite{Hosseini:2016tor} were in the context of 
${\cal N}=2$, non-chiral, quiver gauge theories in $d=3$, with matter fields transforming in the
adjoint and bi-fundamental representations of the gauge group, as well as ``flavour'' fields 
that transform in the 
(anti-)fundamental representations. Below we will restrict to the class of flavoured theories, with quantum corrected moduli spaces, that were considered in \cite{Benini:2009qs}, but we expect that a similar connection between ${\cal I}$ and $\cS$ will  hold more generally, for the 
class of theories considered in  \cite{Hosseini:2016tor} (indeed, the SPP example is not a flavoured model). 
We will also further restrict to cases where the field-divisor map \eqref{fieldtodivisors} is invertible,  
which includes the  $\mathbb{C}\times$Conifold example of section \ref{cconsection} and the
$Q^{1,1,1}$ example of section \ref{gotit}.

For this class of field theories the geometric R-charges, $\Delta_a$, and fluxes, $\fm_a$,
are related to the field theory R-charges, $\Delta_I$, and magnetic fluxes, $\mathfrak{n}_I$,
by invertible linear relationships of the form
\begin{align}
\label{linearchange}
 \Delta_I   =\sum_{a=1}^d c^a_I \Delta_a\, , \quad \qquad \mathfrak{n}_I  =\sum_{a=1}^d c^a_I  \fm_a  \, ,
 \end{align}
 where the index $I$ here runs over the adjoint and bi-fundamental chiral fields in the quiver, as well as the diagonal monopole operators $T$ and $\tilde T$, but \emph{not} the (anti-) fundamentals.
Using \eqref{linearchange}   in \eqref{ssss} one quickly deduces that
\begin{align}\label{ftsumentconj}
\Ssusy = -  4\pi \sum_a \fm_a\frac{\de \mathcal{V}}{\de \Delta_a}  
=-4\pi \sum_I \fm_I\frac{\de \mathcal{V}}{\de \Delta_I}\,.
\end{align}
Furthermore, \emph{if}\footnote{Note that this does not seem to be the case for the example of $Q^{1,1,1}$ that we discussed in 
section \ref{exsgsols}. With the assumptions made in that section
we have $\Delta_m=0$, and the free energy on $S^3$ is given by 
$F=4\pi N^{3/2}\frac{1}{3\sqrt{3}}$ and is only related to $\mathcal{V}$ via \eqref{effmtcalv} when $x=1$, which is the case of the universal twist.}
the master volume $\mathcal{V}$ coincides with the $S^3$ free energy via\footnote{As we have noted several times,
sign ambiguities arise in carrying out the extremal problem which we are not explicitly writing.}
\begin{align}\label{effmtcalv}
\mathcal{V} (\Delta_I)  = \frac{(2\pi\ell_p)^9}{L^9} \frac{1}{64\pi^3}F_{S^3} (\Delta_I)\, .
\end{align}
we can conclude that
\begin{align}\label{myindex}
\cS  =-\frac{1}{2} \sum_I \fm_I\frac{\de F_{S^3}}{\de \Delta_I}\, ,
\end{align}
which has  exactly the form of the index theorem discussed in \cite{Hosseini:2016tor}.

In the field theory result of \cite{Hosseini:2016tor} the sum over the index 
``$I$'' runs  \emph{a priori} over all the chiral fields in the quiver, namely the adjoint, the bi-fiundamentals, as well as the (anti-)fundamental flavours, and also includes contributions from 
the magnetic fluxes 
and fugacities 
associated with the topological symmetry of the theories.  
 However,  the large $N$ free energy 
depends on the topological symmetry charges only through the combination $\Delta_{m} =\frac{1}{2} (\Delta_T - \Delta_{\tilde T})$
and  the contribution  of the (anti-)fundamental fields can always be rewritten 
in terms of adjoint and bi-fundamental fields, using the constraints  imposed by the superpotential  \cite{Jafferis:2011zi}.

Returning to our formula (\ref{myindex}), it is illuminating to extract from the sum the contributions of the 
monopole operators $T$ and $\tilde T$. After defining the linear combinations 
\begin{align}
\Delta_{m} &=\frac{1}{2} (\Delta_T - \Delta_{\tilde T})\,,\qquad
\Delta_{p} =\frac{1}{2} (\Delta_T + \Delta_{\tilde T})\nn\,,\\
\fm_{m} &=\frac{1}{2} (\fm_T - \fm_{\tilde T})\,,\qquad
\fm_{p} =\frac{1}{2} (\fm_T + \fm_{\tilde T})\,,
\end{align}
we can rewrite \eqref{myindex} as
\begin{align}\label{doggy}
\cS&= -  \frac{1}{2} {\sum_I}' \fm_I\frac{\de F_{S^3}}{\de \Delta_I} - \frac{ \fm_m}{2} \frac{\de F_{S^3}}{\de \Delta_m}
- \frac{ \fm_p}{2} \frac{\de F_{S^3}}{\de \Delta_p}~,
\nn\\
&= -  \frac{1}{2} {\sum_I}' \fm_I\frac{\de F_{S^3}}{\de \Delta_I} - \frac{ \fm_m}{2} \frac{\de F_{S^3}}{\de \Delta_m}~,
\end{align}
where the prime in the sum indicates that it now does not include the monopole operators
(nor, as usual, the (anti-)fundamental fields) and the second line follows from the fact that in the large $N$ limit
$F_{S^3}$ is independent \cite{Jafferis:2011zi} of $\Delta_p$ as mentioned above.
This result can be favourably compared with the field theory results of \cite{Hosseini:2016tor} after recalling that
here we are using a set of constrained variables, such that the master volume/free energy is a \emph{homogeneous} function.

Let us now return to the $\C\times$conifold example of  section \ref{cconsection} and   use this general expression to discuss further the restriction on the fluxes (\ref{extrarel}). The geometric fluxes $n_i$ are related to the variables introduced above as
\begin{align}
\fm_p= \frac{1}{2}(n_1-n_2) \, ,   \qquad \fm_{m}= \frac{1}{2} (n_1-n_2)-n_4 \, , 
\end{align}
and to get agreement with \cite{Hosseini:2016ume}  we had to impose $\fm_{m}=1$.  Indeed we find that the first two terms in (\ref{finalentropyccon2}) match the primed sum in \eqref{doggy}, while the remainder term exactly agrees if  $\fm_m=1$. We then conclude that in the field theory calculation in \cite{Hosseini:2016ume}  it has been  assumed that $\fm_m=1$,  but   it should be possible to incorporate a generic value of $\fm_m$ that would then fully agree with our   geometric result.

\section{Discussion}\label{secdisc}

In this paper we have extended the results of \cite{Couzens:2018wnk,Gauntlett:2018dpc} concerning a
geometric extremal problem, analogous to volume minimization in Sasakian geometry \cite{Martelli:2005tp,Martelli:2006yb}, 
that allows one to calculate key properties 
of supersymmetric AdS$_3\times Y_7$ and AdS$_2\times Y_9$ solutions.
Specifically, we have provided a formalism based on a \emph{master volume} that allows one to study
AdS$_3\times Y_7$ solutions with toric $Y_7$ as well as AdS$_2\times Y_9$ solutions
where $Y_9$ is a fibration of a toric $Y_7$ over a Riemann surface $\Sigma_g$. In both cases $Y_7$ can
be non-convex toric \cite{Couzens:2018wnk}.

The results concerning the latter class of solutions comprise a geometric dual of 
${\cal I}$-extremization \cite{Benini:2015eyy}
for the class of $d=1$, ${\cal N}=2$ SCFTs obtained from 
compactifying toric $d=3$,  ${\cal N}=2$ SCFTs ({\it i.e.} dual to AdS$_4\times SE_7$ solutions with toric $SE_7$)
on a Riemann surface $\Sigma_g$, with a 
partial topological twist.  
We expect that this class of AdS$_2\times Y_9$ solutions will generically arise as the near horizon limit of supersymmetric black hole solutions that asymptotically approach AdS$_4\times SE_7$ in the UV.  
The supersymmetric action to be extremized in our procedure  can also be interpreted, after suitable normalization,
as  an \emph{entropy function}, which reduces  to  the Bekenstein-Hawking entropy of the black hole at the critical point. Thus, our results can be used to calculate the entropy of large classes of supersymmetric black holes, independently of a detailed knowledge of the full supergravity solutions, just assuming that they exist,
thus extending \cite{Benini:2015eyy} to a much more general class of black hole solutions. 
Furthermore, when it is possible to carry out a calculation of the associated topological index in the field theory using localization techniques, and assuming that they agree, one will then have a microscopic state counting interpretation of the black hole entropy for this class of black holes.

We illustrated the extremization procedure in various examples. 
We highlighted that the formalism can be used, with care, for toric singularities with
worse-than-orbifold singularities. This is important since many examples in the literature are
associated with such singular $Y_7$.
Most strikingly, in the examples that we studied we are able to match
the off-shell entropy function and the off-shell field theory results for the large $N$ limits of 
the partition functions on $S^1 \times \Sigma_g$ (\emph{i.e.} the topological twisted
index). In addition we were also able to obtain some general results in section \ref{SISI}. It would certainly be very
interesting determine the necessary and sufficient conditions for the relations discussed there to hold,
in order to aim for a general proof of the equivalence of 
${\cal I}$-extremization and $\cS$-extremization, analogous to 
the result of \cite{Hosseini:2019use} who analysed a similar problem for the class of AdS$_3\times Y_7$ solutions
with $Y_7$ a fibration of a toric $Y_5$ over $\Sigma_g$.

In a slightly different direction, it would certainly be interesting to construct 
a master volume formula for AdS$_2\times Y_9$ solutions with toric $Y_9$. This would
provide a direct geometric dual of the $d=1$ version of $F$-extremization for toric $Y_9$ and may well help in
establishing a precise field theory version of $F$-extremization in $d=1$. It would also be very
interesting to determine whether or not this more general class of AdS$_2\times Y_9$ solutions can arise as
the near horizon limit of black holes. 


\subsection*{Acknowledgments}
We thank Noppadol Mekareeya and Alberto Zaffaroni for discussions.
JPG is supported by the European Research Council under the European Union's Seventh Framework Programme (FP7/2007-2013), ERC Grant agreement ADG 339140. JPG is also supported by STFC grant ST/P000762/1, EPSRC grant EP/K034456/1, as a KIAS Scholar and as a Visiting Fellow at the Perimeter Institute. DM is supported in part by the STFC  grant ST/P000258/1.


\appendix

\section{Computation of the master volume in examples}
\label{masterdetails}

In this appendix we give some details of the computation of the master volume formula in the examples that we studied in the paper. We recall the master volume formula is given by
\begin{align}
\label{masterinapp}
  \mathcal{V} (\b;\{\lambda_a\}) =  -\frac{(2\pi)^4}{3!} \sum_{a=1}^d    \lambda_a  \sum_{k=2}^{\n_a-1}   \frac{X_{a,k}^I X_{a,k}^{II}  }{(\v_a,\v_{a,k-1},\v_{a,k},\b )   (\v_{a,\n_a},\v_a,\v_{a,1},\b )  (\v_{a,k},\v_{a,k+1},\v_a,\b) }\,,
\end{align}
where
\begin{align}
X_{a,k}^I  =  &  -\lambda_a (\v_{a,k-1},\v_{a,k},\v_{a,k+1},\b)   +   \lambda_{a,k-1}(\v_{a,k},\v_{a,k+1},\v_a,\b)    \nn\\
& -  \lambda_{a,k} (\v_{a,k+1},\v_a,\v_{a,k-1},\b )   + \lambda_{a,k+1}(\v_a,\v_{a,k-1},\v_{a,k},\b )\,,\nn\\
 X_{a,k}^{II} \  = \ &  -\lambda_a (\v_{a,1},\v_{a,k},\v_{a,\n_a},\b)   +   \lambda_{a,1}(\v_{a,k},\v_{a,\n_a},\v_a,\b)   \nn\\
  &  -  \lambda_{a,k} (\v_{a,\n_a},\v_a,\v_{a,1},\b )   + \lambda_{a,\n_a}(\v_a,\v_{a,1},\v_{a,k},\b )~.
\end{align}
Here $d$ is the number of vertices  $\{\v_a\}$, $a=1,\dots d$, of the toric diagram and for each vertex, $\n_a$ denotes the number of edges meeting there.   

\subsection{$Y_7= Y^{p,k} (\C P^2)$}

The toric diagram for $Y^{p,k} (\C P^2)$ is given in Figure 
\ref{CC2toricdiag}, where we labeled the vertices as in \cite{Martelli:2008rt}. In particular, we have
\begin{align}
\label{Ypkcp2app}
\v_1 \,  &=\,  (1,0,0,0)~, \quad \v_2 \, = \, (1,0,0,p)~, \quad \v_3 \, = \, (1,1,0,0)~, \nn\\
\quad \v_4 \, &= \, (1,0,1,0)~,\quad \v_5 \, = \, (1,-1,-1,k) \,.
\end{align}
The polytope has $d=5$ vertices, with $\n_1  = \n_2=3$ and $\n_3=\n_4=\n_5=4$. The vectors needed to evaluate 
(\ref{masterinapp}) are given in Table \ref{table1}. The ordering of the vectors can be obtained by going counter-clockwise
around a vertex when viewed from outside the toric diagram in Figure \ref{CP1CP1toricdiag}. One can explicitly 
check that this ensures that each term in the sums in (\ref{masterinapp}) is positive for $\b$ inside the Reeb cone. 
\begin{table}[h!]
\begin{center}
\begin{tabular}{|c|c||c|c|c|c|}
\hline
$a$ & $\v_a$ & $\v_{a,1}$ &$\v_{a,2}$& $\v_{a,3}$ &$\v_{a,4}$\\ 
\hline\hline
$1$ & $\v_1$ &$\v_5$ &$\v_4$ &$\v_3$ &$ - $  \\
\hline
$2$ & $\v_2$ &$\v_3$ &$\v_4$ &$\v_5$ &$ - $  \\
\hline
$3$ & $\v_3$ &$\v_1$ &$\v_4$ &$\v_2$ &$\v_5$  \\
\hline
$4$ & $\v_4$ &$\v_3$ &$\v_1$ &$\v_5$ &$\v_2$  \\
\hline
$5$ & $\v_5$ &$\v_1$ &$\v_3$ &$\v_2$ &$\v_4$  \\
\hline
\end{tabular}
\caption{Vectors used to compute the master volume of $Y^{p,k} (\C P^2)$.}
\label{table1}\end{center}
\end{table}
The resulting formula fits in a few lines and we do not report it here.

\subsection{$Y_7 = Y^{p,k} (\C P^1\times \C P^1)$}

The toric diagram for $Y^{p,k} (\C P^1\times \C P^1)$ is given in Figure 
\ref{CP1CP1toricdiag}, where recall the vertices are given by
\begin{align}
\label{Ypkcp1cp1app}
\v_1 \,  &=\,  (1,0,0,0)~, \quad \v_2 \, = \, (1,0,0,p)~, \quad \v_3 \, = \, (1,-1,0,0)~, \nn\\
\quad \v_4 \, &= \, (1,1,0,k)~,\quad \v_5 \, = \, (1,0,-1,0)\, , \quad \v_6 \, = \, (1,0,1,k)\,.
\end{align}
The polytope has $d=6$ vertices, with $\n_1=\n_2=\n_3=\n_4=\n_5=\n_6=4$. 
The vectors needed to evaluate 
(\ref{masterinapp}) are given in Table \ref{table2}. Again the ordering of the vectors can be obtained by going counter-clockwise
around a vertex when viewed from outside the toric diagram in Figure 3 of  \cite{Martelli:2008rt}. 
\begin{table}[h!]
\begin{center}
\begin{tabular}{|c|c||c|c|c|c|}
\hline
$a$ & $\v_a$ & $\v_{a,1}$ &$\v_{a,2}$& $\v_{a,3}$ &$\v_{a,4}$\\ 
\hline\hline
$1$ & $\v_1$ &$\v_6$ &$\v_4$ &$\v_5$ &$\v_3$  \\
\hline
$2$ & $\v_2$ &$\v_3$ &$\v_5$ &$\v_4$ &$\v_6$  \\
\hline
$3$ & $\v_3$ &$\v_2$ &$\v_6$ &$\v_1$ &$\v_5$  \\
\hline
$4$ & $\v_4$ &$\v_5$ &$\v_1$ &$\v_6$ &$\v_2$  \\
\hline
$5$ & $\v_5$ &$\v_4$ &$\v_2$ &$\v_3$ &$\v_1$  \\
\hline
$6$ & $\v_6$ &$\v_1$ &$\v_3$ &$\v_2$ &$\v_4$  \\
\hline
\end{tabular}
\caption{Vectors used to compute the master volume of $Y^{p,k} (\C P^1\times \C P^1)$.}
\label{table2}\end{center}
\end{table}
The formula of the master volume is lengthy, so we do not write it down.

\subsection{$Y_7$ = Link of $\mathbb{C}\times$ Conifold}\label{secctcon}
The toric diagram for the toric K\"ahler cone $\mathbb{C}\times$ Conifold
is given in Figure \ref{Cconfigure} and the vertices are given by
\begin{align}
\label{Cconapp}
\v_1 \,  &=\,  (1,0,0,0)~, \quad \v_2 \, = \, (1,0,0,1)~, \quad \v_3 \, = \, (1,0,1,1)~, \nn\\
\quad \v_4 \, &= \, (1,0,1,0)~,\quad \v_5 \, = \, (1,1,0,0) \,.
\end{align}
An important feature of this example is that the square face in the toric diagram shows
that the link $Y_7$ has worse-than-orbifold singularities. Indeed,
the toric divisor associated with $\v_5$ is a copy of the conifold.
The polytope has $d=5$ vertices, with $\n_1  = \n_2=\n_3=\n_4=3$ and $\n_5=4$. The vectors needed to evaluate the master volume
(\ref{masterinapp}) are given in Table \ref{tableccon}. As usual, the ordering of the vectors can be obtained by going counter-clockwise
around a vertex when viewed from outside the toric diagram in Figure \ref{Cconfigure}. 
\begin{table}[h!]
\begin{center}
\begin{tabular}{|c|c||c|c|c|c|}
\hline
$a$ & $\v_a$ & $\v_{a,1}$ &$\v_{a,2}$& $\v_{a,3}$ &$\v_{a,4}$\\ 
\hline\hline
$1$ & $\v_1$ &$\v_5$ &$\v_2$ &$\v_4$ &$ - $  \\
\hline
$2$ & $\v_2$ &$\v_5$ &$\v_3$ &$\v_1$ &$ - $  \\
\hline
$3$ & $\v_3$ &$\v_5$ &$\v_4$ &$\v_2$ &$-$  \\
\hline
$4$ & $\v_4$ &$\v_5$ &$\v_1$ &$\v_3$ &$-$  \\
\hline
$5$ & $\v_5$ &$\v_4$ &$\v_3$ &$\v_2$ &$\v_1$  \\
\hline
\end{tabular}
\caption{Vectors used to compute the master volume of the link of $\mathbb{C}\times$ Conifold.}
\label{tableccon}\end{center}
\end{table}
It is important to note that the relation \eqref{keyvRrel} is \emph{not} satisfied for this example in general. However, it is satisfied when
we impose $\lambda_1-\lambda_2+\lambda_{3}-\lambda_{4}=0$. This is precisely a consequence of the fact that
$Y_7$ has worse-than-orbifold singularities. Specifically, when 
$\lambda_{1}-\lambda_{2}+\lambda_{3}-\lambda_{4}\ne 0$ the master volume formula is no longer calculating
a volume. 

\begin{figure}[h!]
\begin{center}
  \includegraphics[height=4.2cm]{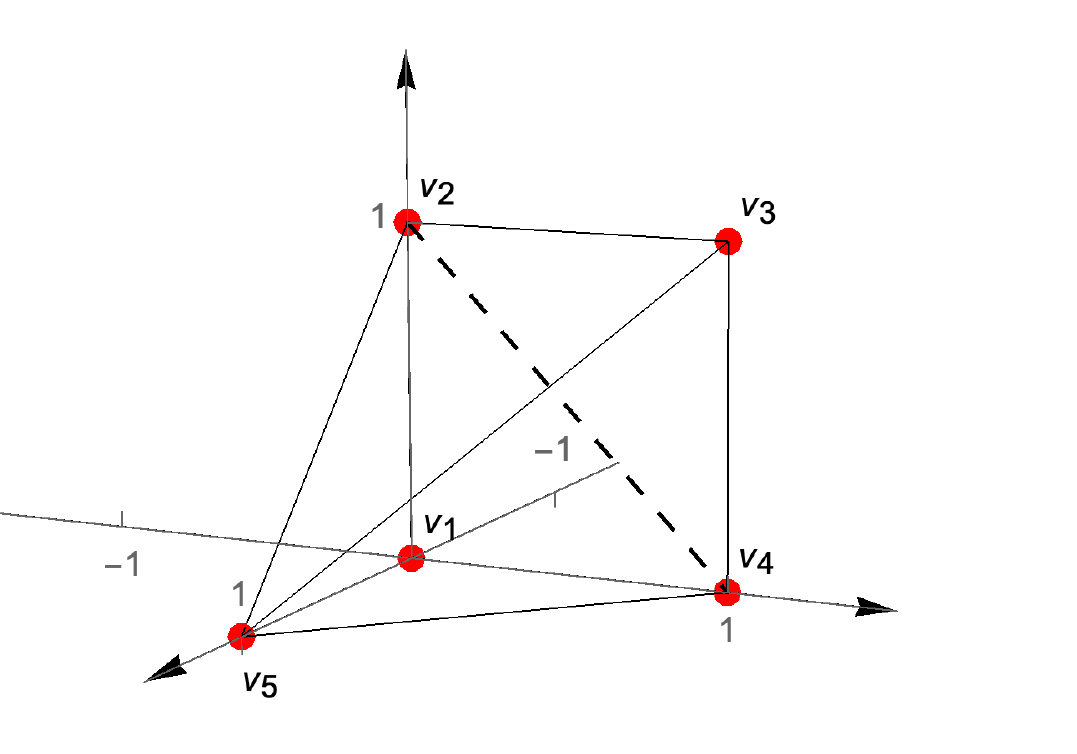}$\qquad$
    \includegraphics[height=4cm]{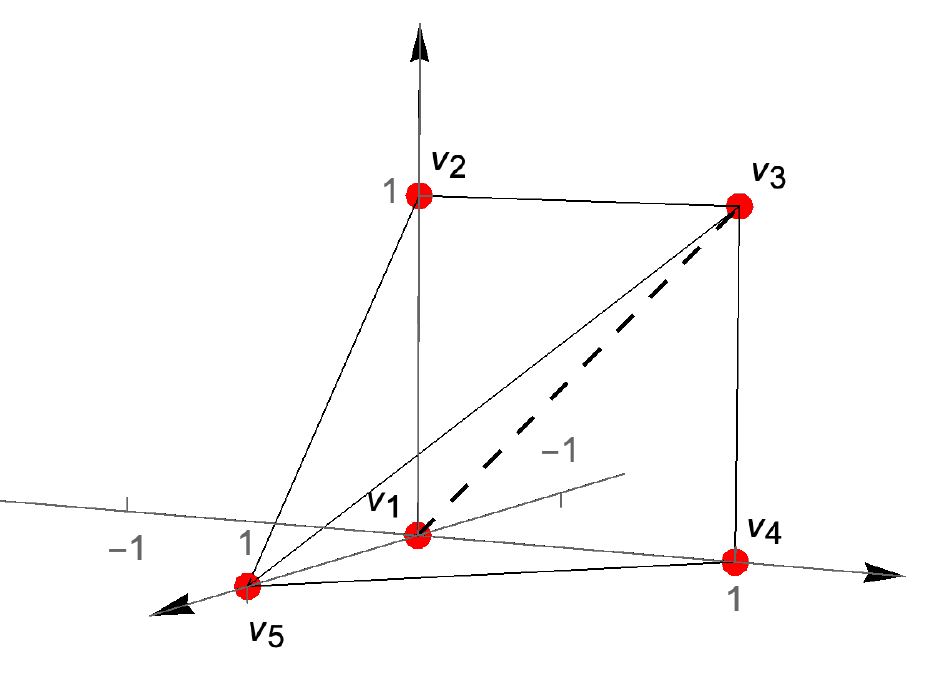}
 \caption{\label{Cconfigurediagres} Toric diagrams for two resolutions of the link of the $\mathbb{C}\times$ Conifold singularity.}
\end{center}
\end{figure}

Further insight can be obtained by resolving the conifold singularity on $Y_7$. This can be done in two 
different ways, associated with a flop transition of the conifold. In each case we 
keep the same vertices but add in an extra line, either stretching from $\v_2$ to $\v_4$
or from $\v_1$ to $\v_3$ as in Figure \ref{Cconfigurediagres}.
Notice that all faces of the toric diagram are triangles
and so there are now at worst, only orbifold singularities. To calculate the master volume for these two cases
we note that in the first we have $\n_1  =\n_3=3$ and $\n_2=\n_4=\n_5=4$ 
while in the second we have $\n_2  =\n_4=3$ and $\n_1=\n_3=\n_5=4$.
The vectors needed to evaluate (\ref{masterinapp}) are given in the left and right hand tables in Table \ref{tableccon2}, respectively. 
\begin{table}[h!]
\begin{center}
\begin{tabular}{|c|c||c|c|c|c|}
\hline
$a$ & $\v_a$ & $\v_{a,1}$ &$\v_{a,2}$& $\v_{a,3}$ &$\v_{a,4}$\\ 
\hline\hline
$1$ & $\v_1$ &$\v_5$ &$\v_2$ &$\v_4$ &$ - $  \\
\hline
$2$ & $\v_2$ &$\v_5$ &$\v_3$ &$\v_4$ &$ \v_1 $  \\
\hline
$3$ & $\v_3$ &$\v_5$ &$\v_4$ &$\v_2$ &$-$  \\
\hline
$4$ & $\v_4$ &$\v_5$ &$\v_1$ &$\v_2$ &$\v_3$  \\
\hline
$5$ & $\v_5$ &$\v_4$ &$\v_3$ &$\v_2$ &$\v_1$  \\
\hline
\end{tabular}
\qquad\qquad
\begin{tabular}{|c|c||c|c|c|c|}
\hline
$a$ & $\v_a$ & $\v_{a,1}$ &$\v_{a,2}$& $\v_{a,3}$ &$\v_{a,4}$\\ 
\hline\hline
$1$ & $\v_1$ &$\v_5$ &$\v_2$ &$\v_3$ &$ \v_4 $  \\
\hline
$2$ & $\v_2$ &$\v_5$ &$\v_3$ &$\v_1$ &$ - $  \\
\hline
$3$ & $\v_3$ &$\v_5$ &$\v_4$ &$\v_1$ &$\v_2$  \\
\hline
$4$ & $\v_4$ &$\v_5$ &$\v_1$ &$\v_3$ &$-$  \\
\hline
$5$ & $\v_5$ &$\v_4$ &$\v_3$ &$\v_2$ &$\v_1$  \\
\hline
\end{tabular}
\caption{Vectors used to compute the master volume for the two resolved geometries given in Figure \ref{Cconfigurediagres}, respectively.}
\label{tableccon2}\end{center}
\end{table}
For each of these two cases, we find that the expression for $\mathcal{V} (\b;\{\lambda_a\})$ now satisfies \eqref{keyvRrel}.
By analysing \eqref{lamgt} we can obtain two gauge-invariant combinations of the $\{\lambda_a\}$ given by
\begin{align}
X&=\lambda_{1}-\lambda_{2}+\lambda_{3}-\lambda_{4}\,,\nn\\
Y&=(b_1-b_2-b_4) \lambda_{1}+(b_4-b_3) \lambda_{2}+b_3 \lambda_{3}+b_2 \lambda_{5}\,.
\end{align}
In terms of these variables, the volumes $\mathcal{V} (\b;\{\lambda_a\})$ take the form
\begin{align}
\mathcal{V}^{(1)}  &=
8 \pi ^4\tfrac{ b_3 (b_1-b_2-b_4) \left[\left(b_3^2+b_4^2-(b_1-b_2+b_3) b_4\right) X^3-3 (b_3-b_4) X^2 Y+3 X Y^2\right]+(b_2-b_1) Y^3}{3 b_2 b_3 b_4 (-b_1+b_2+b_3) (-b_1+b_2+b_4)}\,,\nn\\
\mathcal{V}^{(2)}  &=
8 \pi ^4\tfrac{ b_3 (b_1-b_2-b_4) \left[(b_3-b_4)^2 X^3-3 (b_3-b_4) X^2 Y+3 X Y^2\right]+(b_2-b_1) Y^3}{3 b_2 b_3 (-b_1+b_2+b_3) b_4 (-b_1+b_2+b_4)}\,,
\end{align}
for the two different resolved geometries, respectively, and $\mathcal{V}^{(2)}-\mathcal{V}^{(1)}=\frac{8 \pi ^4 X^3}{3 b_2}$. Furthermore, one can check that for the special K\"ahler class with $X=0$, then 
we get
 \begin{align}
\mathcal{V}^{(1)} =\mathcal{V}^{(2)}   
=  \frac{8\pi ^4 Y^3(b_2-b_1)}{3 b_2 b_3 b_4(-b_1+b_2+b_3)  (-b_1+b_2+b_4)}\,,\quad\text{when $X=0$}
 \end{align}
and, importantly, this agrees with the expression for $\mathcal{V} (\b;\{\lambda_a\})$ for 
the link of $\mathbb{C}\times$ Conifold, with its worse-than-orbifold singularities, when $X=0$.

The Sasaki K\"ahler class is obtained when $\lambda_a = - \tfrac{1}{2b_1}$, or equivalently when $X=0$ and $Y=-1/2$. Extremizing this volume while holding $b_1=4$ fixed, at the critical point we recover the Sasakian volume formula\footnote{One can check that this volume is twice that given in equation (3.33) of \cite{Martelli:2008rt} for the case of $Y^{p,k}(\C P^1\times \C P^1)$ with $p=1$ $k=2$, consistent
with the fact that this is a $\mathbb{Z}_2$ orbifold of the link of $\C\times$ Conifold.}
 \begin{align}
 \mathrm{Vol}(Y_7)   =  \frac{16 \pi ^4}{81}\,,
 \end{align}
with the critical Reeb vector given by
\begin{align}
\label{SPPcritReeb}\b =  (4,1,\frac{3}{2},\frac{3}{2})\,.
\end{align} 

\subsection{$Y_7$ = Link of SPP}\label{secsppapp}
The toric diagram for the SPP 4-fold singularity comprises of the following six vertices 
\begin{align}
\label{Labaapp}
\v_1 \,  &=\,  (1,0,0,0)~, \,\,\quad \qquad \v_2 \, = \, (1,1,1,0)~, \quad\quad \v_3 \, = \, (1,1,-1,0)~, \nn\\
\quad \v_4 \, &= \, (1,2,0,0)~,\,\quad \qquad \v_5 \, = \, (1,0,0,1)\, , \,\,\quad\quad \v_6 \, = \, (1,1,0,1)\,.
\end{align}
The toric diagram, obtained by projecting the vertices in (\ref{Labaapp}) onto $\R^3$, is shown in Figure
\ref{spptoricdiag}. Notice that this has a manifest $\Z_2$ reflection symmetry along the $y$ axis (third entry). 
Just like the case of $\C\times $ Conifold that we considered in the previous subsection the link of this singularity
has worse-than-orbifold singularities as revealed by the presence of a non-triangular face bounded by
$\v_1$, $\v_2$, $\v_3$ and $\v_4$.

The polytope has $d=6$ vertices, with $\n_1=\n_4=3$ and $\n_2=\n_3=\n_5=\n_6=4$. 
The vectors needed to evaluate 
(\ref{masterinapp}) are given in Table \ref{table3}.
\begin{table}[h!]
\begin{center}
\begin{tabular}{|c|c||c|c|c|c|}
\hline
$a$ & $\v_a$ & $\v_{a,1}$ &$\v_{a,2}$& $\v_{a,3}$ &$\v_{a,4}$\\ 
\hline\hline
$1$ & $\v_1$ &$\v_3$ &$\v_5$ &$\v_2$ & $ - $   \\
\hline
$2$ & $\v_2$ &$\v_4$ &$\v_1$ &$\v_5$ & $ \v_6 $    \\
\hline
$3$ & $\v_3$ &$\v_4$ &$\v_6$ &$\v_5$ & $ \v_1 $    \\
\hline
$4$ & $\v_4$ &$\v_2$ &$\v_6$ &$\v_3$ & $ - $    \\
\hline
$5$ & $\v_5$ &$\v_6$ &$\v_2$ &$\v_1$ & $ \v_3 $    \\
\hline
$6$ & $\v_6$ &$\v_2$ &$\v_5$ &$\v_3$ &$ \v_4 $    \\
\hline
\end{tabular}
\caption{Vectors used to compute the master volume of the link of the SPP singularity.}
\label{table3}\end{center}
\end{table}

In general the relation \eqref{keyvRrel} is not satisfied for this example, but it is satisfied when
we impose $\lambda_{1}-\lambda_{2}+\lambda_{3}-\lambda_{4}=0$. As in the case
of $\C\times $ Conifold this is again due to the presence of
worse-than-orbifold singularities and we can proceed in a similar manner.
One can consider two resolutions obtained by adding in an extra line in the toric diagram, either 
from $\v_1$ to $\v_4$ or from $\v_2$ to $\v_3$. In each case one gets a master volume formula,
$\mathcal{V}^{(1)} $ and $\mathcal{V}^{(2)}$, respectively, both of which satisfy the relation \eqref{keyvRrel}.
For these cases we can then construct three gauge invariant variables given by
\begin{align}\label{xyzspp}
X&=\lambda_1-\lambda_2-\lambda_3+\lambda_4\,,\nn\\
Y&=b_3 (\lambda_3-\lambda_2)+2 b_2 (\lambda_5-\lambda_6)+b_4 (\lambda_2+\lambda_3-2 \lambda_6)-b_1 (\lambda_2+\lambda_3+2 \lambda_5-2 \lambda_6)
\,,\nn\\
Z&=b_1 \lambda_1+b_4 (\lambda_5-\lambda_1)+b_2 (\lambda_6-\lambda_5)+b_3 (\lambda_1-\lambda_3-\lambda_5+\lambda_6)\,.
\end{align}
We find that $\mathcal{V}^{(2)}-\mathcal{V}^{(1)}=-\frac{4 \pi ^4 X^3}{3 b_4}$ and moreover when
$X=0$ then $\mathcal{V}^{(1)}=\mathcal{V}^{(2)}$ is precisely the same as the master volume calculated from the toric diagram in Figure \ref{spptoricdiag} after setting $X=0$.

An expression that we found useful is when $\lambda_1=\lambda_2=\lambda_3=\lambda_4$ and $\lambda_5=\lambda_6$ in which case $X=0$, $Y=-2Z=b_1 \lambda_1+b_4 (\lambda_5-\lambda_1)$
\begin{align}
\mathcal{V}\to
-16 \pi ^4\tfrac{ \left[-4 b_1^3+8 b_4 b_1^2+\left(4 b_3^2-b_4 (2 b_2+5 b_4)\right) b_1+b_4 \left(b_2^2+b_4 b_2-3 b_3^2+b_4^2\right)\right] Z^3}{3b_4  (b_2^2-b_3^2) (b_1+b_3-b_4) (-2 b_1+b_2-b_3+b_4) (-b_1+b_3+b_4) (-2 b_1+b_2+b_3+b_4)}\,.
\end{align}
The case of the Sasakian volume is obtained by further setting $Z=-1/2$, and hence
$\lambda_a = - \tfrac{1}{2b_1}$. Further
extremizing while holding $b_1=4$  we recover the Sasakian volume formula given in equation (5.5) of \cite{Martelli:2011qj}, namely 
\begin{align}
 \mathrm{Vol}(Y_7)   =  \frac{ \pi ^4 (4- 3 \Delta )}{96\Delta  (2-\Delta )^2 (1-\Delta )^2  }\, ,
 \label{sasvolumeapp}
\end{align}
with the Reeb vector parameterised as 
\begin{align}
\label{SPPcritReeb2}\b =  (4,2(2-\Delta), 0,4\Delta)\,,
\end{align}and $\Delta \simeq 0.319$ at the critical point.

\section{Homology relations and twisting}\label{apptop}

In this appendix we give a proof of the homology relation (\ref{homrel}) that leads to
 (\ref{vMrelation}), extending some of the arguments used in 
\cite{Franco:2005sm}.

 For simplicity we assume that $Y_7$ is simply connected, which since 
$b_1(Y_7)=0$ we may always do by passing to a finite covering space. 
Recall that the cone $C(Y_7)$ may be realized 
as a K\"ahler quotient $C(Y_7)=\C^d//U(1)^{d-4}$. 
The torus $U(1)^{d-4}$ arises as follows. 
Define the linear map  
\begin{align}\label{Alinear}
\mathcal{A}: \R^d \rightarrow \R^4~, \qquad \mbox{where}\ \mathcal{A}(e_a) = v_a~.
\end{align}
Here $\{e_a\}$ denotes the standard orthonormal basis of $\R^d$, with components $e_a^b=\delta_{ab}$. 
Since $\mathcal{A}$ also maps $\Z^d$ to $\Z^4$, 
 (\ref{Alinear}) induces a corresponding map of tori $U(1)^d=\R^d/2\pi \Z^d \rightarrow \R^4/2\pi\, \mathrm{Span}_{\Z}\{v_a\}$, 
and the torus $U(1)^{d-4}$ is precisely the kernel of this map. 
It is generated by an integer matrix $Q_I^a$, $I=1,\ldots,d-4$, satisfying
\begin{align}\label{Qv}
\sum_{a=1}^d Q_I^a v_a^i = 0~,
\end{align}
which specifies the embedding $U(1)^{d-4}\subset U(1)^d$. 
The toric $U(1)^4$ action on $C(Y_7)$ is then via the quotient $U(1)^4=U(1)^d/U(1)^{d-4}$. More physically, 
the above construction may be viewed as a gauged linear sigma model with $d$ complex fields and $U(1)^4$ charges 
specified by $Q^I_a$, with $C(Y_7)$ being the vacuum moduli space of this theory.

In order to fibre $C(Y_7)$ (or equivalently $Y_7$) over a Riemann surface $\Sigma_g$, we may first fibre $\C^d$ over $\Sigma_g$. To do so
we must first lift the $U(1)^4$ action on $C(Y_7)$ to $\C^d$, which means specifying $\alpha^i\in\Z^d$, $i=1,2,3,4$, satisfying
\begin{align}\label{alphai}
\mathcal{A}(\alpha^i)= e_i\in \Z^4~,
\end{align}
where $\{e_i\}$ denotes the standard orthonormal basis of $\R^4$. In components (\ref{alphai}) reads
\begin{align}\label{valpha}
\sum_{a=1}^d v_a^j \alpha_a^i = \delta_{ij}~.
\end{align}
Of course, the choice of each $\alpha^i\in \Z^d$ is unique only up 
to the kernel of $\mathcal{A}$, generated by $Q^a_I$. Geometrically, this is because $C(Y_7)$ is precisely a K\"ahler quotient of $\C^d$ via the 
torus $U(1)^{d-4}$ generated by this kernel. By construction, the  charge of the $a$th coordinate $z_a$ of $\C^d$ under the $i$th $U(1)\subset U(1)^4$ is 
$\alpha_a^i$.
We then construct the associated bundle
\begin{align}\label{Xfibration}
\mathcal{X} \equiv  O(\vec{n})_{\Sigma_g}\times_{U(1)^4} \C^d~.
\end{align}
The space $\mathcal{X}$ is the total space of a $\C^d$ fibration over $\Sigma_g$, where we twist 
the $i$th $U(1)$ action on $\C^d$ via the line bundle $O(n_i)_{\Sigma_g}$, $i=1,2,3,4$. 
This means that $z_a$ may be regarded as a coordinate on the fibre of
 $O((\alpha_a,\vec{n}))_{\Sigma_g}$, where 
$(\alpha_a,\vec{n})=\sum_{i=1}^4 \alpha_a^i n_i$. The fibred geometry we are interested in is
\begin{align}
Y_9 = \left.\mathcal{X}//U(1)^{d-4}\right|_{r=1}~,
\end{align}
where we take a K\"ahler quotient of the fibres $\C^d$ in (\ref{Xfibration}), and set $r=1$ to obtain $Y_7=C(Y_7)\mid_{r=1}$. 

Next we may define the 
 torus-invariant seven-manifolds $\Sigma_a\subset Y_9$ via $\Sigma_a\equiv \{z_a=0\}$ in the above construction. 
In $Y_9$ we may view the $z_a$ as sections of complex line bundles $\mathcal{L}_a$ over $Y_9$. These are sections of line bundles, 
rather than functions, because  $z_a$ is charged both under the torus $U(1)^{d-4}$ that we quotient by, and is also fibred 
over $\Sigma_g$ as $O((\alpha_a,\vec{n}))_{\Sigma_g}$. Consider now the 4 line bundles
\begin{align}\label{Mi}
\mathcal{M}_i \equiv  \bigotimes_{a=1}^d \mathcal{L}_a^{v_a^i}~, \qquad i=1,2,3,4~.
\end{align}
The  restriction of $\mathcal{M}_i$ to each fibre $Y_7$ is a trivial line bundle over that fibre. This follows from (\ref{Qv}), 
where recall that $Q_I^a$ generates the torus action $U(1)^{d-4}$ on $\C^d$. This implies that $\prod_{a=1}^d z_a^{v_a^i}$ are 
invariant under $U(1)^{d-4}$, and so the sections of $\mathcal{M}_i$ are simply complex-valued functions on each fibre $Y_7=\C^d//U(1)^{d-4}\mid_{r=1}$. On the other hand, 
(\ref{valpha}) says that these sections of $\mathcal{M}_j$ have charge $\delta_{ij}$ under the $i$th toric $U(1)\subset U(1)^4$. 
As such, we may identify $\mathcal{M}_i=\pi^{-1}\left[O(-n_i)_{\Sigma_g}\right]$, where $\pi:Y_9\rightarrow\Sigma_g$ is the projection 
to the base.  Taking the first Chern class of (\ref{Mi}) and applying Poincar\'e duality then precisely 
gives
\begin{align}\label{homrel}
\sum_{a=1}^d v_a^i [\Sigma_a] = -n_i [Y_7]\in H_7(Y_9;\Z)~, \qquad i=1,2,3,4~,
\end{align}
 where $n_i [Y_7]$ is the Poincar\'e dual to $n_i\vol_{\Sigma_g}=c_1(O(n_i)_{\Sigma_g})$. Integrating the 
seven-form flux of these cycles and using (\ref{homrel}) then immediately leads to  (\ref{vMrelation}).

Finally, let us comment on (\ref{frakmarel}), which recall arises in the field theory analysis. 
Here by definition $\fm_a\in \Z$ is precisely the twisting of the $a$th gauged linear sigma model field 
over the Riemann surface $\Sigma_g$. On the other hand, in our geometric construction above the 
$a$th gauged linear sigma model field is precisely the coordinate $z_a$ on $\C^d$, and thus we may identify
\begin{align}
\fm_a = \sum_{i=1}^4 \alpha_a^in_i~, \qquad a=1,\ldots,d~.
\end{align}
Thus in the bundle $\mathcal{X}$ defined by (\ref{Xfibration}), the $\C^d$ fibre coordinate
$z_a$ is precisely a section of $O(\fm_a)_{\Sigma_g}$. 
On the other hand, (\ref{valpha}) then implies
\begin{align}
\sum_{a=1}^d v_a^i \fm_a = \sum_{a=1}^d\sum_{i=1}^4 v_a^i \alpha_a^j n_j = n_i~,
\end{align}
which is precisely (\ref{frakmarel}).

\section{Explicit supergravity solutions}\label{explicsol}
Here we further analyse a class of explicit supergravity solutions of the form AdS$_2\times Y_9$ that were first 
discussed in section 6.3 of \cite{Gauntlett:2006ns} and were recently discussed in section 4.3.1 of 
\cite{Azzurli:2017kxo}. 
The eight-dimensional K\"ahler base space used for
the construction of the solutions is given by a product of K\"ahler-Einstein metrics
\begin{align}
\diff s^2_8= \diff s^2(KE_2^{(1)})+\diff s^2(KE_4^+)+\diff s^2(KE_2^{(4)})\,,
\end{align}
where $\diff s^2(KE_4^+)$ is taken to have positive curvature. The Ricci form is given by
\begin{align}
\mathcal{R}= l_1J_{KE_2^{(1)}}+l_2J_{KE_4^+}+l_4 J_{KE_2^{(4)}}\,,
\end{align}
where the $J$'s are the associated K\"ahler forms, $l_i$ are constants and, without loss of generality, we can take $l_2=1$. In order to solve the equation $\Box R-\tfrac{1}{2}R^2+R_{ij}R^{ij}=0$ on the eight-dimensional manifold
we should take $l_4=-\frac{1+2 l_1}{2+l_1}$. One of the two-dimensional K\"ahler-Einstein spaces is
always a Riemann surface of genus $g>1$, which we take to be $KE_2^{(4)}$. The range of $l_1$ is then
$-2+\sqrt{3}\le l_1<\infty$. There are three cases to consider. First when $l_1\in [-2+\sqrt{3},0)$ and 
$l_4\in [-2+\sqrt{3},-1/2)$ then $KE_2^{(1)}$ is also a Riemann surface with genus $g'>1$. Second
when $l_1=0$ and $l_4=-1/2$, then $KE_2^{(1)}$ is a Riemann surface with genus $g'=1$ and finally
when $l_1>0$ and $l_4\in (-2,-1/2)$, then $KE_2^{(1)}$ is a Riemann surface with genus $g'=0$. For simplicity
of presentation we only present details of the analysis for the latter case, which is the case relevant for the analysis
in section \ref{ads2examples}.

We relabel $l_1\equiv x$ and continue with $x>0$. 
The eight-dimensional K\"ahler base space can then be written
\begin{align}\label{kahlereight}
\diff s^2_8= \frac{1}{x}\diff s^2(S^2)+\diff s^2(KE_4^+)+\frac{2+x}{1+2x}\diff s^2(\Sigma_g)\,,
\end{align}
where $\diff s^2(S^2)$ is the metric on the unit radius round two-sphere, so that $\int_{S^2} \vol_{S^2}=4\pi$,
and $\Sigma_g$ is a Riemann surface of genus $g>1$ with $\int_{\Sigma_g} \vol_{S^2}=4\pi(g-1)$. It will
also be useful to note that since the metric on $KE_4^+$ satisfies $\mathcal{R}_{KE_4^+}=J_{KE_4^+}$, we have
$\int_{KE_4^+}\vol_{KE_4^+}=\tfrac{1}{2}(2\pi)^2M$ where $M$ is a topological integer for 
$KE_4^+$. For further discussion of this result see, for example, appendix B of \cite{Gauntlett:2006qw}, where
one can also find a discussion of the Fano index, $m$, of $KE_4^+$. Here we will need the fact that
if we consider a set of two-cycles $\Sigma_i\subset KE_4^+$ that generate $H_2(KE_4^+, \mathbb{Z})$, then we
have $\int_{\Sigma_i}J_{KE_4^+}=2\pi m n_i$ for $n_i\in\mathbb{Z}$. For $KE_4^+=
\C P^1 \times \C P^1$ we have $(m,M)=(2,8)$ and $n_1=n_2=1$. For $\mathbb{C}P^2$
we have $(m,M)=(3,9)$ and $n_1=1$. For the del Pezzos, $dP_k$, $k=3,\dots 8$, we have $m=1$ and $M=9-k$, 
as well as $n_i=1$, $i=1,\dots,k$ and $n_{k+1}=3$.

The metric on $Y_9$ appearing in the AdS$_2$ solution as in \eqref{ansatzd11}
is given by
\begin{align}
\diff s^2_9 = (\diff z+P)^2+\ex^B \diff s^2_8\,,
\end{align}
where $P$ is a local one-form satisfying 
$\diff P= \mathcal{R}_{S^2}+\mathcal{R}_{KE_4^+}+\mathcal{R}_{\Sigma_g}$ and $\ex^B=R/2=\frac{3+2x+x^2}{2+x}$. By defining $Y_7$ to be a circle fibration
over $S^2\times KE_4^+$, we can then view $Y_9$ in the solutions as being obtained by fibreing $Y_7$ over 
$\Sigma_g$. When $x=1$ we have special examples of the universal twist solutions, discussed in section \ref{unitwist}; the cases $KE_4^+= \C P^1 \times \C P^1$ and $\mathbb{C}P^2$ correspond to $Q^{1,1,1}$ and $M^{3,2}$, respectively. The two-form $F$ appearing in the four-form flux \eqref{ansatzd11} is given by
\begin{align}
F= -\frac{3}{x(3+2x+x^2)}\vol_{S^2}-\frac{1+x+x^2}{3+2x+x^2}J_{KE_4^+}-\frac{(2+x)^3}{(1+2x)(3+2x+x^2)}\vol_{\Sigma_g}\,.
\end{align}
For flux quantization we need the seven-form $*G_4$, which takes the form
\begin{align}
*G_4= L^6(\diff z+P)\wedge \Big(\Big[\frac{2+x}{x} \vol_{S^2}+ &\frac{3}{(1+2x)} \vol_{\Sigma_g}\Big]\wedge  \vol_{KE_4^+}\nn\\
&+\frac{1+x+x^2}{x(1+2x)}\vol_{S^2}\wedge J_{KE_4^+}\wedge \vol_{\Sigma_g}\Big)\,,
\end{align}
where $\vol_{KE_4^+}=\frac{1}{2}J_{KE_4^+}\wedge J_{KE_4^+}$.

Regularity of the metric is ensured if $z$ parametrizes a circle with period $2\pi h$, where $h=hcf(2,2(g-1), m)$. Thus, if $m$ is even then $h=2$ and if $m$ is odd then $h=1$.
Next we calculate the flux through the various seven-cycles.
We first consider the seven-cycle obtained by fixing a point on $\Sigma_g$, {\it i.e.} the $z$ circle fibred over 
$S^2\times KE_4^+$, which is a copy of $Y_7$, and we obtain
\begin{align}\label{ennfirst}
N= \left(\frac{L}{\ell_p}\right)^6\frac{hM}{2^2\pi^2}\frac{2+x}{x}\,,
\end{align}
with $N\in\mathbb{Z}$. 
We can also consider the cycle obtained by fixing a point on the $S^2$ as well as the cycle obtained as 
the fibration of $z$ over $S^2\times \Sigma_g\times \Sigma_i$, where $\Sigma_i\subset KE_4^+$ generate 
$H_2(KE_4^+, \mathbb{Z})$. 
After using \eqref{ennfirst}, for these cycles we find, respectively,
\begin{align}\label{enssother}
\tilde N& = \frac{3x}{(1+2 x)(2+x)}(g-1)N\nn\,,\\
N_{i}& = 
\frac{2^2m}{M}\frac{1+x+x^2}{(2+x)(1+2x)}(g-1)N n_i\,,
\end{align}
with $\tilde N, N_i\in\mathbb{Z}$. We need to ensure that $N,\tilde N, N_i\in\mathbb{Z}$. By considering the ratio $N_i/\tilde N$
we conclude that we must have
\begin{align}\label{xrest}
x+\frac{1}{x}\in\mathbb{Q}\,,
\end{align}
which, interestingly, can be achieved for irrational $x$ ({\it e.g.} $x=2+\sqrt{3}$). We 
can then suitably choose $\frac{L}{\ell_p}$ and hence $N$ so that
$\tilde N, N_i\in\mathbb{Z}$.
We also note that for the universal twist, when $x=1$, we have $\tilde N=(1/3)(g-1)N$ and $N_i=(4m)/(3M)(g-1)Nn_i$.
We next calculate
\begin{align}
\cS\ \equiv \  \frac{1}{4G_2}& = \frac{1}{(2\pi)^8}\left(\frac{L}{\ell_p}\right)^9 4 \pi\int \ex^B (\diff z+P) \wedge \mathrm{vol}_8\,,\nn\\
& = \frac{2^3}{(hM)^{1/2}}\pi (g-1)\frac{3+2x+x^2}{(1+2x)}\frac{x^{1/2}}{(2+x)^{3/2}}N^{3/2}\,.
\end{align}
When $x=1$ this expression can be recast in the form 
\begin{align}
\cS  |_{x=1}
\   = \    \frac{(g-1)N^{3/2}\pi^3 \sqrt{2}}{\sqrt{27\mathrm{Vol} (Y_7)}}~,
\label{entropytovolumeuniversal2}
\end{align}
using the fact that the volume of the regular Sasaki-Einstein metrics associated with circle fibrations over
$S^2\times KE_4^+$ can be expressed as $\mathrm{Vol} (Y_7)=h M\pi^4/128$.

In the special case that $KE_4^+$ is toric, {\it i.e.} $KE_4^+=\C P^1\times \C P^1$, $\mathbb{C}P^2$ or $dP_3$,
from \eqref{defwrapped}, the R-charges associated with 
M5-branes wrapping toric divisors associated with five-cycles $T_a$ on $Y_7$ ({\it i.e.} at a fixed point on $\Sigma_g$), are given by,
\begin{align}
R_a= R[T_a]= \frac{2L^6}{(2\pi)^5\ell_p^6}\int_{T_a}(\diff z+P)\wedge \left(\frac{1}{x}\vol_{S^2}\wedge J_{KE_4^+}
+\tfrac{1}{2!}J_{KE_4^+}^2\right)\,.
\end{align}

We now consider the case of $KE_4^+=\C P^1\times \C P^1$, with $(m,M)=(2,8)$ and hence $h=2$. In this case\footnote{Note that if we took $h=1$ then we would have the regular orbifold $Q^{1,1,1}/\mathbb{Z}_2$.}
$Y_7$ is $Q^{1,1,1}$.
Two of the $T_a$, which we take to be $T_5, T_6$ are associated with the circle fibration over $KE_4^+=\C P^1\times \C P^1$, while sitting at
the north and the south pole of the generic $S^2$ in \eqref{kahlereight}, respectively. Similarly,  we can consider sitting at the north and  south pole
of each of the two $\C P^1$ factors in $KE_4^+$ leading to four more $T_a$ with $a=1,\dots, 4$. 
We then find
\begin{align}
R_1\, = \, R_2 \,  = \,  R_3 \, = \,  R_4  \, =\ \frac{1}{2+x}N~,\qquad R_5 \, =  \, R_6  \, = \,  \frac{x}{2+x}N\,,
\end{align}
and one can check that $\sum_a R_a=2N$. 
The labelling we have chosen is consistent with the toric data given by
\begin{align}
\label{Q111toricdata2}
\v_1 \,  &=\,  (1,0,0,0)~, \quad \v_2 \, = \, (1,0,0,1)~, \quad\   \ \v_3 \, = \, (1,-1,0,0)~, \nn\\
\quad \v_4 \, &= \, (1,1,0,1)~,\quad \v_5 \, = \, (1,0,-1,0)\, , \quad \v_6 \, = \, (1,0,1,1)\,,
\end{align} 
as used in \eqref{Ypkvcp1cp1}. 
One can directly obtain the R-symmetry vector $\vec{b}$ and the integers $\vec{n}$ defining the fibration, in the toric basis
\eqref{Q111toricdata2}, by further analysing the associated Killing vectors in the explicit metric. 
However, it is more convenient to obtain them via the following method. For this example, we have verified that 
the identity \eqref{keyvRrel}, which we give again here:
\begin{align}\label{keyvRrel2appendix}
\sum_{a=1}^6 v_a^iR_a= 2 b^i N\,,
\end{align}
holds. Therefore, we can immediately conclude that $\vec{b}=(1,0,0,\frac{1}{2})$ for any $x$.
In addition the fluxes $M_1,\dots,M_4$ are associated with the $N_i$, $i=1,2$ in \eqref{enssother} with $n_i=1$, while $M_5,M_6$ are associated with $\tilde N$. Thus we have
\begin{align}
M_1 \, = \, M_2 \, = \,  M_3& \, =  \,  M_4  \,  = \, \ \frac{1+x+x^2}{(2+x)(1+2x)}(g-1)N\,,\nn\\
M_5& \, = \,  M_6 \, = \, \frac{3x}{(1+2 x)(2+x)}(g-1)N\,.
\end{align}
Similarly, we can use the condition \eqref{vMrelation},
\begin{align}\label{vMrelation2}
\sum_{a=1}^6 v_a^iM_a= -n_i N\,,
\end{align}
to conclude that
$\vec{n}=2(1-g)(1,0,0,\frac{1}{2})$. Notice that for all values of $x$ we have
$n_i  \ =   \  \frac{n_1}{b_1} b_i$, as in the universal twist
\eqref{twistb}. For the special value of $x=1$ we also have $R_a=M_a/(g-1)=N/3$, which is an additional
condition that is required in order to obtain the universal twist solutions in section \ref{unitwist}.

We now consider the case of $KE_4^+=\mathbb{C}P^2$, with $(m,M)=(3,9)$ and hence $h=1$. 
In this case $Y_7$ is $M^{3,2}$.
Two of the $T_a$, which we take to be $T_1,T_2$, are associated with the circle fibration over $KE_4^+=\mathbb{C}P^2$, while sitting at the north and the south pole of the generic $S^2$ in \eqref{kahlereight}. Similarly,  
we can consider the product of the generic $S^2$ with the three two-spheres associated with the toric divisors
of $\mathbb{C}P^2$, leading to three more $T_a$ which we label $T_4,T_5,T_6$.
We then find
\begin{align}R_1 \, = \, R_2 \, =  \,  \frac{x}{2+x}N\,,\qquad
R_3 \,  = \,  R_4 \,  = \,   R_5 \, = \,  \frac{4}{3}\frac{1}{2+x}N\,,
\end{align}
and again we have $\sum_a R_a=2N$. 
The labelling we have chosen is consistent with the toric data given by
\begin{align}
\label{M32toricdata2}
\v_1 \,  &=\,  (1,0,0,0)~, \quad \v_2 \, = \, (1,0,0,2)~, \quad \v_3 \, = \, (1,1,0,0)~, \nn\\
\quad \v_4 \, &= \, (1,0,1,0)~,\quad \v_5 \, = \, (1,-1,-1,3)\, \,.
\end{align} 
We find that the condition \eqref{keyvRrel2}, which is valid for this example,
implies $\vec{b}=(1,0,0,1)$ for any $x$.
In addition the fluxes $M_1,M_2$ are associated with $\tilde N$ in \eqref{enssother} while 
$M_a$, $a=3,4,5$ are associated with the $N_i$, $i=1$
with $n_1=1$. 
Thus we have
\begin{align}
M_1& \, =  \, M_2 \, = \, \frac{3x}{(1+2 x)(2+x)}(g-1)N\,,\nn\\
M_3 \, = \, M_4& \, = \, M_5 \, = \, \frac{4}{3}\frac{1+x+x^2}{(2+x)(1+2x)}(g-1)N\,.
\end{align}
The condition \eqref{vMrelation2}
then implies
$\vec{n}=2(1-g)(1,0,0,1)$. Notice that for all values of $x$ we again have
$n_i  \ =   \  \frac{n_1}{b_1} b_i$, as in the universal twist
\eqref{twistb}. For the special value of $x=1$ we also have $R_a=M_a/(g-1)=N/3$, for $a=1,2$ and
$R_a=M_a/(g-1)=4N/9$, for $a=3,4,5$; the proportionality of the R-charges and the fluxes
is an additional condition for the universal twist solutions in section \ref{unitwist}.

In both of the above examples, the R-symmetry vector $\vec{b}$ had rational entries and hence, as far as the
geometry is concerned, the R-symmetry foliation is (quasi-)regular. Furthermore, when $x$ is rational all of the R-charges
are also rational. However, when $x$ is irrational, and satisfying \eqref{xrest}, the R-charges are irrational numbers. 


\providecommand{\href}[2]{#2}\begingroup\raggedright\endgroup

\end{document}